\documentstyle[aps,epsf,floats]{revtex}

\begin{document}
\author{Y.M. Vilk$^{*}$}
\address{Materials Science Division, Bldg. 223, 9700 S. Case Ave.}
\address{Argonne National Laboratory, Argonne IL 60439\\
$^{*}$ yvilk@hexi.msd.anl.gov}
\address{and}
\address{D\'{e}partement de physique and Centre de recherche en physique du solide.}
\address{Universit\'{e} de Sherbrooke, Sherbrooke, Qu\'{e}bec, Canada J1K 2R1}
\author{A.--M.S. Tremblay$^{+}$}
\address{D\'{e}partement de physique and Centre de recherche en physique du solide.}
\address{Universit\'{e} de Sherbrooke, Sherbrooke, Qu\'{e}bec, Canada J1K 2R1\\
$^{+}$ tremblay@physique.usherb.ca}
\title{Non-perturbative many-body approach to the Hubbard model and single-particle
pseudogap.}
\date{20 February 1997}
\maketitle

\begin{abstract}
A new approach to the single-band Hubbard model is described in the general
context of many-body theories. It is based on enforcing conservation laws,
the Pauli principle and a number of crucial sum-rules. More specifically,
spin and charge susceptibilities are expressed, in a conserving
approximation, as a function of two irreducible vertices whose values are
found by imposing the local Pauli principle $\left\langle n_{\uparrow
}^2\right\rangle =\left\langle n_{\uparrow }\right\rangle $ as well as the
local-moment sum-rule and consistency with the equations of motion in a
local-field approximation. The Mermin-Wagner theorem in two dimensions is
automatically satisfied. The effect of collective modes on single-particle
properties is then obtained by a paramagnon-like formula that is consistent
with the two-particle properties in the sense that the potential energy
obtained from $Tr\Sigma G$ is identical to that obtained using the
fluctuation-dissipation theorem for susceptibilities. Since there is no
Migdal theorem controlling the effect of spin and charge fluctuations on the
self-energy, the required vertex corrections are included. It is shown that
the theory is in quantitative agreement with Monte Carlo simulations for
both single-particle and two-particle properties. The theory predicts a
magnetic phase diagram where magnetic order persists away from half-filling
but where ferromagnetism is completely suppressed. Both quantum-critical and
renormalized-classical behavior can occur in certain parameter ranges. It is
shown that in the renormalized classical regime, spin fluctuations lead to
precursors of antiferromagnetic bands (shadow bands) and to the destruction
of the Fermi-liquid quasiparticles in a wide temperature range above the
zero-temperature phase transition. The upper critical dimension for this
phenomenon is three. The analogous phenomenon of pairing pseudogap can occur
in the attractive model in two dimensions when the pairing fluctuations
become critical. Simple analytical expressions for the self-energy are
derived in both the magnetic and pairing pseudogap regimes. Other
approaches, such as paramagnon, self-consistent fluctuation exchange
approximation (FLEX), and pseudo-potential parquet approaches are critically
compared. In particular, it is argued that the failure of the FLEX
approximation to reproduce the pseudogap and the precursors AFM bands in the
weak coupling regime and the Hubbard bands in the strong coupling regime is
due to inconsistent treatment of vertex corrections in the expression for
the self-energy. Treating the spin fluctuations as if there was a Migdal's
theorem can lead not only to quantitatively wrong results but also to
qualitatively wrong predictions, in particular with regard to the the
single-particle pseudogap.
\end{abstract}


\section{Introduction}

Understanding all the consequences of the interplay between band structure
effects and electron-electron interactions remains one of the present-day
goals of theoretical solid-state Physics. One of the simplest model that
contains the essence of this problem is the Hubbard model. In the more than
thirty years\cite{Hubbard}\cite{Kanamori} since this model was formulated,
much progress has been accomplished. In one dimension,\cite{VoitR}\cite
{Haldane} various techniques such as diagrammatic resummations,\cite{dzyalo}
bosonization,\cite{emery} renormalization group\cite{Bourbon2}\cite{geo} and
conformal approaches\cite{frahm}\cite{Senechal} have lead to a very detailed
understanding of correlation functions, from weak to strong coupling.
Similarly, in infinite dimensions a dynamical mean-field theory\cite{Georges}
leads to an essentially exact solution of the model, although many results
must be obtained by numerically solving self-consistent integral equations.
Detailed comparisons with experimental results on transition-metal oxides
have shown that three-dimensional materials can be well described by the
infinite-dimensional self-consistent mean-field approach.\cite{Georges}
Other methods, such as slave-boson\cite{SlaveBoson} or slave-fermion\cite
{SlaveFermion} approaches, have also allowed one to gain insights into the
Hubbard model through various mean-field theories corrected for
fluctuations. In this context however, the mean-field theories are not based
on a variational principle. Instead, they are generally based on expansions
in the inverse of a degeneracy parameter,\cite{RevueBickers} such as the
number of fermion flavors $N$, where $N$ is taken to be large despite the
fact that the physical limit corresponds to a small value of this parameter,
say $N=2$. Hence these theories must be used in conjunction with other
approaches to estimate their limits of validity.\cite{BoiesNagaoka}
Expansions around solvable limits have also been explored.\cite{Boies}
Finally, numerical solutions,\cite{DagottoRevue} with proper account of
finite-size effects, can often provide a way to test the range of validity
of approximation methods independently of experiments on materials that are
generally described by much more complicated Hamiltonians.

Despite all this progress, we are still lacking reliable theoretical methods
that work in arbitrary space dimension. In two dimensions in particular, it
is believed that the Hubbard model may hold the key to understanding normal
state properties of high-temperature superconductors. But even the simpler
goal of understanding the magnetic phase diagram of the Hubbard model in two
dimensions is a challenge. Traditional mean-field techniques, or even
slave-boson mean-field approaches, for studying magnetic instabilities of
interacting electrons fail in two dimensions. The Random Phase Approximation
(RPA) for example does not satisfy the Pauli principle, and furthermore it
predicts finite temperature antiferromagnetic or spin density wave (SDW)
transitions while this is forbidden by the Mermin-Wagner theorem. Even
though one can study universal critical behavior using various forms of
renormalization group treatments\cite{Chakravarty}\cite{Sachdev}\cite
{Chubukov}\cite{Hertz}\cite{Millis} or through the
self-consistent-renormalized approach of Moriya\cite{moriya} which all
satisfy the Mermin-Wagner theorem in two dimensions, cutoff-dependent scales
are left undetermined by these approaches. This means that the range of
interactions or fillings for which a given type of ground-state magnetic
order may appear is left undetermined.

Amongst the recently developed theoretical methods for understanding both
collective and single-particle properties of the Hubbard model, one should
note the fluctuation exchange approximation\cite{FLEX} (FLEX) and the
pseudo-potential parquet approach.\cite{parquet} The first one, FLEX, is
based on the idea of conserving approximations proposed by Baym and Kadanoff.%
\cite{Baym}\cite{BaymKadanoff} This approach starts with a set of skeleton
diagrams for the Luttinger-Ward functional\cite{Luttinger} to generate a
self-energy that is computed self-consistently. The choice of initial
diagrams however is arbitrary and left to physical intuition. In the
pseudo-potential parquet approach, one parameterizes response functions in
all channels, and then one iterates crossing-symmetric many-body integral
equations. While the latter approach partially satisfies the Pauli
principle, it violates conservation laws. The opposite is true for FLEX.

In this paper, we present the formal aspects of a new approach that we have
recently developed for the Hubbard model \cite{Vilk}\cite{Vilk2}. The
approach is based on enforcing sum rules and conservation laws, rather than
on diagrammatic perturbative methods that are not valid for interaction $U$
larger than hopping $t$. We first start from a Luttinger-Ward functional
that is parameterized by two irreducible vertices $U_{sp}$ and $U_{ch}$ that
are local in space-time. This generates RPA-like equations for spin and
charge fluctuations that are conserving. The local-moment sum rule, local
charge sum rule, and the constraint imposed by the Pauli principle, $%
\left\langle n_{\uparrow }^2\right\rangle =\left\langle n_{\uparrow
}\right\rangle $ then allow us to find the vertices as a function of double
occupancy $\left\langle n_{\uparrow }n_{\downarrow }\right\rangle $ (see
Eqs.(\ref{Charge}) and (\ref{Spin})). Since $\left\langle n_{\uparrow
}n_{\downarrow }\right\rangle $ is a local quantity it depends very little
on the size of the system and, in principle, it could be obtained reliably
using numerical methods, such as for example Monte Carlo simulations. Here,
however, we adopt another approach and find $\left\langle n_{\uparrow
}n_{\downarrow }\right\rangle $ self-consistently\cite{Vilk} without any
input from outside the present theory. This is done by using an {\it ansatz}
Eq. (\ref{Usp}) for the double-occupancy $\left\langle n_{\uparrow
}n_{\downarrow }\right\rangle $ that has been inspired by ideas from the
local field approach of Singwi et al.\cite{singwi}. Once we have the spin
and charge fluctuations, the next step is to use them to compute a new
approximation, Eq.(\ref{param}), for the single-particle self-energy. This
approach to the calculation of the effect of collective modes on
single-particle properties\cite{Vilk2} is similar in spirit to paramagnon
theories.\cite{Enz} Contrary to these approaches however, we {\it do}
include vertex corrections in such a way that, if $\Sigma ^{\left( 1\right)
} $ is our new approximation for the self-energy while $G^{\left( 0\right) }$
is the initial Green's function used in the calculation of the collective
modes, and $\left\langle n_{\uparrow }n_{\downarrow }\right\rangle $ is the
value obtained from spin and charge susceptibilities, then $\frac 12Tr\left[
\Sigma ^{\left( 1\right) }G^{\left( 0\right) }\right] =U\left\langle
n_{\uparrow }n_{\downarrow }\right\rangle $ is satisfied exactly. The extent
to which $\frac 12Tr\left[ \Sigma ^{\left( 1\right) }G^{\left( 1\right)
}\right] $ (computed with $G^{\left( 1\right) }$ instead of $G^{\left(
0\right) }$) differs from $U\left\langle n_{\uparrow }n_{\downarrow
}\right\rangle $ can then be used both as an internal accuracy check and as
a way to improve the vertex corrections.

If one is interested only in two-particle properties, namely spin and charge
fluctuations, then this approach has the simple physical appeal of RPA but
it satisfies key constraints that are always violated by RPA, namely the
Mermin-Wagner theorem and the Pauli principle. To contrast it with usual
RPA, that has a self-consistency only at the single-particle level, we call
it the Two-Particle Self-Consistent approach (TPSC).\cite{Vilk}\cite{Vilk2}%
\cite{Dare} The TPSC gives a {\it quantitative} description of the Hubbard
model not only far from phase transitions, but also upon entering the
critical regime. Indeed we have shown quantitative agreement with Monte
Carlo simulations of the nearest-neighbor\cite{Vilk} and next-nearest
neighbor\cite{Veilleux} Hubbard model in two dimensions. Quantitative
agreement is also obtained as one enters the narrow critical regime
accessible in Monte Carlo simulations. We also have shown\cite{Dare} in full
generality that the TPSC approach gives the $n\rightarrow \infty $ limit of
the $O\left( n\right) $ model, while $n=3$ is the physically correct
(Heisenberg) limit. In two dimensions, we then recover both quantum-critical%
\cite{Sachdev} and renormalized classical\cite{Chakravarty} regimes to
leading order in $1/n$. Since there is no arbitrariness in cutoff, given a
microscopic Hubbard model no parameter is left undetermined. This allows us
to go with the same theory from the non-critical to the beginning of the
critical regime, thus providing quantitative estimates for the magnetic
phase diagram of the Hubbard model, not only in two dimensions but also in
higher dimensions\cite{Dare}.

The main limitation of the approach presented in this paper is that it is
valid only from weak to intermediate coupling. The strong-coupling case
cannot be treated with frequency-independent irreducible vertices, as will
become clear later. However, a suitable ansatz for these irreducible
vertices in a Luttinger-Ward functional might allow us to apply our general
scheme to this limit as well.

Our approach predicts\cite{Vilk2} that in two dimensions, Fermi liquid
quasiparticles disappear in the renormalized classical regime $\xi
_{AFM}\propto \exp (const/T)$, which always precedes the zero-temperature
phase transition in two-dimensions. In this regime the antiferromagnetic
correlation length becomes larger than the single-particle thermal de
Broglie wave length $\xi _{th}(=v_F/T),$ leading to the destruction of Fermi
liquid quasiparticles with a concomitant appearance of precursors of
antiferromagnetic bands (``shadow bands'') with no quasi-particle peak
between them. We stress the crucial role of the classical thermal spin
fluctuations and low dimensionality for the existence of this effect and
contrast our results with the earlier results of Kampf and Schrieffer\cite
{Kampf} who used a susceptibility separable in momentum and frequency $\chi
_{sp}=f\left( {\bf q}\right) g(\omega )$. The latter form of $\chi
_{sp}=f\left( {\bf q}\right) g(\omega )$ leads to an artifact that
dispersive precursors of antiferromagnetic bands can exist at $T=0$ (for
details see \cite{Yury3}). We also contrast our results with those obtained
in the fluctuation exchange approximation (FLEX), which includes
self-consistency in the single particle propagators but neglects the
corresponding vertex corrections. The latter approach predicts only the
so-called ``shadow feature'' \cite{Yury3,Langer} which is an enhancement in
the incoherent background of the spectral function due to antiferromagnetic
fluctuations. However, it does not predict\cite{Serene} the existence of
``shadow bands'' in the renormalized classical regime. These bands occur
when the condition $\omega -\epsilon _{{\bf k}}-\Sigma _\sigma ({\bf k,}%
\omega )+\mu =0$ is satisfied. FLEX also predicts no pseudogap in the
spectral function $A({\bf k}_F,\omega )$ at half-filling \cite{Serene}. By
analyzing temperature and size dependence of the Monte Carlo data and
comparing them with the theoretical calculations, we argue that the Monte
Carlo data supports our conclusion that the precursors of antiferromagnetic
bands and the pseudogap do appear in the renormalized classical regime. We
believe that the reason for which the FLEX approximation fails to reproduce
this effect is essentially the same reason for which it fails to reproduce
Hubbard bands in the strong coupling limit. More specifically, the failure
is due to an inconsistent treatment of vertex corrections in the self-energy 
{\it ansatz}. Contrary to the electron-phonon case, these vertex corrections
have a strong tendency to cancel the effects of using dressed propagators in
the expression for the self-energy.

Recently, there have been very exciting developments in photoemission
studies of the High-$T_c$ materials\cite{Shen,Campusano} that show the
opening of the pseudogap in single particle spectra above the
superconducting phase transition. At present, there is an intense debate
about the physical origin of this phenomena and, in particular, whether it
is of magnetic or of pairing origin. From the theoretical point of view
there are a lot of formal similarities in the description of
antiferromagnetism in repulsive models and superconductivity in attractive
models. In Sec.(\ref{SecDestruction}) we use this formal analogy to obtain a
simple analytical expressions for the self-energy in the regime dominated by
critical pairing fluctuations. We then point out on the similarities and
differences in the spectral function in the case of magnetic and pairing
pseudogaps.

Our approach has been described in simple physical terms in Refs.\cite{Vilk}
and \cite{Vilk2}. The plan of the present paper is as follows. After
recalling the model and the notation, we present our theory in Sec.(\ref
{SecFormal}). There we point out which exact requirements of many-body
theory are satisfied, and which are violated. Before Sec.(\ref{SecFormal}),
the reader is urged to read Appendix~(\ref{SecSumRules}) that contains a
summary of sum rules, conservation laws and other exact constraints.
Although this discussion contains many original results, it is not in the
main text since the more expert reader can refer to the Appendix as need be.
We also illustrate in this Appendix how an inconsistent treatment of the
self-energy and vertex corrections can lead to the violation of a number of
sum rules and inhibit the appearance of the Hubbard bands, a subject also
treated in Sec.(\ref{SecFLEX}). Section (\ref{SecMonteCarlo}) compares the
results of our approach and of other approaches to Monte Carlo simulations.
We study in more details in Sec.(\ref{SecDestruction}) the renormalized
classical regime at half-filling where, in two dimensions, Fermi liquid
quasiparticles are destroyed and replaced by precursors of antiferromagnetic
bands well before the $T=0$ phase transition. We also consider in this
section the analogous phenomenon of pairing pseudogap which can appear in
two dimensions when the pairing fluctuations become critical. The following
section Sec.(\ref{SecFLEX}) explains other attempts to obtain precursors of
antiferromagnetic bands and points out why approaches such as FLEX fail to
see the effect. We conclude in Sec.(\ref{SecValidity}) with a discussion of
the domain of validity of our approach and in Sec.(\ref{SecComparisons})
with a critical comparison with FLEX and pseudo-potential parquet
approaches, listing the weaknesses and strengths of our approach compared
with these. A more systematic description and critique of various many-body
approaches, as well as proofs of some of our results, appear in appendices.

\section{Model and definitions}

\label{SecModel}

We first present the model and various definitions. The Hubbard model is
given by the Hamiltonian, 
\begin{equation}  \label{Hubbard}
H=-\sum_{<ij>\sigma }t_{i,j}\left( c_{i\sigma }^{\dagger }c_{j\sigma
}+c_{j\sigma }^{\dagger }c_{i\sigma }\right) +U\sum_in_{i\uparrow
}n_{i\downarrow \,\,\,\,}.
\end{equation}
In this expression, the operator $c_{i\sigma }$ destroys an electron of spin 
$\sigma $ at site $i$. Its adjoint $c_{i\sigma }^{\dagger }$ creates an
electron and the number operator is defined by $n_{i\sigma }=$ $c_{i\sigma
}^{\dagger }c_{i\sigma }$. The symmetric hopping matrix $t_{i,j}$ determines
the band structure, which here can be arbitrary. Double occupation of a site
costs an energy $U$ due to the screened Coulomb interaction. We work in
units where $k_B=1$, $\hbar =1$ and the lattice spacing is also unity, $a=1$%
. As an example that occurs later, the dispersion relation in the $d$%
-dimensional nearest-neighbor model is given by 
\begin{equation}
\epsilon _{{\bf k}}=-2t\sum_{i=1}^d\left( \cos k_i\right) .
\end{equation}

\subsection{Single-particle propagators, spectral weight and self-energy.}

We will use a ``four''-vector notation $k\equiv \left( {\bf k,}ik_n\right) $
for momentum-frequency space, and $1\equiv \left( {\bf r}_1{\bf ,}\tau
_1\right) $ for position-imaginary time. For example, the definition of the
single-particle Green's function can be written as 
\begin{equation}
G_\sigma \left( 1{\bf ,}2\right) \equiv -\left\langle T_\tau c_{1\sigma
}\left( \tau _1\right) c_{2\sigma }^{\dagger }\left( \tau _2\right)
\right\rangle \equiv -\left\langle T_\tau c_\sigma \left( 1\right) c_\sigma
^{\dagger }(2)\right\rangle  \label{DefG}
\end{equation}
where the brackets $\left\langle {}\right\rangle $ represent a thermal
average in the grand canonical ensemble, $T_\tau $ is the time-ordering
operator, and $\tau $ is imaginary time. In zero external field and in the
absence of the symmetry breaking $G_\sigma \left( 1{\bf ,}2\right) =G_\sigma
\left( 1{\bf -}2\right) $ and the Fourier-Matsubara transforms of the
Green's function are

\begin{equation}
G_\sigma \left( k\right) =\sum_{{\bf r}_1}e^{-i{\bf k\cdot r}_1}\int_0^\beta
d\tau \ e^{ik_n\tau _1}G_\sigma \left( {\bf r}_1{\bf ,}\tau _1\right) \equiv
\int d\left( 1\right) e^{-ik(1)}G_\sigma \left( 1\right)  \label{Gfur}
\end{equation}
\begin{equation}
G_\sigma \left( 1\right) =\frac TN\sum_ke^{ik\left( 1\right) }G_\sigma
\left( k\right) .
\end{equation}

As usual, experimentally observable retarded quantities are obtained from
the Matsubara ones by analytical continuation $ik_n\rightarrow \omega +i\eta
.$ In particular, the single-particle spectral weight $A\left( {\bf k,}%
\omega \right) $ is related to the single-particle propagator by 
\begin{equation}
G_\sigma \left( {\bf k,}ik_n\right) =\int \frac{d\omega }{2\pi }\frac{%
A_\sigma \left( {\bf k,}\omega \right) }{ik_n-\omega }  \label{SpectralG}
\end{equation}
\begin{equation}
A_\sigma \left( {\bf k,}\omega \right) =-2\text{Im}G_\sigma ^R\left( {\bf k,}%
\omega \right) .
\end{equation}
The self-energy obeys Dyson's equation, leading to 
\begin{equation}
G_\sigma \left( {\bf k,}ik_n\right) =\frac 1{ik_n-\left( \epsilon _{{\bf k}%
}-\mu \right) -\Sigma _\sigma \left( {\bf k,}ik_n\right) }.  \label{Gself}
\end{equation}
It is convenient to use the following notation for real and imaginary parts
of the analytically continued retarded self-energy 
\begin{equation}
\Sigma _\sigma ^R\left( {\bf k,}ik_n\rightarrow \omega +i\eta \right)
=\Sigma _\sigma ^{\prime }\left( {\bf k,}\omega \right) +i\Sigma _\sigma
^{\prime \prime }\left( {\bf k,}\omega \right) .  \label{SelfReIm}
\end{equation}
Causality and positivity of the spectral weight imply that 
\begin{equation}
\Sigma _\sigma ^{\prime \prime }\left( {\bf k,}\omega \right) <0.
\end{equation}

Finally, let us point out that for nearest-neighbor hopping, the Hamiltonian
is particle-hole symmetric at half-filling, $\left( c_{{\bf k}\sigma
}\rightarrow c_{{\bf k+Q}\sigma }^{\dagger }\ ;\ c_{{\bf k}\sigma }^{\dagger
}\rightarrow c_{{\bf k+Q}\sigma }\right) $ with ${\bf Q=}\left( \pi ,\pi
\right) ,$ implying that $\mu =U/2$ and that, 
\begin{equation}
G_\sigma \left( {\bf k},\tau \right) =-G_\sigma \left( {\bf k+Q},-\tau
\right)
\end{equation}
\begin{equation}
\left[ \Sigma \left( {\bf k},ik_n\right) -\frac U2\right] =-\left[ \Sigma
\left( {\bf k+Q},-ik_n\right) -\frac U2\right] .  \label{phs}
\end{equation}

\subsection{Spin and charge correlation functions}

We shall be primarily concerned with spin and charge fluctuations, which are
the most important collective modes in the repulsive Hubbard model. Let the
charge and $z$ components of the spin operators at site $i$ be given
respectively by 
\begin{equation}
\rho _i\left( \tau \right) \equiv n_{i\uparrow }\left( \tau \right)
+n_{i\downarrow }\left( \tau \right)
\end{equation}
\begin{equation}
S_i^z\equiv n_{i\uparrow }\left( \tau \right) -n_{i\downarrow }\left( \tau
\right) .
\end{equation}
The time evolution here is again that of the Heisenberg representation in
imaginary time.

The charge and spin susceptibilities in imaginary time are the responses to
perturbations applied in imaginary-time. For example, the linear response of
the spin to an external field that couples linearly to the $z$ component 
\begin{equation}
e^{-\beta H}\rightarrow e^{-\beta H}T_\tau e^{\int d\tau S_i^z\left( \tau
^{\prime }\right) \phi _i^S\left( \tau ^{\prime }\right) }
\end{equation}
is given by 
\begin{equation}
\chi _{sp}\left( {\bf r}_i-{\bf r}_j,\tau _i-\tau _j\right) =\frac{\delta
\left\langle S_j\left( \tau _j\right) \right\rangle }{\delta \phi _i^S\left(
\tau _i\right) }=\left\langle T_\tau S_i^z\left( \tau _i\right) S_j^z\left(
\tau _j\right) \right\rangle .  \label{SusSpin}
\end{equation}
In an analogous manner, for charge we have 
\begin{equation}
\chi _{ch}\left( {\bf r}_i-{\bf r}_j,\tau _i-\tau _j\right) =\frac{\delta
\left\langle \rho _j\left( \tau _j\right) \right\rangle }{\delta \phi
_i^\rho \left( \tau _i\right) }=\left\langle T_\tau \rho _i\left( \tau
_i\right) \rho _j\left( \tau _j\right) \right\rangle -n^2.  \label{SusCharge}
\end{equation}
Here $n\equiv \left\langle \rho _i\right\rangle $ is the filling so that the
disconnected piece is denoted $n^2$. It is well known that when analytically
continued, these susceptibilities give physical retarded and advanced
response functions. In fact, the above two expressions are the
imaginary-time version of the fluctuation-dissipation theorem.

The expansion of the above functions in Matsubara frequencies uses even
frequencies. Defining the subscript $ch,sp$ to mean either charge or spin,
we have 
\begin{equation}
\chi _{ch,sp}\left( {\bf q},iq_n\right) =\int \frac{d\omega ^{\prime }}\pi 
\frac{\chi _{ch,sp}^{\prime \prime }\left( {\bf q,}\omega ^{\prime }\right) 
}{\omega ^{\prime }-iq_n}
\end{equation}
\begin{equation}
\chi _{ch}^{\prime \prime }\left( {\bf q,}t\right) =\frac 12\left\langle
\left[ \rho _{{\bf q}}\left( t\right) ,\rho _{-{\bf q}}\left( 0\right)
\right] \right\rangle \quad ;\quad \chi _{sp}^{\prime \prime }\left( {\bf q,}%
t\right) =\frac 12\left\langle \left[ S_{{\bf q}}^z\left( t\right) ,S_{-{\bf %
q}}^z\left( 0\right) \right] \right\rangle .
\end{equation}
The fact that $\chi _{ch,sp}^{\prime \prime }\left( {\bf q,}\omega ^{\prime
}\right) \ $is real and odd in frequency in turn means that $\chi
_{ch,sp}\left( {\bf q},iq_n\right) $ is real 
\begin{equation}
\chi _{ch,sp}\left( {\bf q},iq_n\right) =\int \frac{d\omega ^{\prime }}\pi 
\frac{\omega ^{\prime }\chi _{ch,sp}^{\prime \prime }\left( {\bf q,}\omega
^{\prime }\right) }{\left( \omega ^{\prime }\right) ^2+\left( q_n\right) ^2}
\label{SpectCh}
\end{equation}
a convenient feature for numerical calculations. The high-frequency
expansion has $1/q_n^2$ as a leading term so that there is no discontinuity
in $\chi _{ch,sp}\left( {\bf q},\tau \right) $as $\tau \rightarrow 0,$
contrary to the single-particle case.

\section{Formal derivation}

\label{SecFormal}

To understand how to satisfy as well as possible the requirements imposed on
many-body theory by exact results, such as those in Appendix (\ref
{SecSumRules}), it is necessary to start from a general non-perturbative
formulation of the many-body problem. We thus first present a general
approach to many-body theory that is set in the framework introduced by
Martin and Schwinger\cite{Martin}, Luttinger and Ward\cite{Luttinger} and
Kadanoff and Baym\cite{BaymKadanoff}\cite{Baym}. This allows one to see
clearly the structure of the general theory expressed in terms of the
one-particle irreducible self-energy and of the particle-hole irreducible
vertices. These quantities represent projected propagators and there is a
great advantage in doing approximations for these quantities rather than
directly on propagators.

Our own approximation to the Hubbard model is then described in the
subsection that follows the formalism. In our approach, the irreducible
quantities are determined from various consistency requirements. The reader
who is interested primarily in the results rather than in formal aspects of
the theory can skip the next subsection and refer back later as needed.

\subsection{General formalism}

Following Kadanoff and Baym,\cite{BaymKadanoff} we introduce the generating
function for the Green's function 
\begin{equation}
\ln Z\left[ \phi \right] =\ln \left\langle T_\tau e^{-c_{\overline{\sigma }%
}^{\dagger }\left( \overline{1}\right) c_{\overline{\sigma }}\left( 
\overline{2}\right) \phi _{\overline{\sigma }}\left( \overline{1},\overline{2%
}\right) }\right\rangle  \label{Z}
\end{equation}
where, as above, a bar over a number means summation over position and
imaginary time and, similarly, a bar over a spin index means a sum over that
spin index. The quantity $Z$ is a functional of $\phi _\sigma $, the
position and imaginary-time dependent field. $Z$ reduces to the usual
partition function when the field $\phi _\sigma $ vanishes. The one-particle
Green's function in the presence of this external field is given by 
\begin{equation}
G_\sigma \left( 1,2;\left[ \phi \right] \right) =-\frac{\delta \ln Z\left[
\phi \right] }{\delta \phi _\sigma \left( 2,1\right) }  \label{GdePhi}
\end{equation}
and, as shown by Kadanoff and Baym, the inverse Green's function is related
to the self-energy through 
\begin{equation}
G^{-1}=G_0^{-1}-\phi -\Sigma .  \label{Dyson3}
\end{equation}
The self-energy in this expression is a functional of $\phi $.

Performing a Legendre transform on the generating functional $\ln Z\left[
\phi \right] $ in Eq.(\ref{Z}) with the help of the last two equations, one
can find a functional $\Phi \left[ G\right] $ of $G$ that acts as a
generating function for the self-energy 
\begin{equation}
\Sigma _\sigma \left( 1,2;\left[ G\right] \right) =\frac{\delta \Phi \left[
G\right] }{\delta G_\sigma \left( 2,1\right) }.  \label{SelfLutt}
\end{equation}
The quantity $\Phi \left[ G\right] $ is the Luttinger-Ward functional.\cite
{Luttinger} Formally, it is expressed as the sum of all connected skeleton
diagrams, with appropriate counting factors. Conserving approximations start
from a subset of all possible connected diagrams for $\Phi \left[ G\right] $
to generate both the self-energy and the irreducible vertices entering the
integral equation obeyed by response functions. These response functions are
then guaranteed to satisfy the conservation laws. They obey integral
equations containing as irreducible vertices 
\begin{equation}
\Gamma _{\sigma \sigma ^{\prime }}^{ir}(1,2;3,4)\equiv \frac{\delta \Sigma
_\sigma \left( 1,2;\left[ G\right] \right) }{\delta G_{\sigma ^{\prime
}}\left( 3,4\right) }=\frac{\delta ^2\Phi \left[ G\right] }{\delta G_\sigma
\left( 2,1\right) \delta G_{\sigma ^{\prime }}\left( 3,4\right) }=\Gamma
_{\sigma ^{\prime }\sigma }^{ir}(4,3;2,1)  \label{IrrLutt}
\end{equation}

A complete and exact picture of one- and two-particle properties is obtained
then as follows. First, the generalized susceptibilities $\chi _{\sigma
\sigma ^{\prime }}(1,3;2)\equiv -\delta G_\sigma \left( 1,3\right) /\delta
\phi _{\sigma ^{\prime }}\left( 2^{+},2\right) $ are calculated by taking
the functional derivative of $GG^{-1}$ and using the Dyson equation (\ref
{Dyson3}) to compute $\delta G^{-1}/\delta \phi $. One obtains\cite
{BaymKadanoff} 
\begin{equation}
\chi _{\sigma \sigma ^{\prime }}(1,3;2)=-G_\sigma \left( 1,2\right) \delta
_{\sigma ,\sigma ^{\prime }}G_\sigma \left( 2,3\right) +G_\sigma \left( 1,%
\overline{2}\right) \Gamma _{\sigma \overline{\sigma }}^{ir}(\overline{2},%
\overline{3};\overline{4},\overline{5})\chi _{\overline{\sigma }\sigma
^{\prime }}(\overline{4},\overline{5};2)G_\sigma \left( \overline{3},3\right)
\label{1a}
\end{equation}
where one recognizes the Bethe-Salpeter equation for the three-point
susceptibility in the particle-hole channel. The second equation that we
need is automatically satisfied in an exact theory. It relates the
self-energy to the response function just discussed through the equation 
\begin{equation}
\Sigma _\sigma \left( 1,2\right) =Un_{-\sigma }\delta \left( 1-2\right)
+UG_\sigma \left( 1,\overline{2}\right) \Gamma _{\sigma \sigma ^{\prime
}}^{ir}(\overline{2},2;\overline{4},\overline{5})\chi _{\sigma ^{\prime
}-\sigma }(\overline{4},\overline{5};1)  \label{2a}
\end{equation}
which is proven in Appendix (\ref{Proofs}).

\begin{figure}%
\centerline{\epsfxsize 6cm \epsffile{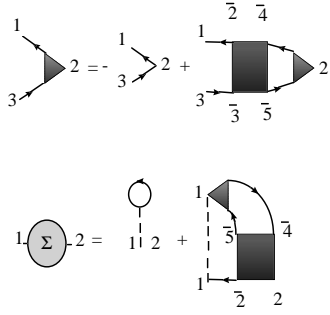}}%
\caption{The first line is a diagramatic representation of  the Bethe-Salpeter equation 
Eq.(\protect\ref{1a}) for the three point susceptibility and the second line is the corresponding
equation Eq.(\protect\ref{2a}) for the self-energy. In the Hubbard model, the Fock contribution is absent, 
but in general it should be there.  
Solid lines are Green's functions
and dashed lines represent the contact interaction U. The triangle is the
three point vertex, while the three-point susceptibility $\chi(1,3;2)$ is the 
triangle along with the attached Green's function.  The usual two-point susceptibility is
obtained by identifying points $1$ and $3$ in the Bethe-Salpeter equation. The 
rectangular box is the irreducible four-point vertex in the selected particle-hole
channel.}%
\label{FigKadanoffBaym}%
\end{figure}

The diagrammatic representation of these two equations Eqs.(\ref{1a})(\ref
{2a}) appearing in Fig.(\ref{FigKadanoffBaym}) may make them look more
familiar. Despite this diagrammatic representation, we stress that this is
only for illustrative purposes. The rest of our discussion will not be
diagrammatic.

Because of the spin-rotational symmetry the above equations Eqs.(\ref{1a})
and (\ref{2a}) can be decoupled into symmetric (charge) and antisymmetric
(spin) parts, by introducing spin and charge irreducible vertices and
generalized susceptibilities: 
\begin{equation}
\Gamma _{ch}\equiv \Gamma _{\uparrow \downarrow }^{ir}+\Gamma _{\uparrow
\uparrow }^{ir}\quad ;\quad \Gamma _{sp}\equiv \Gamma _{\uparrow \downarrow
}^{ir}-\Gamma _{\uparrow \uparrow }^{ir}
\end{equation}
\begin{equation}
\chi _{ch}\equiv 2(\chi _{_{\uparrow \downarrow }}+\chi _{_{\uparrow
\uparrow }})\quad ;\quad \chi _{sp}\equiv 2(\chi _{_{\uparrow \uparrow
}}-\chi _{_{\uparrow \downarrow }})
\end{equation}
The usual two-point susceptibilities are obtained from the generalized ones
as $\chi _{sp,ch}(1,2)=\chi _{sp,ch}(1,1^{+};2)$. The equation Eq.(\ref{1a})
for the generalized spin susceptibility leads to

\begin{equation}
\chi _{sp}(1,3;2)=-2G\left( 1,2\right) G\left( 2,3\right) -\Gamma _{sp}(%
\overline{2},\overline{3};\overline{4},\overline{5})G\left( 1,\overline{2}%
\right) G\left( \overline{3},3\right) \chi _{sp}(\overline{4},\overline{5};2)
\label{1}
\end{equation}
and similarly for charge, but with the plus sign in front of the second term.

Finally, one can write the exact equation Eq.(\ref{2a}) for the self-energy
in terms of the response functions as 
\begin{equation}
\Sigma _\sigma \left( 1,2\right) =Un_{-\sigma }\delta \left( 1-2\right) +%
\frac U4\left[ \Gamma _{sp}(\overline{2},2;\overline{4},\overline{5})\chi
_{sp}(\overline{4},\overline{5};1)+\Gamma _{ch}(\overline{2},2;\overline{4},%
\overline{5})\chi _{ch}(\overline{4},\overline{5};1)\right] G_\sigma \left(
1,\overline{2}\right) .  \label{2}
\end{equation}

Our two key equations are thus those for the three-point susceptibilities,
Eq.(\ref{1}), and for the self-energy, Eq.(\ref{2}). It is clear from the
derivation in Appendix~(\ref{Proofs}) that these equations are intimately
related.

\subsection{Approximations through local irreducible vertices.}

\subsubsection{Conserving approximation for the collective modes.}

In formulating approximation methods for the many-body problem, it is
preferable to confine our ignorance to high-order correlation functions
whose detailed momentum and frequency dependence is not singular and whose
influence on the low energy Physics comes only through averages over
momentum and frequency. We do this here by parameterizing the Luttinger-Ward
functional by two constants $\Gamma _{\uparrow \downarrow }^{ir}$ and $%
\Gamma _{\uparrow \uparrow }^{ir}$. They play the role of particle-hole
irreducible vertices that are eventually determined by enforcing sum rules
and a self-consistency requirement at the two-particle level. In the present
context, this functional can be also considered as the interacting part of a
Landau functional. The {\it ansatz }is, 
\begin{equation}
\Phi \left[ G\right] =\frac 12G_{\overline{\sigma }}\left( \overline{1},%
\overline{1}^{+}\right) \Gamma _{\overline{\sigma }\overline{\sigma }%
}^{ir}G_{\overline{\sigma }}\left( \overline{1},\overline{1}^{+}\right) +%
\frac 12G_{\overline{\sigma }}\left( \overline{1},\overline{1}^{+}\right)
\Gamma _{\overline{\sigma }-\overline{\sigma }}^{ir}G_{-\overline{\sigma }%
}\left( \overline{1},\overline{1}^{+}\right) .  \label{Trial}
\end{equation}
As in every conserving approximation, the self-energy and irreducible
vertices are obtained from functional derivatives as in Eq.(\ref{SelfLutt})
and Eq.(\ref{IrrLutt}) and then the collective modes are computed from the
Bethe-Salpeter equation Eq.(\ref{1}). The above Luttinger-Ward functional
gives a momentum and frequency independent self-energy\cite{NoteSelf0}, that
can be absorbed in a chemical potential shift. From the Luttinger-Ward
functional, one also obtains two local particle-hole irreducible vertices $%
\Gamma _{\sigma \sigma }^{ir}$ and $\Gamma _{\sigma -\sigma }^{ir}$ 
\begin{equation}
\Gamma _{\sigma \overline{\sigma }}^{ir}(2,3;4,5)\equiv \frac{\delta \Sigma
_\sigma \left( 2,3\right) }{\delta G_{\sigma ^{\prime }}\left( 4,5\right) }%
=\delta \left( 2-5\right) \delta \left( 3-4\right) \delta \left(
4^{+}-5\right) \Gamma _{\sigma \sigma ^{\prime }}^{ir}.  \label{Irreducible}
\end{equation}
We denote the corresponding local spin and charge irreducible vertices as 
\begin{equation}
U_{sp}\equiv \Gamma _{\sigma -\sigma }^{ir}-\Gamma _{\sigma \sigma
}^{ir}\quad ;\quad U_{ch}\equiv \Gamma _{\sigma -\sigma }^{ir}+\Gamma
_{\sigma \sigma }^{ir}.  \label{defVertex}
\end{equation}

Notice now that there are only two equal-time, equal-point ({\it i.e.}
local) two-particle correlation functions in this problem, namely $\langle
n_{\uparrow }n_{\downarrow }\rangle $ and $\langle n_{\uparrow }^2\rangle
=\langle n_{\downarrow }^2\rangle =\langle n_{\downarrow }\rangle =n/2.$ The
last one is completely determined by the Pauli principle and by the known
filling factor, while $U\langle n_{\uparrow }n_{\downarrow }\rangle $ is the
expectation value of the interaction term in the Hamiltonian. Only one of
these two correlators, namely $U\langle n_{\uparrow }n_{\downarrow }\rangle $%
, is unknown. Assume for the moment that it is known$.$ Then, we can use the
two sum rules Eqs.(\ref{SusChargeSum}) and (\ref{SusSpinSum}) that follow
from the fluctuation-dissipation theorem and from the Pauli principle to
determine the two trial irreducible vertices from the known value of this
one key local correlation functions. In the present notation, these two sum
rules are 
\begin{equation}
\chi _{ch}\left( 1,1^{+}\right) =\frac TN\sum_{{\bf q}}\sum_{iq_n}\chi _{ch}(%
{\bf q,}iq_n)=\left\langle n_{\uparrow }\right\rangle +\left\langle
n_{\downarrow }\right\rangle +2\left\langle n_{\uparrow }n_{\downarrow
}\right\rangle -n^2  \label{sumCharge}
\end{equation}
\begin{equation}
\chi _{sp}\left( 1,1^{+}\right) =\frac TN\sum_{{\bf q}}\sum_{iq_n}\chi _{sp}(%
{\bf q,}iq_n)=\left\langle n_{\uparrow }\right\rangle +\left\langle
n_{\downarrow }\right\rangle -2\left\langle n_{\uparrow }n_{\downarrow
}\right\rangle  \label{sumSpin}
\end{equation}
and since the spin and charge susceptibilities entering these equations are
obtained by solving the Bethe-Salpeter Eq.(\ref{1}) with the constant
irreducible vertices Eqs.(\ref{Irreducible})(\ref{defVertex}) we have one
equation for each of the irreducible vertices 
\begin{equation}
n+2\langle n_{\uparrow }n_{\downarrow }\rangle -n^2=\frac TN\sum_q\frac{\chi
_0(q)}{1+\frac 12U_{ch}\chi _0(q)},  \label{Charge}
\end{equation}
\begin{equation}
n-2\langle n_{\uparrow }n_{\downarrow }\rangle =\frac TN\sum_{\widetilde{q}}%
\frac{\chi _0(q)}{1-\frac 12U_{sp}\chi _0(q)}.  \label{Spin}
\end{equation}
We used our usual short-hand notation for wave vector and Matsubara
frequency $q=({\bf q,}iq_n)$. Since the self-energy corresponding to our
trial Luttinger-Ward functional is constant, the irreducible
susceptibilities take their non-interacting value $\chi _0(q).$

The local Pauli principle $\langle n_{\downarrow }^2\rangle =\langle
n_{\downarrow }\rangle $ leads to the following important sum-rule 
\begin{equation}
\frac TN\sum_{{\bf q}}\sum_{iq_n}\left[ \chi _{sp}\left( {\bf q},iq_n\right)
+\chi _{ch}\left( {\bf q},iq_n\right) \right] =2n-n^2,
\label{SpinChargePauli0}
\end{equation}
which can be obtained by adding Eqs.(\ref{Spin}),(\ref{Charge}). This
sum-rule implies that effective interactions for spin $U_{sp}$ and charge $%
U_{ch}$ channels must be different from one another and hence that ordinary
RPA is inconsistent with the Pauli principle (for details see Appendix~(\ref
{SecConstTwoPart})).

Eqs.(\ref{Charge}) and (\ref{Spin}) determine $U_{sp}$ and $U_{ch}$ as a
function of double occupancy $\langle n_{\uparrow }n_{\downarrow }\rangle $.
Since double occupancy is a local quantity it depends little on the size of
the system. It could be obtained reliably from a number of approaches, such
as for example Monte Carlo simulations. However, there is a way to obtain
double-occupancy self-consistently\cite{Vilk} without input from outside of
the present theory. It suffices to add to the above set of equations the
relation 
\begin{equation}
U_{sp}=g_{\uparrow \downarrow }(0)\,U\quad ;\quad g_{\uparrow \downarrow
}(0)\equiv \frac{\langle n_{\uparrow }n_{\downarrow }\rangle }{\langle
n_{\downarrow }\rangle \langle n_{\uparrow }\rangle }.  \label{Usp}
\end{equation}
Eqs.(\ref{Spin}) and (\ref{Usp}) then define a set of self-consistent
equations for $U_{sp}$ that involve only two-particle quantities. This {\it %
ansatz} is motivated by a similar approximation suggested by Singwi et al.%
\cite{singwi} in the electron gas, which proved to be quite successful in
that case. On a lattice we will use it for $n\leq 1$. The case $n>1$ can be
mapped on the latter case using particle-hole transformation. In the context
of the Hubbard model with on-site repulsion, the physical meaning of Eq. (%
\ref{Usp}) is that the effective interaction in the most singular spin
channel, is reduced by the probability of having two electrons with the
opposite spins on the same site. Consequently, the {\it ansatz} reproduces
the Kanamori-Brueckner screening that inhibits ferromagnetism in the weak to
intermediate coupling regime (see also below). We want to stress, however,
that this {\it ansatz} is not a rigorous result like sum rules described
above. The plausible derivation of this {\it ansatz} can be found in Refs.%
\cite{singwi}, \cite{Vilk} as well as, in the present notation, in Appendix (%
\ref{Ansatz}).

We have called this approach Two-Particle Self-Consistent to contrast it
with other conserving approximations like Hartree-Fock or Fluctuation
Exchange Approximation (FLEX)\cite{FLEX} that are self-consistent at the
one-particle level, but not at the two-particle level. This approach\cite
{Vilk} to the calculation of spin and charge fluctuations satisfies the
Pauli principle $\langle n_\sigma ^2\rangle =\langle n_\sigma \rangle =n/2$
by construction, and it also satisfies the Mermin-Wagner theorem in two
dimensions.

To demonstrate that this theorem is satisfied, it suffices to show that $%
\langle n_{\uparrow }n_{\downarrow }\rangle =g_{\uparrow \downarrow }\left(
0\right) \langle n_{\uparrow }\rangle \langle n_{\downarrow }\rangle $ does
not grow indefinitely. (This guarantees that the constant $\widetilde{C}$
appearing in Eq.(\ref{IntegLorentz=C}) is finite.) To see how this occurs,
write the self-consistency condition Eq.(\ref{Spin}) in the form 
\begin{equation}
n-2\langle n_{\uparrow }n_{\downarrow }\rangle =\frac TN\sum_{\widetilde{q}}%
\frac{\chi _0(q)}{1-\frac 12U\frac{\langle n_{\uparrow }n_{\downarrow
}\rangle }{\langle n_{\uparrow }\rangle \left\langle n_{\downarrow
}\right\rangle }\chi _0(q)}.  \label{Spin2}
\end{equation}
Consider increasing $\langle n_{\uparrow }n_{\downarrow }\rangle $ on the
right-hand side of this equation. This leads to a decrease of the same
quantity on the left-hand side. There is thus negative feedback in this
equation that will make the self-consistent solution finite. A more direct
proof by contradiction has been given in Ref.\cite{Vilk}: suppose that there
is a phase transition, in other words suppose that $\langle n_{\uparrow
}\rangle \left\langle n_{\downarrow }\right\rangle =\frac 12U\langle
n_{\uparrow }n_{\downarrow }\rangle \chi _0(q).$ Then the zero-Matsubara
frequency contribution to the right-hand side of Eq.(\ref{Spin2}) becomes
infinite and positive in two dimensions as one can see from phase-space
arguments (See Eq.(\ref{IntegLorentz=C})). This implies that $\langle
n_{\uparrow }n_{\downarrow }\rangle $ on the left-hand side must become
negative and infinite, but that contradicts the starting hypothesis since $%
\langle n_{\uparrow }\rangle \left\langle n_{\downarrow }\right\rangle =%
\frac 12U\langle n_{\uparrow }n_{\downarrow }\rangle \chi _0(q)$ means that $%
\langle n_{\uparrow }n_{\downarrow }\rangle $ is positive.

Although there is no finite-temperature phase transition, our theory shows
that sufficiently close to half-filling (see Sec.(\ref{SecPD})) there is a
crossover temperature $T_X$ below which the system enters the so-called
renormalized classical regime, where antiferromagnetic correlations grow
exponentially. This will be discussed in detail in Sec.(\ref{SecRC}).

Kanamori-Brueckner screening is also included as we already mentioned above.
To see how the screening occurs, consider a case away from half-filling,
where one is far from a phase transition. In this case, the denominator in
the self-consistency condition can be expanded to linear order in $U$ and
one obtains 
\begin{equation}
g_{\uparrow \downarrow }\left( 0\right) =\frac{\langle n_{\uparrow
}n_{\downarrow }\rangle }{\langle n_{\uparrow }\rangle \left\langle
n_{\downarrow }\right\rangle }=\frac 1{1+\Lambda U}
\end{equation}
where 
\begin{equation}
\Lambda =\frac 2{n^2}\frac TN\sum_q\chi _0(q)^2.
\end{equation}
Clearly, quantum fluctuations contribute to the sum appearing above and
hence to the renormalization of $U_{sp}=$ $g_{\uparrow \downarrow }\left(
0\right) U.$ The value of $\Lambda $ is found to be near $0.2$ as in
explicit numerical calculations of the maximally crossed Kanamori-Brueckner
diagrams.\cite{Chen} At large $U$, the value of $U_{sp}=g_{\uparrow
\downarrow }\left( 0\right) U\sim 1/\Lambda $ saturates to a value of the
order of the inverse bandwidth which corresponds to the energy cost for
creating a node in the two-body wave function, in agreement with the Physics
described by Kanamori.\cite{Kanamori}

\begin{figure}%
\centerline{\epsfxsize 6cm \epsffile{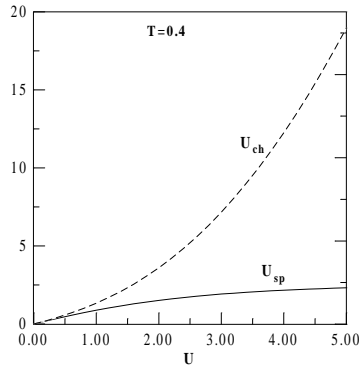}}%
\caption{ Dependence on $U$ of the charge and spin effective interactions 
(irreducible vertices). The temperature is chosen so that for all $U$, it is above 
the crossover temperature. In this case, temperature dependence is not significant. 
The filling is $n=1$}%
\label{FigUspUch}%
\end{figure}

\begin{figure}%
\centerline{\epsfxsize 6cm \epsffile{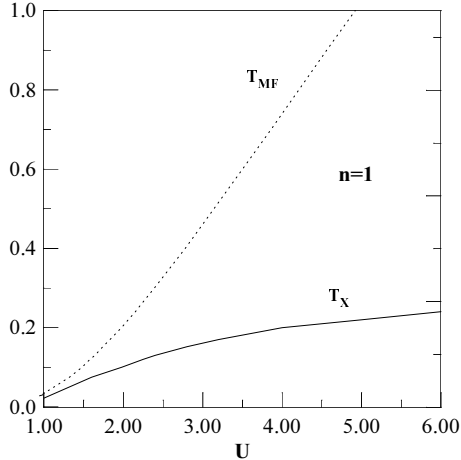}}%
\caption{Crossover temperature at half-filling as function of $U$
 compared with the mean-field transition temperature}%
\label{FigTXU}%
\end{figure}

To illustrate the dependence of $U_{sp},U_{ch}$ on bare $U$ we give in Fig.(%
\ref{FigUspUch}) a plot of these quantities at half-filling where the
correlation effects are strongest. The temperature for this plot is chosen
to be above the crossover temperature $T_X$ to the renormalized classical
regime, in which case the dependence of $U_{sp}$ and $U_{ch}$ on temperature
is not significant. As one can see, $U_{sp}$ rapidly saturates to a fraction
of the bandwidth, while $U_{ch}$ rapidly increases with $U$, reflecting the
tendency to the Mott transition. We have also shown previously in Fig.(2) of
Ref.\cite{Vilk} that $U_{sp}$ depends only weakly on filling. Since $U_{sp}$
saturates as a function of $U$ due to Kanamori-Brueckner screening, the
crossover temperature $T_X$ also saturates as a function of $U$. This is
illustrated in Fig.(\ref{FigTXU}) along with the mean-field transition
temperature that, by contrast, increases rapidly with $U.$

Quantitative agreement with Monte Carlo simulations on the nearest-neighbor%
\cite{Vilk} and next-nearest-neighbor models\cite{Veilleux} is obtained\cite
{Vilk} for all fillings and temperatures in the weak to intermediate
coupling regime $U<8t$. This is discussed further below in Sec.(\ref
{SecMonteCarlo}). We have also shown that the above approach reproduces both
quantum-critical and renormalized-classical regimes in two dimensions to
leading order in the $1/n$ expansion (spherical model)\cite{Dare}.

As judged by comparisons with Monte Carlo simulations\cite{Dare1}, the
particle-particle channel in the repulsive two-dimensional Hubbard model is
relatively well described by more standard perturbative approaches. Although
our approach can be extended to this channel as well, we do not consider it
directly in this paper. It manifests itself only indirectly through the
renormalization of $U_{sp}$ and $U_{ch}$ that it produces.

\subsubsection{Single-particle properties}

\label{Single-particle}

As in any implementation of conserving approximations, the initial guess for
the self-energy, $\Sigma ^{\left( 0\right) }$, obtained from the trial
Luttinger-Ward functional is inconsistent with the exact self-energy formula
Eq.(\ref{2}). The latter formula takes into account the feedback of the spin
and charge collective modes actually calculated from the conserving
approximation. In our approach, we use this self-energy formula Eq.(\ref{2})
in an iterative manner to improve on our initial guess of the self-energy.
The resulting formula for an improved self-energy $\Sigma ^{\left( 1\right)
} $ has the simple physical interpretation of paramagnon theories.\cite
{Stamp}

As another way of Physically explaining this point of view, consider the
following: The bosonic collective modes are weakly dependent on the precise
form of the single-particle excitations, as long as they have a
quasiparticle structure. In other words, zero-sound or paramagnons exist,
whether the Bethe-Salpeter equation is solved with non-interacting particles
or with quasiparticles. The details of the single-particle self-energy by
contrast can be strongly influenced by scattering from collective modes
because these bosonic modes are low-lying excitations. Hence, we first
compute the two-particle propagators with Hartree-Fock single-particle
Green's functions, and then we improve on the self-energy by including the
effect of collective modes on single-particle properties. The fact that
collective modes can be calculated first and self-energy afterwards is
reminiscent of renormalization group approaches,\cite{BourbonnaisCaron}\cite
{geo} where collective modes are obtained at one-loop order while the
non-trivial self-energy comes out only at two-loop order.

The derivation of the general self-energy formula Eq.(\ref{2}) given in
Appendix~(\ref{Proofs}) shows that it basically comes from the definition of
the self-energy and from the equation for the collective modes Eq.(\ref{1}).
This also stands out clearly from the diagrammatic representation in Fig.(%
\ref{FigKadanoffBaym}). By construction, these two equations Eqs.(\ref{1})
and (\ref{2}) satisfy the consistency requirement $\frac 12Tr\Sigma
G=U\left\langle n_{\uparrow }n_{\downarrow }\right\rangle $ (see Appendix~(%
\ref{Proofs})), which in momentum and frequency space can be written as 
\begin{equation}
\lim_{\tau \rightarrow 0^{-}}\frac TN\sum_k\Sigma _\sigma \left( k\right)
G_\sigma \left( k\right) e^{-ik_n\tau }=U\left\langle n_{\uparrow
}n_{\downarrow }\right\rangle ,  \label{Consistency}
\end{equation}
The importance of the latter sum rule, or consistency requirement, for
approximate theories should be clear from the appearance of the correlation
function $\langle n_{\uparrow }n_{\downarrow }\rangle $ that played such an
important role in determining the irreducible vertices and in obtaining the
collective modes. Using the fluctuation dissipation theorem Eqs.(\ref
{sumSpin}),(\ref{sumCharge}) this sum-rule can be written in form that
explicitly shows the relation between the self-energy and the spin and
charge susceptibilities 
\begin{equation}
\frac TN\sum_k\left[ \Sigma _\sigma (k)-Un_{-\sigma }\right] G_\sigma (k)=%
\frac U4\frac TN\sum_q\left[ \chi _{ch}(q)-\chi _{sp}(q)\right] .
\label{sumSigma1}
\end{equation}

To keep as much as possible of this consistency, we use on the right-hand
side of the self-energy expression Eq.(\ref{2}) the same irreducible
vertices and Green's functions as those that appear in the collective-mode
calculation Eq.(\ref{1}). Let us call $G^{\left( 0\right) }$ the initial
Green's function corresponding to the initial Luttinger-Ward self-energy $%
\Sigma ^{\left( 0\right) }.$ Our new approximation for the self-energy $%
\Sigma ^{\left( 1\right) }$ then takes the form

\begin{equation}
\Sigma _\sigma ^{\left( 1\right) }\left( k\right) =Un_{-\sigma }+\frac U4%
\frac TN\sum_q\left[ U_{sp}\chi _{sp}(q)+U_{ch}\chi _{ch}(q)\right] G_\sigma
^{\left( 0\right) }(k+q).  \label{param}
\end{equation}
Note that $\Sigma _\sigma ^{\left( 1\right) }\left( k\right) $ satisfies
particle-hole symmetry Eq.(\ref{phs}) where appropriate. This self-energy
expression (\ref{param}) is physically appealing since, as expected from
general skeleton diagrams, one of the vertices is the bare one $U$, while
the other vertex is dressed and given by $U_{sp}$ or $U_{ch}$ depending on
the type of fluctuation being exchanged. It is because Migdal's theorem does
not apply for this problem that $U_{sp}$ and $U_{ch}$ are different from the
bare $U$ at one of the vertices. $U_{sp}$ and $U_{ch}$ here take care of
vertex corrections.\cite{StampVertex}

The use of the full $G_\sigma (k+q)$ instead of $G_\sigma ^0(k+q)$ in the
above expression Eq.(\ref{param}) would be inconsistent with
frequency-independent irreducible vertices. For the collective mode Eq.(\ref
{1}) this is well known to lead to the violation of the conservation laws as
was discussed in detail in the previous subsection. Here we insist that the
same is true in the calculation of the effect of {\em electronic} collective
modes on the single-particle properties. Formally, this is suggested by the
similarity between the equation for the susceptibility Eq.(\ref{1}) and that
for the self-energy Eq.(\ref{2}) in terms of irreducible vertices. More
importantly, two physical effects would be absent if one were to use full $G$
and frequency independent irreducible vertices. First, upper and lower
Hubbard bands would not appear because the $U^2/\omega $ high-frequency
behavior in Eq.(\ref{SelfHaut}) that is necessary to obtain the Hubbard
bands would set in too late, as we discuss in Sec.(\ref{SubSecConstSingle})
and in Sec.(\ref{HubbardBands}). This result is also apparent from the fact
that FLEX calculations in infinite dimension do not find upper and lower
Hubbard bands\cite{MengeMH} where the exact numerical solution does. The
other physical effect that would be absent is precursors of
antiferromagnetic bands, Sec.(\ref{SecDestruction}) and the pseudogap in $A(%
{\bf k_F},\omega )$, that would not appear for reasons discussed in Sec.(\ref
{SecFLEX}). We also will see in Sec.(\ref{SecMonteCarlo}) below that FLEX
calculations of the single-particle Green's function, significantly disagree
with Monte Carlo data, even away from half-filling, as was already shown in
Fig.1 of Ref.\cite{Vilk2}.

The chemical potential for interacting electrons $\mu $ is found from the
usual condition on particle number 
\begin{equation}
n=\frac TN\sum_kG_\sigma ^{\left( 1\right) }(k)\exp (-ik_n0^{-})=\frac TN%
\sum_k\frac{\exp (-ik_n0^{-})}{i\omega _n-\varepsilon _{{\bf k}}+\mu
^{(1)}-\Sigma ^{(1)}({\bf k},k_n)}.  \label{mu}
\end{equation}
This chemical potential $\mu $ is, of course, different from $\mu _0$ but
the Luttinger sum rule $\sum \theta (-\varepsilon _{{\bf k}}+\mu -\Sigma
^{\left( 1\right) })=n_\sigma $ is satisfied to a high accuracy (about few
percent) for all fillings and temperatures $T_X\leq T\ll W$. As usual this
occurs because the change in $\mu ^{(1)}-\mu _0$ is compensated by the
self-energy shift on the Fermi surface $\Sigma ^{(1)}({\bf k}_F,0)$. For $%
T<T_X$ there is some deviation from the Luttinger sum rule which is due to
the appearance of the precursors of the antiferromagnetic bands below $T_X$
(Sec. \ref{SecDestruction}) which develop into true SDW bands at $T=0$.

It is important to realize that $G^{(0)}$ on the right hand side of the
equation for the self-energy $\Sigma $ cannot be calculated as $%
G^{(0)}=1/(\omega -\varepsilon _{{\bf k}}+\mu ^{(1)})$, because otherwise it
would not reduce to zero-temperature perturbation theory when it is
appropriate. As was pointed out by Luttinger, (see also section \ref{Lat})
the ``non-interacting'' Green's function used in the calculation for $\Sigma 
$ should be calculated as $G^{(0)}=1/(\omega -\varepsilon _{{\bf k}}-\Sigma
^{(n)}({\bf k}_F,0)+\mu ^{(n)})$, where $\mu ^{(n)}$ is calculated on the
same level of accuracy as $\Sigma ^{(n)}({\bf k}_F,0)$, i.e. from Eq.(\ref
{mu}) with $\Sigma ^{(n)}({\bf k},ik_n)$. In our calculation below, we
approximate $\mu ^{(1)}-\Sigma ^{(1)}({\bf k}_F,0)$ by $\mu _0$ because for
the coupling strength and temperatures considered in this paper ($U\leq W/2$%
, $T_X\leq T\ll W$) the Luttinger theorem is satisfied to high accuracy and
the change of the Fermi surface shape is insignificant. In addition, at
half-filling the condition $\mu -\Sigma ({\bf k}_F,0)=\mu _0$ is satisfied
exactly at any $U$ and $T$ because of particle-hole symmetry. For somewhat
larger coupling strengths and away from half-filling, one may try to improve
the theory by using $G^{(0)}=1/(\omega -\varepsilon _{{\bf k}}-\Sigma ^{(1)}(%
{\bf k}_F,0)+\mu ^{(1)})$, with $\Sigma ^{(1)}$ and $\mu $ found
self-consistently. However, the domain of validity of our approach is
limited to the weak-to-intermediate coupling regime since the
strong-coupling regime requires frequency-dependent pseudopotentials (see
below).

Finally, let us note that, in the same spirit as Landau theory, the only
vertices entering our theory are of the type $\Gamma _{\uparrow \downarrow }$
and $\Gamma _{\uparrow \uparrow },$ or, through Eq.(\ref{defVertex}), $%
U_{sp} $ and $U_{ch}$. In other words, we look at the problem from the
longitudinal spin and charge particle-hole channel. Consequently, in the
contact pseudopotential approximation the exact equation for the self-energy
Eq.(\ref{2}) reduces to our expression Eq. (\ref{param}) which does {\em not}
have the factor $3$ in the front of the spin susceptibility. This is
different from some paramagnon theories, in which such factor was introduced
to take care of rotational invariance. However, we show in Appendix (\ref
{Paramagnon theories}) that these paramagnon theories are inconsistent with
the sum-rule Eq.(\ref{sumSigma1}) which relates one and two-particle
properties. In our approach, questions about transverse spin fluctuations
are answered by invoking rotational invariance $\chi _{sp}^{xx}=\chi
_{sp}^{yy}=\chi _{sp}^{zz}$. In particular, one can write the expression for
the self-energy Eq.(\ref{param}) in an explicitly rotationally invariant
form by replacing $\chi _{sp}$ by $(1/3)Tr[\chi _{sp}^{\nu \nu }]$. If
calculations had been done in the transverse channel, it would have been
crucial to do them while simultaneously enforcing the Pauli principle in
that channel. In functional integration methods, it is well known that
methods that enforce rotational invariance without enforcing the Pauli
principle at the same time give unphysical answers, such as the wrong factor 
$2/3$ in the RPA susceptibility\cite{moriya} $\chi _{sp}=\chi
^0/(1-(2/3)U\chi ^0)$ or wrong Hartree-Fock ground state.\cite{Bresil}

\subsubsection{Internal accuracy check}

\label{SubSecAccuracy}

The quantitative accuracy of the theory will be discussed in detail when we
compare with Monte Carlo calculations in the next section. Here we show that
we can use the consistency requirement between one- and two-particle
properties Eq.(\ref{Consistency}) to gauge the accuracy of the theory from
within the theory itself.

The important advantage of the expression for the self-energy $\Sigma
_\sigma ^{\left( 1\right) }\left( k\right) $ given by Eq.(\ref{param}) is
that, as shown in Appendix~(\ref{Proofs}), it satisfies the consistency
requirement between one- and two-particle properties Eq.(\ref{Consistency}),
in the following sense 
\begin{equation}
\lim_{\tau \rightarrow 0^{-}}\frac TN\sum_k\Sigma _\sigma ^{\left( 1\right)
}\left( k\right) G_\sigma ^{\left( 0\right) }\left( k\right) e^{-ik_n\tau
}=U\left\langle n_{\uparrow }n_{\downarrow }\right\rangle ,  \label{sumS}
\end{equation}

Let $G_\sigma ^{\left( 1\right) }$ be defined by $\left[ G_\sigma
^{(1)}\right] ^{-1}\equiv G_0^{-1}-\Sigma ^{\left( 1\right) }.$ We can use
the fact that in an exact theory we should have $Tr\left[ \Sigma _\sigma
^{\left( 1\right) }G_\sigma ^{\left( 1\right) }\right] $ in the above
expression instead of $Tr\left[ \Sigma _\sigma ^{\left( 1\right) }G_\sigma
^{\left( 0\right) }\right] $ to check the accuracy of the theory. It
suffices to compute by how much $Tr\left[ \Sigma _\sigma ^{\left( 1\right)
}G_\sigma ^{\left( 0\right) }\right] $ differs from $Tr\left[ \Sigma _\sigma
^{\left( 1\right) }G_\sigma ^{\left( 1\right) }\right] .$ In the parameter
range $U<4t$ and $n$,$T$ arbitrary but not too deep in the, soon to be
described, renormalized-classical regime, we find that $Tr\left[ \Sigma
_\sigma ^{\left( 1\right) }G_\sigma ^{\left( 0\right) }\right] $ differs
from $Tr\left[ \Sigma _\sigma ^{\left( 1\right) }G_\sigma ^{\left( 1\right)
}\right] $ by at most $15\%.$ Another way to check the accuracy of our
approach is to evaluate the right-hand side of the $f-$sum rule Eqs.(\ref{f2}%
) with $n_{{\bf k}\sigma }=G_\sigma ^{\left( 1\right) }\left( {\bf k}%
,0^{-}\right) $ and to compare with the result that had been obtained with $%
f_{{\bf k},\sigma }$. Again we find the same $15\%$ disagreement, at worse,
in the same parameter range. As one can expect, this deviation is maximal at
half-filling and becomes smaller away from it.

Eq.(\ref{param}) for the self-energy $\Sigma ^{\left( 1\right) }$ already
gives good agreement with Monte Carlo data but the accuracy can be improved
even further by using the general consistency condition Eq.(\ref{Consistency}%
) on $Tr\left[ \Sigma _\sigma ^{\left( 1\right) }G_\sigma ^{\left( 1\right)
}\right] $ to improve on the approximation for vertex corrections. To do so
we replace $U_{sp}$ and $U_{ch}$ on the right-hand side of Eq.(\ref{param})
by $\alpha U_{sp}$ and $\alpha U_{ch}$ with $\alpha $ determined
self-consistently in such a way that Eq.(\ref{sumS}) is satisfied with $%
G_\sigma ^{\left( 0\right) }\left( k\right) $ replaced by $G_\sigma ^{\left(
1\right) }\left( k\right) $. For $U<4$, we have $\alpha <1.15$. The slight
difference between the irreducible vertices entering the collective modes
and the vertex corrections entering the self-energy formula can be
understood from the fact that the replacement of irreducible vertices by
constants is, in a way, justified by the mean-value theorem for integrals.
Since the averages are not taken over the same variables, it is clear that
the vertex corrections in the self-energy formula and irreducible vertices
in the collective modes do not need to be strictly identical when they are
approximated by constants.

Before we move on to comparisons with Monte Carlo simulations, we stress
that $\Sigma ^{\left( 1\right) }$ given by Eq. (\ref{param}) cannot be
substituted back into the calculation of $\chi _{sp,ch}$ by simply replacing 
$\chi _0=G_0G_0$ with the dressed bubble $\tilde{\chi}_0=GG$. Indeed, this
would violate conservation of spin and charge and $f$-sum rule. In
particular, the condition $\chi _{sp,ch}({\bf q}=0,iq_n\neq 0)=0$ that
follows from the Ward identity (\ref{Ward}) would be violated as we see in
Eq.(\ref{WardTriste}). In the next order, one is forced to work with
frequency-dependent irreducible vertices that offset the unphysical behavior
of $\tilde{\chi}_0$ at non-zero frequencies.

\section{Numerical results and comparisons with Monte Carlo simulations}

\label{SecMonteCarlo}

In this section, we present a few numerical results and comparisons with
Monte Carlo simulations. We divide this section in two parts. In the first
one we discuss data sufficiently far from half-filling, or at high enough
temperature, where size effects are unimportant for systems sizes available
in Monte Carlo simulations. In the second part, we discuss data at
half-filling. There, size effects become important below the crossover
temperature $T_X$ where correlations start to grow exponentially (Sec.(\ref
{SecDestruction})). All single-particle properties are calculated with our
approximation Eq.(\ref{param}) for the self-energy using the vertex
renormalization $\alpha $ explained in the previous section. The results
would differ at worse by $15\%$ if we had used $\alpha =1.$

\subsection{Far from the crossover temperature $T_X$}

\subsubsection{\em Two-particle properties}

We have shown previously in Fig.4a-d of Ref.\cite{Vilk} and in Figs.2-4 and
Fig.6 of Ref.\cite{Veilleux} that both spin and charge structure factor
sufficiently away from the crossover temperature $T_X$ are in quantitative
agreement with Monte Carlo data for values of $U$ as large as the bandwidth.
On the qualitative level, the decrease in charge fluctuations as one
approaches half-filling has been explained\cite{Vilk} as a consequence of
the Pauli principle embodied in the calculation of the irreducible vertex $%
U_{ch}.$\cite{HottaFujimoto}

\begin{figure}%
\centerline{\epsfxsize 6cm \epsffile{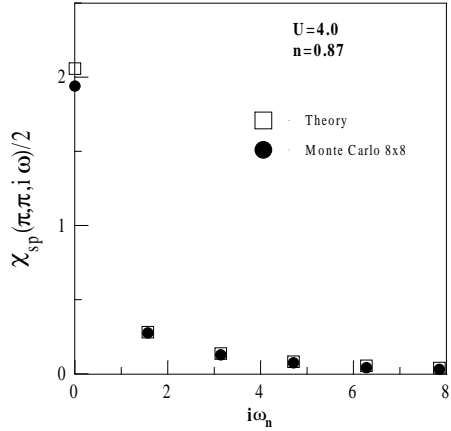}}%
\caption{Comparisons between Monte Carlo simulations \protect\cite{BWS} and our theory 
for the spin susceptibility at $Q=(\pi,\pi)$ as a function of Matsubara frequency. 
The temperature is $T=0.25$, and the system size $8\times8$. The factor 1/2 on the vertical axis 
is due to the fact that the susceptibility in \protect\cite{BWS} is $\chi_{+-}$ a quantity
that is by definition twice smaller then ours and that of \protect\cite{FLEX-parquet}}%
\label{FigChiOmega}%
\end{figure}

\begin{figure}%
\centerline{\epsfxsize 6cm \epsffile{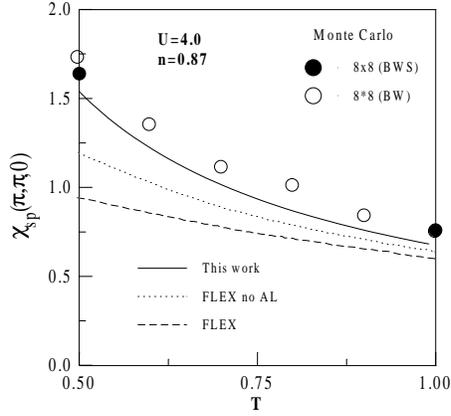}}%
\caption{Comparisons between the Monte Carlo simulations (BW) and FLEX calculations 
presented in Fig.19 of Ref. \protect\cite{FLEX-parquet} 
and our theory for the spin susceptibility at $Q=(\pi,\pi)$ as a function of temperature 
at zero Matsubara frequency. The filled circles (BWS) are from Ref. \protect\cite{BWS}  }%
\label{FigChiT}%
\end{figure}

Here we present on Fig.(\ref{FigChiOmega}) and Fig.(\ref{FigChiT})
comparisons with a dynamical quantity, namely the spin susceptibility.
Similar comparisons, but with a phenomenological value of $U_{sp}$, have
been done by Bulut {\it et al.} Ref.\cite{BulutParamagnon}. Fig.(\ref
{FigChiOmega}) shows the staggered spin susceptibility as a function of
Matsubara frequencies for $n=0.87,$ $T=0.25$ and $U=4$. The effect of
interactions is already quite large for the zero-frequency susceptibility.
It is enhanced by a factor of over $5$ compared with the non-interacting
value. Nevertheless, one can see that the theory and Monte Carlo simulations
are in good agreement.

Fig.(\ref{FigChiT}) shows the temperature dependence of the zero-frequency
staggered spin susceptibility for the same filling and interaction as in the
previous figure. Symbols represent Monte Carlo simulations from Refs.\cite
{BWS} and \cite{FLEX-parquet}, the solid line is for our theory while dotted
and dashed lines are for two versions of FLEX. Surprisingly, the fully
conserving FLEX theory, (dashed line) compares worse with Monte Carlo data
than the non-conserving version of this theory that neglects the so-called
Aslamasov-Larkin diagrams (dotted line). By contrast, our theory is in
better agreement with the Monte Carlo data than FLEX for the staggered
susceptibility $\chi _{sp}\left( {\bf q}=\left( \pi ,\pi \right) ,i\omega
_n=0\right) $, and at the same time it agrees exactly with the conservation
law that states that $\chi _{sp,ch}\left( {\bf q}=0,i\omega _n\neq 0\right)
=0$.

\begin{figure}%
\centerline{\epsfxsize 6cm \epsffile{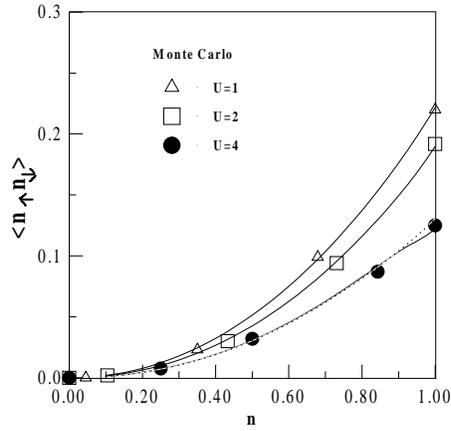}}%
\caption{Comparisons between the Monte Carlo simulations of  Ref.\protect\cite{Moreo} 
and our theory (solid lines) for the filling dependence of the double occupancy. The results are 
are for $T=1/6$ as a function of filling and for various values of $U$ except for $U=4$ where 
the dotted line shows the results of our theory at the crossover temperature
$T=T_X$.}%
\label{FigGUpDown}%
\end{figure}

Finally, Fig.(\ref{FigGUpDown}) shows the double occupancy $\left\langle
n_{\uparrow }n_{\downarrow }\right\rangle $ as a function of filling for
various values of $U$. The symbols again represent Monte Carlo data for $%
T=1/6$, and the lines are the results of our theory. Everywhere the
agreement is very good, except for $n=1,U=4.$ In the latter case, the system
is already below the crossover temperature $T_X$ to the renormalized
classical regime. As explained in Sec.(\ref{SecValidity}), the appropriate
procedure for calculating double occupancy in this case is to take for $%
\left\langle n_{\uparrow }n_{\downarrow }\right\rangle $ its value (dotted
line) at $T_X$ instead of using the {\it ansatz }Eq.(\ref{Usp}). In any
case, the difference is not large.

\subsubsection{\em Single-particle properties}

Fig.1(a) of Ref.\cite{Vilk2} shows $G\left( {\bf k,}\tau \right) $ for
filling $n=0.875$, temperature $T=0.25$ and $U=4$ for the wave vector on the 
$8\times 8$ lattice which is closest to the Fermi surface, namely $\left(
\pi ,0\right) $. Our theory is in agreement with Monte Carlo data and with
the parquet approach\cite{FLEX-parquet} but in this regime second-order
perturbation theory for the self-energy gives the same result. Surprisingly,
FLEX is the only theory that disagrees significantly with Monte Carlo data.
The good performance of perturbation theory (see also \cite{Pertub}) can be
explained in part by compensation between the renormalized vertices and
susceptibilities ($U_{sp}<U$, $\chi _{sp}(q)>\chi _0(q)$; $U_{ch}>U$, $\chi
_{ch}(q)<\chi _0(q)$ ).

We have also calculated Re$(\Sigma \left( ik_n\right) /ik_n)$ and compared
with the Monte Carlo data in Fig.2a of Ref.\cite{BulutParamagnon} obtained
at $n=0.87$, $U=4$, $\beta =6.$ Our approach agrees with Monte Carlo data
for all frequencies, but again second-order perturbation theory gives
similar results.

\subsection{Close to crossover temperature $T_X$ at half-filling}

\subsubsection{\em Two-particle properties}

\begin{figure}%
\centerline{\epsfxsize 10cm \epsffile{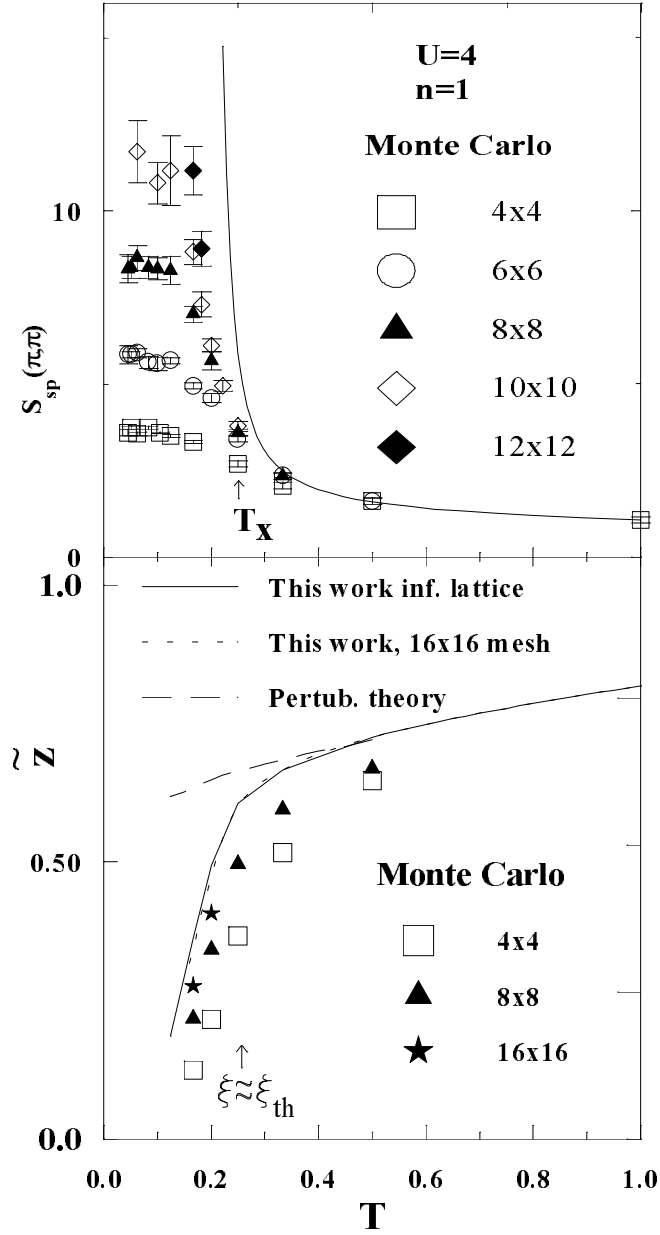}}%
\caption{The upper part of the figure, adapted from Ref.\protect\cite{Vilk}, shows the 
temperature dependence of $S_{sp}(\pi ,\pi )$ at half-filling $n=1$. 
The solid line is our theory for an infinite system while
symbols are Monte Carlo data from Ref.\protect\cite{White89}. 
The bottom part of the figure, adapted from Ref.\protect\cite{Vilk2}, shows
the behavior of $\tilde{z(T)}=-2G({\bf k}_F,\beta /2)$ in Eq.(\ref{z}),  
as a function of temperature as obtained from 
Monte Carlo\protect\cite{FLEX-parquet} simulations (symbols), 
from second order pertrubation theory (dashed line) 
and from our theory for an infinite system (solid line) 
and for a $16 \times 16$ lattice (dashed line).}%
\label{FigJoint}%
\end{figure}

The occurrence of the crossover temperature $T_X$ at half-filling is perhaps
best illustrated in the upper part of Fig.(\ref{FigJoint}) by the behavior
of the static structure factor $S_{sp}\left( \pi ,\pi \right) $ for $U=4$ as
a function of temperature. When the correlation length becomes comparable to
the size of the system used in Monte Carlo simulations,\cite{White} the
static structure factor starts to increase rapidly, saturating to a value
that increases with system size. The solid line is calculated from our
theory for an infinite lattice. The Monte Carlo data follow our theoretical
curve (solid line) until they saturate to a size-dependent value. The theory
correctly describes the static structure factor not only above $T_X$ but
also as we enter the renormalized classical regime at $T_X$. Analytical
results for this regime are given in Sec.(\ref{SecRC}). Note that the RPA
mean-field transition temperature for this value of $U$ is more than three
times larger than $T_X\sim 0.2$. The size-dependence of Monte Carlo data for 
$S_{sp}\left( {\bf q}\right) $ at all other values of ${\bf q\neq }\left(
\pi ,\pi \right) $ available in simulations is negligible and our
calculation for infinite system reproduces this data (not shown).

\subsubsection{\em Single-particle properties}

Equal-time (frequency integrated) single-particle properties are much less
sensitive to precursor effects than dynamical quantities as we now proceed
to show. For example, $n\left( {\bf k}\right) =G\left( {\bf k},0^{-}\right) $
is a sum of $G\left( {\bf k},ik_n\right) $ over all Matsubara frequencies.
We have verified (figure not shown) that $\frac 1N\sum_{{\bf k}\sigma }n_{%
{\bf k}\sigma }\partial ^2\epsilon _{{\bf k}}/\partial k_x^2$ obtained from
Monte Carlo simulations\cite{White89} is given quite accurately by either
second-order perturbation theory or by our theory. This has very important
consequences since, for this quantity, the non-interacting value differs
from second-order perturbation theory by at most $15\%.$ This means that the
numerical value of the right-hand side of the $f$ sum-rule Eq.(\ref{f2}) is
quite close to that obtained from the left-hand side using our expression
for the spin and charge susceptibility.

\begin{figure}%
\centerline{\epsfxsize 6cm \epsffile{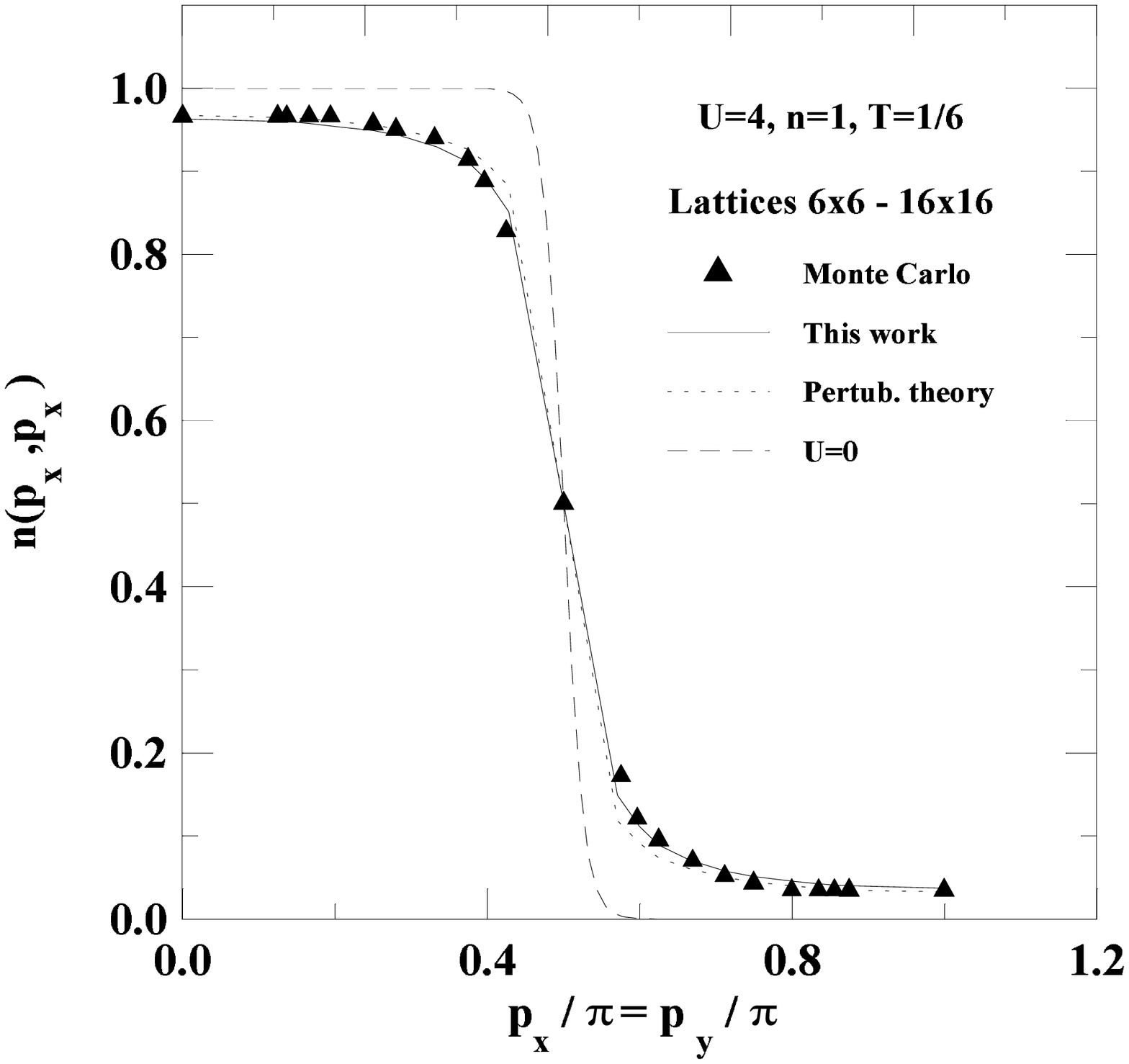}}%
\caption{Occupation number $n(\vec k)$ as a function of wave vector $\vec k$ at half-filling
for $T=1/6$, $U=4$, and system sizes $6\times 6$ to $16\times 16$. The symbols
are Monte Carlo results from
Ref.\protect\cite{Moreo} while the solid line is our theory and the dotted line 
is the prediction from second order perturbation theory. The dashed line shows the
result for $U=0$ as a reference.}%
\label{FigN(k)}%
\end{figure}

One can also look in more details at $n\left( {\bf k}\right) $ itself
instead of focusing on a sum rule. Fig.(\ref{FigN(k)}) shows a comparison of
our theory and of second order perturbation theory with Monte Carlo data for 
$n\left( {\bf k}\right) $ obtained for a set of lattice sizes from $6\times
6 $ to $16\times 16$ at $n=1,$ $T=1/6$, $U=4.$ Size effects appear
unimportant for this quantity at this temperature. These Monte Carlo data
have been used in the past\cite{Moreo} to extract a gap by comparison with
mean field SDW theory. Our theory for the same set of lattice sizes is in
excellent agreement with Monte Carlo data and predicts a pseudogap at this
temperature, as we will discuss below. However, for available values of $%
{\bf k}$ on finite lattices, second order perturbation theory is also in
reasonable agreement with Monte Carlo data for $n\left( {\bf k}\right) $.
Since second order perturbation theory does not predict a pseudogap, this
means that $n\left( {\bf k}\right) $ is not really sensitive to the opening
of a pseudogap. This is so both because of the finite temperature and
because the wave vectors closest to the Fermi surface are actually quite far
on the appropriate scale. For this filling, the value of $n\left( {\bf k}%
\right) $ is fixed to $1/2$ on the Fermi surface itself.

It is thus necessary to find a dynamical quantity defined on the Fermi
surface whose temperature dependence will allow us to unambiguously identify
the pseudogap regime in both theory and in Monte Carlo data. The most
dramatic effect is illustrated in the lower part of Fig.(\ref{FigJoint})
where we plot the quantity $\tilde{z}\left( T\right) $ defined by\cite{Vilk2}%
\cite{Nandini}

\begin{equation}
\tilde{z}\left( T\right) =-2G\left( {\bf k}_F,\beta /2\right) =\int \frac{%
d\omega }{2\pi }\frac{A\left( {\bf k}_F,\omega \right) }{\cosh \left( \beta
\omega /2\right) }.  \label{z}
\end{equation}
The physical meaning of this quantity $\tilde{z}\left( T\right) $ is that it
is an average of the single-particle spectral weight $A\left( {\bf k}%
_F,\omega \right) $ within $T\equiv 1/\beta $ of the Fermi level ($\omega =0$%
). When quasiparticles exist, this is the best estimate of the usual
zero-temperature quasiparticle renormalization factor $z\equiv 1/(1-\partial
\Sigma /\partial \omega )$ that can be obtained directly from imaginary-time
Monte Carlo data. For non-interacting particles $\tilde{z}\left( T\right) $
is unity. For a normal Fermi liquid it becomes equal to a constant less than
unity as the temperature decreases since the width of the quasiparticle peak
scales as $T^2$ and hence lies within $T$ of the Fermi level. However,
contrary to the usual $z\equiv 1/(1-\partial \Sigma /\partial \omega )$ this
quantity gives an estimate of the spectral weight $A\left( {\bf k}_F,\omega
\right) $ around the Fermi level, even if quasiparticles disappear and a
pseudogap forms, as in the present case, (see Sec.(\ref{SecDestruction})).

One can clearly see from the lower part of Fig.(\ref{FigJoint}) that while
second-order perturbation theory exhibits typical Fermi-liquid behavior for $%
\tilde{z}\left( T\right) $, both Monte Carlo data\cite{FLEX-parquet} and a
numerical evaluation of our expression for the self-energy lead to a rapid
fall-off of $\tilde{z}\left( T\right) $ below $T_X$ (for $U=4$, $T_X\approx
0.2$\cite{Vilk}). The rapid decrease of $\tilde{z}\left( T\right) $ clearly
suggests non Fermi-liquid behavior. We checked also that our theory
reproduces the Monte Carlo size-dependence. This dependence is explained
analytically in Sec.(\ref{SubSecEffectOnQP}). In Ref.\cite{Vilk2} we have
shown that at half-filling, our theory gives better agreement with Monte
Carlo data\cite{FLEX-parquet} for $G\left( {\bf k}_F,\tau \right) $ than
FLEX, parquet or second order perturbation theory.

\begin{figure}%
\centerline{\epsfxsize 6cm \epsffile{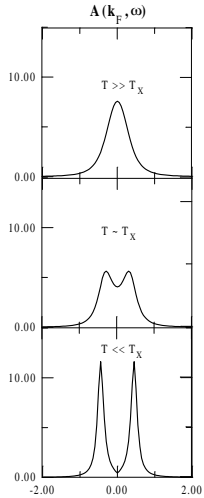}}%
\caption{Qualitative sketch of the spectral weight at the Fermi wave vector at half-filling
for three temperatures. This plot is obtained by retainig only the classical contribution to 
the self-energy using parameters corresponding to the typical $U=4$ of Monte Carlo 
simulations. The top plot is for $T>T_X$, the middle one for $T \sim T_X$ and the 
bottom one for $T<T_X$. The precursors of antiferromagnetic bands would look like this
last figure.}%
\label{FigSpectralWeight}%
\end{figure}

To gain a qualitative insight into the meaning of this drop in $\tilde{z}%
\left( T\right) $, we use the analytical results of the next section to plot
in Fig.(\ref{FigSpectralWeight}) the value of $A\left( {\bf k}_F,\omega
\right) .$ This plot is obtained by retaining only the contribution of
classical fluctuations Eq.(\ref{Somega}) to the self-energy. One sees that
above $T_X$, there is a quasiparticle but that at $T\sim T_X$ a minimum
instead of a maximum starts to develop at the Fermi surface $\omega =0.$
Below $T_X$, the quasiparticle maximum is replaced by two peaks that are the
precursors of antiferromagnetic bands. This is discussed in detail in much
of the rest of this paper.

\subsection{Phase diagram}

\label{SecPD}

\begin{figure}%
\centerline{\epsfxsize 6cm \epsffile{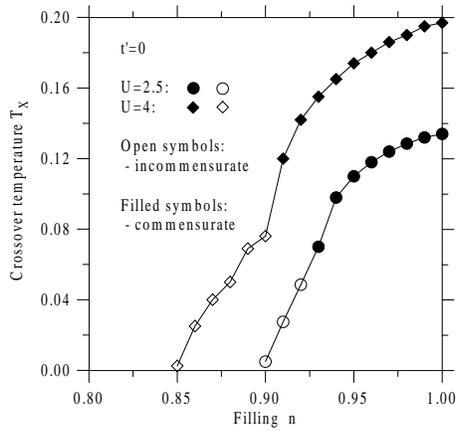}}%
\caption{Crossover temperature $T_X$ as a function of filling for $U=4$ and $U=2.5$.
On this crossover line, $\xi^2$ is enhanced by a factor of 500 over the bare value. 
Filled symbols indicate that the crossover is at the antiferromagnetic wave vector, while
open symbols indicate a crossover at an incommensurate wave vector. Reproduced with
permission from Ref. \protect\cite{JacksonMemoire} }%
\label{FigJacksonPhaseDiagram}%
\end{figure}

The main features predicted by our approach for the magnetic phase diagram
of the nearest-neighbor hopping model have been given in Ref.\cite{Vilk}.
Needless to say, all our considerations apply in the weak to intermediate
coupling regime. Note also that both quantum critical and renormalized
classical properties of this model have been studied in another publication%
\cite{Dare}. The shape of the phase diagram that we find is illustrated in
Fig.(\ref{FigJacksonPhaseDiagram}) for $U=2.5$ and $U=4$.

At zero temperature and small filling, the system is a paramagnetic Fermi
liquid, whatever the value of the interaction $U$ $\left( <W\right) $. Then,
as one moves closer to half-filling, one hits a quantum critical point at a
value of filling $n_c$. Since, $U_{sp}$ in our theory saturates with
increasing $U$, the value of $n_c$ is necessarily larger than about $%
n_c(U=\infty )=0.68$. At this point, incommensurate order sets in at a wave
vector $\left( q_c,\pi \right) $ or at symmetry-related points. Whatever the
value of $U$, the value of $q_c$ is contained\cite{Vilk} in the interval $%
0.74\pi <q_c\leq \pi $, increasing monotonously towards $0.74\pi $ as $U$
increases. Since our approach applies only in the paramagnetic phase, at
zero temperature we cannot move closer to half-filling. Starting from
finite-temperature then, the existence of long-range order at low
temperature is signaled by the existence of a crossover temperature $%
T_X\left( n,U\right) $ below which correlations start to grow exponentially.
We have already discussed the meaning of $T_X\left( n,U\right) $ at
half-filling. This crossover temperature becomes smaller and smaller as one
moves away from half-filling, until it reaches the quantum-critical point
that we just discussed. The correlations that start to grow at $T_X\left(
n,U\right) $ when $n=1$ are at the antiferromagnetic wave vector, and they
stay at this wave vector for a range of fillings $n$. Finally, at some
filling, the correlations that start to grow at $T_X\left( n,U\right) $ are
at an incommensurate value until the quantum-critical point is reached.

Note that the above phase diagram is quite different from the predictions of
Hartree-Fock theory mostly because of the strong renormalization of $U_{sp}$%
. This quantitative change leads to qualitative changes in the Hartree-Fock
phase diagram since, for example, Stoner ferromagnetism never occurs in our
picture. While the existence of ferromagnetism in the {\it strong coupling}
limit has been proven only recently\cite{Tasaki}, the absence of Stoner
ferromagnetism in the Hubbard model was already suggested by Kanamori\cite
{Kanamori} a long time ago and was verified by more recent studies.\cite
{Chen}\cite{MullerHartman}\cite{HirschDiagram} More relevant to the present
debate though, is the fact that SDW order persists away from half-filling
for a finite range of dopings. While this is in agreement with slave-boson
approaches\cite{Wolfle} and studies\cite{Jarrel} using the
infinite-dimension methodology,\cite{Georges} it is in clear disagreement
with Monte Carlo simulations\cite{ImadaDiagram}. Our approach certainly
fails sufficiently below $T_X$, but given the successes described above, we
believe that it can correctly predict the exponential growth of fluctuations
at $T_X$. It would be difficult to imagine how one could modify the theory
in such a way that the growth of magnetic fluctuations does not occur even
at incommensurate wave vectors. Also, such an approach would also need to
stop the growth of fluctuations that we find as we approach the quantum
critical point along the zero temperature axis, from the low-filling,
paramagnetic side, where $T_X\left( n<n_c,U\right) =0$.

It could be that Monte Carlo simulations\cite{ImadaDiagram} fail to see
long-range order at zero temperature away from half-filling because at zero
temperature, in the nearest-neighbor model, this order has a tendency to
being incommensurate everywhere except at $n=1$. Furthermore, as we saw
above, this incommensuration is in general far from one of the available
wave vectors on an $8\times 8$ lattice. It comes close to $\left( 0.75\pi
,\pi \right) $ only for the largest values of $U$ available by Monte Carlo.
Hence, incommensurate order on small lattices is violently frustrated not
only by the boundary conditions, but also by the fact that there is no wave
vector on what would be the Fermi surface of the infinite system. This means
that the electron-electron interaction scatters the electrons at
wave-vectors that are not those where the instability would show up,
rendering these scatterings not singular. This is clearly an open problem.

\section{Replacement of Fermi liquid quasiparticles by a pseudogap in two
dimensions below $T_X.$}

\label{SecDestruction}

One of the most striking consequences of the results discussed in the
context of Monte Carlo simulations is the fall of the spectral weight below
the temperature $T_X$ where antiferromagnetic fluctuations start to grow
exponentially in two dimensions. We have already shown in a previous
publication\cite{Vilk2} that this corresponds to the disappearance of Fermi
liquid quasiparticles at the Fermi surface, well above the zero temperature
phase transition. We also found that, simultaneously, precursors of the
antiferromagnetic bands develop in the single-particle spectrum. Given the
simplicity of our approach, it is possible to demonstrate this phenomenon
analytically. This is particularly important here because size effects and
statistical errors make numerical continuation of the Monte Carlo data to
real frequencies particularly difficult. Such analytic continuations using
the maximum entropy method\cite{White} have, in the past, lead to a
conclusion different from the one obtained later using singular value
decomposition\cite{SingValue}.

In this section then, we will consider the conditions for which Fermi liquid
quasiparticles can be destroyed and replaced by a pseudogap in two
dimensions. The major part of this section will be concerned with the single
particle pseudogap and the precursors of antiferromagnetic bands in the
vicinity of the zero temperature antiferromagnetic phase transition in the
positive $U$ Hubbard model. However, it is well known that the problem of
superconductivity is formally related to the problem of antiferromagnetism,
in particular at half-filling where the nearest-neighbor hopping positive $U$
Hubbard model maps exactly onto the nearest-neighbor negative $U$ Hubbard
model. The corresponding canonical transformation maps the ${\bf q=}\left(
\pi ,\pi \right) $ spin correlations of the repulsive model onto the ${\bf %
q=0}$ pairing and ${\bf q=}\left( \pi ,\pi \right) $ charge correlations of
the attractive model while the single-particle Green's functions of both
models are identical. Thus all our results below concerning the opening of
the pseudogap in $A\left( {\bf k}_F,\omega \right) $ in the repulsive $U$
half-filled Hubbard model are directly applicable to the attractive $U$
model at half-filling, the only difference being in the physical
interpretation. While in the case of repulsive interaction the pseudogap is
due to the critical thermal spin fluctuation, in the case of attractive
interactions it is, obviously, due to the critical thermal pairing and
charge fluctuations. Away from half-filling the mapping between two models
is more complicated and the single particle spectra in the pairing pseudogap
regime $A\left( {\bf k}_F,\omega \right) $ have important qualitative
differences with the single particle spectra in the magnetic pseudogap
regime. However, even in this case there are very useful formal similarities
between two problems so that in Subsec.(\ref{Pairing}) we will give some
simple analytical results for the self-energy in the regime dominated by
critical pairing fluctuations.

The problem of precursor effects in the repulsive Hubbard model has been
first studied by Kampf and Schrieffer\cite{Kampf}. Their analysis however
was done at zero temperature and although the precursor effect that they
found, called ``shadow bands'', looks similar to what we find, there are a
number of important differences. For example, they find a quasiparticle
between the precursors of antiferromagnetic bands, while we do not. Also,
one does not obtain precursors at {\it zero} temperature when one uses our
more standard expression for the dynamical susceptibility instead of the
phenomenological form $\chi _{KShr}=f({\bf q})g(\omega )$ that they use. The
physical reason why a function that is separable in both momentum and
frequency, such as $\chi _{KShr}$, leads to qualitatively different results
than the conventional one has been explained in Ref.\cite{Yury3}. The
microscopic justification for $\chi _{KShr}$ is unclear. We comment below on
this problem as well as on some of the large related literature that has
appeared lately.

Repeating some of the arguments of Ref.\cite{Vilk2}, we first show by
general phase space arguments that the feedback of antiferromagnetic
fluctuations on quasiparticles has the potential of being strong enough to
destroy the Fermi liquid only in low enough dimension, the upper critical
dimension being three. Then we go into more detailed analysis to give
explicit analytic expressions for the quasi-singular part of the
self-energy, first in Matsubara frequency. The analysis of the self-energy
expression directly in real-frequencies is in Appendix~(\ref{AppRealFrequ}).
The latter analysis is useful to exhibit in the same formalism both the
Fermi liquid limit and the non-Fermi liquid limit.

For simplicity we give asymptotics for $n=1$ at the Fermi wave vector, where 
$\varepsilon ({\bf k}_F{\bf )}=0$, but similar results apply for $n\neq 1$
as long as there is long-range order at $T=0$ and one is below $T_X$. This
case is also discussed briefly, but for more details the reader is referred
to Ref.\cite{Yury3}.

\subsection{Upper critical dimension for the destruction of quasiparticles
by critical fluctuations.}

Before describing the effect of spin fluctuations on quasiparticles, we
first describe the so-called renormalized classical regime of spin
fluctuations that precedes the zero-temperature phase transition in two
dimensions.

\subsubsection{Renormalized classical regime of spin fluctuations.}

\label{SecRC}

The spin susceptibility $\chi _{sp}\left( {\bf q},0\right) $ below $T_X$ is
almost singular at the antiferromagnetic wave vector ${\bf Q}_2=\left( \pi
,\pi \right) $ because the energy scale $\delta U\equiv U_{mf,c}-U_{sp}$ ($%
U_{mf,c}\equiv 2/\chi _0\left( {\bf Q},0\right) $) associated with the
proximity to the SDW instability becomes exponentially small.\cite{Vilk}
This small energy scale, $\delta U<<T$, leads to the so-called renormalized
classical regime for the fluctuations.\cite{Chakravarty88} In this regime,
the main contribution to the sum over Matsubara frequencies entering the
local-moment sum rule Eq.(\ref{Spin}) comes from $iq_n=0$ and wave vectors $(%
{\bf q-Q)}^2\leq \xi ^{-2}$ near ${\bf Q}$. Approximating $\chi _{sp}\left( 
{\bf q},0\right) $ by its asymptotic form 
\begin{equation}
\chi _{sp}\left( {\bf q},0\right) \approx \frac 1{U_{sp}\xi _0^2}\frac 2{((%
{\bf q-Q}_d{\bf )}^2+\xi ^{-2})}  \label{ChiAs}
\end{equation}
where ${\bf Q}_2=\left( \pi ,\pi \right) $, ${\bf Q}_3=\left( \pi ,\pi ,\pi
\right) $ and 
\begin{equation}
\xi _0^2\equiv \frac{-1}{2\chi _0\left( Q\right) }\left. \frac{\partial
^2\chi _0\left( {\bf q}\right) }{\partial q_x^2}\right| _{{\bf q=Q}_d}\quad
;\quad \xi \equiv \xi _0(U_{sp}/\delta U)^{1/2}
\end{equation}
we obtain, in $d$ dimensions

\begin{equation}
\tilde{\sigma}^2=\frac{2T}{U_{sp}\xi _0^2}\int \frac{d^dq}{(2\pi )^d}\frac 1{%
q^2+\xi ^{-2}}  \label{ksi1}
\end{equation}
where $\tilde{\sigma}^2\equiv n-2\langle n_{\uparrow }n_{\downarrow }\rangle
-C<1$ is the left-hand side of Eq.(\ref{Spin}) minus corrections $C$ that
come from the sum over non-zero Matsubara frequencies (quantum effects) and
from $({\bf q-Q)}^2\gg \xi ^{-2}$. There is an upper cutoff to the integral
which is less than or of the order of the Brillouin zone size. The important
point is that the left-hand side of the above equation Eq.(\ref{ksi1}) is
bounded and weakly dependent on temperature. This implies, as discussed in
detail in Ref.\cite{Dare}, that the above equation leads to critical
exponents for the correlation length that are in the spherical model ($%
n\rightarrow \infty )$ universality class. For our purposes, it suffices to
notice that the integral converges even when $\xi \rightarrow \infty $ in
more than two dimensions. This leads to a finite transition temperature. In
two dimensions, the transition temperature is pushed down to zero
temperature and, doing the integral, one is left with a correlation length $%
\xi $ that grows exponentially below $T_X$%
\begin{equation}
\xi \sim \exp \left( \pi \tilde{\sigma}^2\xi _0^2\frac{U_{sp}}T\right) .
\label{ksi}
\end{equation}
The important consequence of this is that, below $T_X$, the correlation
length quickly becomes larger than the single-particle thermal de Broglie
wave length $\xi _{th}=v_F/\left( \pi T\right) $. This has dramatic
consequences on quasiparticles in two dimensions.

\subsubsection{Effect of critical spin fluctuations on quasiparticles.}

\label{SubSecEffectOnQP}

When the classical fluctuations ($iq_{n}=0$) become critical, they also
give, in two dimensions, a dominant contribution to the self-energy {\it at
low frequency}. To illustrate what we mean by the classical frequency
contribution, neglect the contribution of charge fluctuations and single out
the zero Matsubara frequency component from Eq.(\ref{param}) to obtain 
\[
\Sigma \left( {\bf k},ik_{n}\right) \approx Un_{-\sigma }+\frac{U}{4}\frac{T%
}{N}\sum_{{\bf q}}U_{sp}\chi _{sp}\left( {\bf q},0\right) \frac{1}{ik_{n}-%
\tilde{\epsilon}_{{\bf k+q}}} 
\]
\begin{equation}
+\frac{U}{4}\frac{T}{N}\sum_{{\bf q}}\sum_{iq_{n}\neq 0}U_{sp}\chi
_{sp}\left( {\bf q},iq_{n}\right) \frac{1}{ik_{n}+iq_{n}-\tilde{\epsilon}_{%
{\bf k+q}}}  \label{ClassicalContribution}
\end{equation}
Here, $\tilde{\epsilon}_{{\bf k}}$ is measured relative to the chemical
potential. The last term is the contribution from quantum fluctuations. In
this last term, the sum over Matsubara frequencies $iq_{n}$ must be done
before the analytical continuation of $ik_{n}$ to real frequencies otherwise
this analytical continuation would involve going through complex plane poles
of the other terms entering the full sum over $iq_{n}.$ The contribution
from classical fluctuations, $iq_{n}=0,$ does not have this problem and
furthermore it has the correct asymptotic behavior at $ik_{n}\rightarrow
\infty $. Hence the contribution of classical fluctuations to the retarded
self-energy $\Sigma ^{R}\left( {\bf k},\omega \right) $ can obtained from
the $iq_{n}=0$ term by trivial analytical continuation $ik_{n}\rightarrow
\omega +i0.$ Note also that the chemical potential entering $G^{\left(
0\right) }$ in the self-energy formula is $\mu _{0}=\mu =0$ at half-filling.

Doing the same substitution as above for the asymptotic form of the spin
susceptibility Eq.(\ref{ChiAs}) in the equation for the self-energy Eq.(\ref
{param}) one obtains the following contribution to $\Sigma $ from classical
fluctuations 
\begin{equation}
\Sigma _{cl}\left( {\bf k},ik_n\right) \cong \frac{UT}{2\xi _0^2}\int \frac{%
d^dq}{(2\pi )^d}\frac 1{q^2+\xi ^{-2}}\frac 1{ik_n-\tilde{\epsilon}_{{\bf k+Q%
}}-{\bf q\cdot v}_{{\bf k+Q}}}.  \label{Sigm}
\end{equation}
where we have expanded $\tilde{\epsilon}_{{\bf k+Q+q}}\simeq \tilde{\epsilon}%
_{{\bf k+Q}}+{\bf q\cdot v}_{{\bf k+Q}}$. In the case that we consider,
namely half-filling and ${\bf k=k}_F$, we have $\mu _0=\mu =0$ and $\tilde{%
\epsilon}_{{\bf k}_F{\bf +Q}}=0.$ The key point is again that in two
dimensions the integral in this equation Eq.(\ref{Sigm}) is divergent at
small $q$ for $\xi =\infty $. In a Fermi liquid, the imaginary part of the
self-energy at the Fermi surface $\left( \omega =0\right) $ behaves as $%
\Sigma _R^{\prime \prime }({\bf k}_F,0)\sim T^2$. Here instead, we find a
singular contribution 
\begin{equation}
\Sigma _R^{\prime \prime }({\bf k}_F,0)\propto T\int d^{d-1}q_{\perp }\frac 1%
{q_{\perp }^2+\xi ^{-2}}\propto T\xi ^{3-d}
\end{equation}
that is proportional to $\xi $ in $d=2$ and hence is very large $\Sigma
_R^{\prime \prime }({\bf k}_F,0)\approx -U\xi /(\xi _{th}\xi _0^2)>1$ when
the condition $\xi >\xi _{th}$ is realized. By contrast, 0for $d=3$, $\Sigma
_R^{\prime \prime }({\bf k}_F,0)\sim -U\left( \ln \xi \right) /\left( \xi
_0^2\xi _{th}\right) $, so that the Fermi liquid is destroyed only in a very
narrow temperature range close the N\'{e}el temperature $T_N$. Dimensional
analysis again suffices to show that in four dimensions the classical
critical fluctuations do not lead to any singular behavior. Three dimensions
then is the upper critical dimension. As usual, logarithmic corrections
exist at the upper critical dimension. The effect will be very small in
three dimensions not only because it is logarithmic, but also because the
fluctuation regime is very small, extending only in a narrow temperature
range around the N\'{e}el temperature. By contrast, in two dimensions the
effect extends all the way from the crossover temperature, $T_X$, which is
of the order of the mean-field transition temperature, to zero temperature
where the transition is.

Wave vectors near Van Hove singularities are even more sensitive to
classical thermal fluctuations. Indeed, near this point the expansion should
be of the type $\epsilon _{{\bf k}_{VH}{\bf +q+}\left( \pi ,\pi \right)
}\propto q_x^2-q_y^2$. This leads, in two dimensions, to even stronger
divergence in $\Sigma _R^{\prime \prime }({\bf k}_F,0)\propto $ $T\xi ^2\int
dq_y\left[ \left( 2q_y^2+1\right) \left| q_y\right| \right] ^{-1} $\cite
{Yury3}. Even if the logarithmic divergence is cutoff the prefactor is
larger by a factor of $\xi $ compared with points far from the Van Hove
singularities.

\subsection{Precursors of antiferromagnetic bands in two dimensions.}

\label{SubSecPrecursors}

Let us analyze in more details the consequences of this singular
contribution of critical fluctuations to the self-energy in two dimensions.
The integral appearing in the two-dimensional version of the expression for
the self-energy, Eq.(\ref{Sigm}), can be performed exactly\cite{Integrals}

\begin{equation}
\Sigma \left( {\bf k}_F,ik_n\right) =\frac U2-i\frac{UT}{8\pi \xi _0^2\sqrt{%
k_n^2-v_F^2\xi ^{-2}}}\ln \frac{k_n+\sqrt{k_n^2-v_F^2\xi ^{-2}}}{k_n-\sqrt{%
k_n^2-v_F^2\xi ^{-2}}}+{\cal R}.  \label{SigmaM}
\end{equation}
Here ${\cal R}$ is a regular part.

As a first application, we can use this expression to understand
qualitatively both the temperature and size dependence of the Monte Carlo
data for $\tilde{z}(T)\ $appearing in Fig.(2) of Ref.\cite{Vilk2} or in the
lower panel of Fig.(\ref{FigJoint}) . Indeed, $\tilde{z}(T)\ $can be written
as the alternating series $-2G\left( {\bf k}_F,\beta /2\right)
=-4T\sum_{n=1}^\infty \left( -1\right) ^n/\left( k_n-\Sigma ^{\prime \prime
}\left( {\bf k}_F,ik_n\right) \right) $. Even though the series converges
slowly, in the beginning of the renormalized classical regime and for
qualitative purposes it suffices to use the first term of this series. Then,
using the expressions for the correlation length Eqs. (\ref{ksi}) and for
the self-energy (\ref{SigmaM}). One finds, 
\begin{equation}
\tilde{z}(T)\sim \frac{T^2}{\tilde{\sigma}^2UU_{sp}}\sqrt{1-\frac{\xi _{th}^2%
}{\xi ^2}}\quad ,\quad T_X-T\ll T_X.
\end{equation}
On the infinite lattice, $\xi $ starts growing exponentially below $T_X$,
quickly becoming much larger than $\xi _{th}$. This implies $\tilde{z}%
(T)\simeq T^2$. On finite lattices $\xi \sim \sqrt{N}$, which explains the
size effect observed in Monte Carlo {\it i.e.} smaller $\tilde{z}$ for
smaller size $N$, ($\xi _{th}(T_X)\sim 5$ for Fig.(\ref{FigJoint})).

The analytic continuation of $\Sigma \left( {\bf k}_F,ik_n\right) $ in Eq.(%
\ref{SigmaM}) is

\begin{equation}
\Sigma ^R\left( {\bf k}_F,\omega \right) =\frac U2+\frac{UT}{8\pi \xi _0^2%
\sqrt{\omega ^2+v_F^2\xi ^{-2}}}\left[ \ln \left| \frac{\omega +\sqrt{\omega
^2+v_F^2\xi ^{-2}}}{\omega -\sqrt{\omega ^2+v_F^2\xi ^{-2}}}\right| -i\pi
\right] +{\cal R}.  \label{Somega}
\end{equation}
For the wave vectors ${\bf k}$ away from the Fermi surface the anomalous
contribution due to the classical fluctuation has a similar form but with $%
\omega $ replaced by $(\omega -\tilde{\epsilon}_{{\bf k+Q}})$. When $T>T_X$,
the correlation length $\xi $ becomes of order unity and, as we will show in
Appendix (\ref{AppRealFrequ}), the regular part ${\cal R}$ dominates so that
one recovers standard Fermi liquid behavior. Furthermore, even for large
correlation length the regular part cannot be neglected when $\omega \gg T$
since the term exhibited here becomes small. Hence we concentrate on small
frequencies and on $T<T_X$ where the regular part ${\cal R}$ can be
neglected.

Exactly at the Fermi level $\left( \omega =0\right) $ we recover the result
of the previous section, namely that the imaginary part of the self-energy
for $\xi >\xi _{th}$ increases exponentially when the temperature decreases, 
$\Sigma ^{\prime \prime }({\bf k}_{F},0)\sim U\xi /(\xi _{th}\xi
_{0}^{2})\propto T\xi \propto T\exp \left( \pi \tilde{\sigma}^{2}\xi
_{0}^{2}U_{sp}/T\right) $. The above analysis shows by contradiction that in
the paramagnetic state below $T_{X}$ there is no Fermi-liquid quasiparticle
at $k_{F},$ yet the symmetry of the system remains unbroken at any finite $T$%
. Indeed, starting from quasiparticles $\left( G_{\sigma }^{\left( 0\right)
}\right) $ we found that as temperature decreases, $\Sigma _{R}^{\prime
\prime }({\bf k}_{F},0)$ {\it increases} indefinitely instead of {\it %
decreasing}, in direct contradiction with the starting hypothesis. By
contrast, a self-consistent treatment where we use in Eq.(\ref{param}) the
full $G_{\sigma }$ with a large $\Sigma _{R}^{\prime \prime }({\bf k}_{F},0)$
shows that, for $T<T_{X}$ , $\Sigma _{R}^{\prime \prime }({\bf k}_{F},0)$
remains large in $d=2$ and does not vanish as $T\rightarrow 0$, again
confirming that the system is not a Fermi liquid in this regime (See however
Sec.(\ref{SubSecWhyFLEX}) below). Strong modifications to the usual Fermi
liquid picture also persist away from half-filling as long as $T_{X}(n)>0$,
as we discuss later.

One can check that the large $\Sigma _R^{\prime \prime }({\bf k}_F,0)$ in
two dimensions (for $T<T_X)$ leads to a pseudogap in the infinite lattice,
contrary to the conclusion reached in Ref.\cite{White}. Indeed, instead of a
quasiparticle peak, the spectral weight $A\left( {\bf k}_F,\omega \right)
\equiv -2$Im$G_R\left( {\bf k}_F,\omega \right) $ has a {\it minimum} at the
Fermi level $\omega =0$ and two symmetrically located maxima away from it.
More specifically, for $v_F/\xi <\left| \omega \right| <T$ we have 
\begin{equation}
A\left( {\bf k}_F,\omega \right) \cong \frac{2\left| \omega \right| UT/(8\xi
_0^2)}{[\omega ^2-UU_{sp}\tilde{\sigma}^2/4]^2+[UT/(8\xi _0^2)]^2}.
\label{SpectralWeight}
\end{equation}
The maxima are located at $\omega =\pm \sqrt{UU_{sp}}\tilde{\sigma}/2$.
These two maxima away from zero frequency correspond to precursors of the
zero-temperature antiferromagnetic (or SDW) bands (shadow bands\cite{Kampf}%
).There is no quasiparticle peak between these two maxima when $\xi >\xi
_{th}$. This remains true in the case of no perfect nesting as well\cite
{Yury3} (see also Sec.(\ref{SubSecAway})). We note that this is different
from the results of the zero-temperature $(\xi _{th}=\infty )$ calculations
of Kampf and Schrieffer\cite{Kampf} that were based on a phenomenological
susceptibility separable in momentum and frequency $\chi _{K.Sh.}=f({\bf q}%
)g(\omega ).$ As was explained in Ref.\cite{Yury3}, the existence of
precursors of antiferromagnetic bands (shadow bands in the terminology of
Ref.\cite{Kampf}) {\it at zero temperature} is an artifact of the separable
form of the susceptibility. The third peak between the two precursors of
antiferromagnetic bands that was found in Ref.\cite{Kampf} is due to the
fact that at zero temperature the imaginary part of the self-energy $\Sigma
^{\prime \prime }({\bf k},\omega =0,T=0)$ is strictly zero at all ${\bf k}$.
In our calculations, precursor bands appear only at finite temperature when
the system is moving towards a zero-temperature phase transition. In this
case, the imaginary part of the self-energy goes to infinity for ${\bf k}$
on the ``shadow Fermi surface'' $\lim_{T\rightarrow 0}\Sigma ^{\prime \prime
}({\bf k}_F+{\bf Q},0)\propto T\xi \propto T\exp \left( Cst/T\right)
\rightarrow \infty $ and to zero at all other wave vectors. This is
consistent with the SDW result which we should recover at $T=0$. Indeed, the
latter result can be described by the self-energy $\Sigma ^R({\bf k,}\omega
)=\Delta ^2/(\omega -\tilde{\varepsilon}({\bf k}+{\bf Q})+i\eta )$ which
implies that the imaginary is a delta function $\Sigma ^{\prime \prime }(%
{\bf k,}\omega )=-\pi \delta (\omega -\tilde{\varepsilon}({\bf k}+{\bf Q}))$
instead of zero at all ${\bf k}$ as in a Fermi liquid. We note also that
analyticity and the zero value of $\Sigma ^{\prime \prime }({\bf k},\omega
=0)$ in Ref.\cite{Kampf} automatically implies that the slope of the real
part of the self-energy $\partial \Sigma ^{\prime }({\bf k},\omega
)/\partial \omega |_{\omega =0}$ is negative. By contrast, in our case $%
\partial \Sigma ^{\prime }({\bf k}_F+{\bf Q},\omega )/\partial \omega
|_{\omega =0}$ is positive and increases with decreasing temperature,
eventually diverging at the zero-temperature phase transition. The real part
of the self-energy obtained using the asymptotic form Eq.(\ref{Somega}) is
at the bottom left corner of Fig.(\ref{Akom}) with the corresponding
spectral function $A\left( {\bf k}_F,\omega \right) $ $\,$shown above it. In
Fig.(\ref{FigSpectralWeight}) we have already shown the evolution of the
spectral function $A\left( {\bf k}_F,\omega \right) $ with temperature. The
positions of the precursors of antiferromagnetic bands scale like $\tilde{%
\sigma}/2$ which itself, at small coupling in two dimensions, scales like
the mean field SDW transition temperature or gap (see Appendix B of Ref.\cite
{Dare}). As $U$ increases, the predicted positions of the maxima obtained
from the asymptotic form Eq.(\ref{SpectralWeight}) will be less accurate
since they will be at intermediate frequencies and the regular quantum
contribution to the self-energy will affect more and more the position of
the peaks.

We have predicted\cite{Vilk2} that the exponential growth of the magnetic
correlation length $\xi $ below $T_X$ will be accompanied by the appearance
of precursors of SDW bands in $A\left( {\bf k}_F,\omega \right) $ with no
quasiparticle peak between them. By contrast with isotropic materials, in
quasi-two-dimensional materials this effect should exist in a wide
temperature range, from $T_X$ $\left( T_X<<U<E_F\right) $ to the N\'{e}el
temperature $T_N$ ($T_X-T_N\sim 10^2\,\,K$).

\subsection{Contrast between magnetic precursor effects and Hubbard bands}

Although there are some formal similarities between the precursors of
antiferromagnetic bands and the Hubbard bands (see Sec.\ref{SecFLEX}) we
would like to stress that these are two different physical phenomena. A
clear illustration of this is when a four peak structure exists in the
spectral function $A({\bf k},\omega ),$ two peaks being precursors of
antiferromagnetic bands, and two peaks being upper and lower Hubbard bands.
The main differences between these bands are in the ${\bf k}${\em $-$%
dependence} of the self-energy $\Sigma ({\bf k},\omega )$ and in the
conditions for which these bands develop. Precursors of antiferromagnetic
bands appear even for small $U$ in the renormalized classical regime $T<T_X$%
, and their dispersion has the quasi-periodicity of the magnetic Brillouin
zone. In contrast, upper and lower Hubbard bands are high-frequency features
that appear only for sufficiently large $U>W$ and $T<U$ and have the
periodicity of the whole Brillouin zone in the paramagnetic state.
Furthermore, the existence of Hubbard bands is not sensitive to
dimensionality so they exist even in infinite dimension where the
self-energy does not depends on momentum ${\bf k}$ at all. In contrast, the
upper critical dimension for the precursors of antiferromagnetic bands is
three (see Sec.(\ref{SubSecAbsIn3d})).

In our theory the precursors of antiferromagnetic bands come from the almost
singular behavior of the zero Matsubara frequency susceptibility $\chi
_{sp}\left( {\bf q}{,0}\right) $, which leads to the characteristic behavior
of $\Sigma ({\bf k},\omega )=\Delta _{Sh.B}^2/\left( \omega -\varepsilon (%
{\bf k+Q})\right) $ with $\Delta _{Sh.B}^2\propto T\ln (\xi )$. On another
hand, the Hubbard bands appear in our theory because the high-frequency
asymptotics $\Sigma ({\bf k},\omega )\propto \Delta _{H.B}^2/\omega $ has
already set in for $\omega >W$, and this leads to the bands at $\omega =\pm
\Delta _{H.B}$ for $\Delta >W$. (see for more details Sec.(\ref{SecFLEX})).
The coefficient $\Delta _{H.B}^2$ is determined by the sum over all
Matsubara frequencies and ${\bf q}$: $\Delta _{H.B}^2=TUN^{-1}\sum_{{\bf q}%
,n}\left[ U_{sp}\chi _{sp}({\bf q},i\omega _n)+U_{ch}\chi _{ch}({\bf q}%
,i\omega _n)\right] $.

It was noticed in Monte Carlo simulations\cite{Dagotto}\cite{Hanke} that for
intermediate $U$, the spectral weight has four maxima. We think that peaks
at $\omega \sim \pm U/2$ are Hubbard bands, while the peaks closer to $%
\omega =0$ are precursors of antiferromagnetic bands. If this interpretation
is correct, then the latter peaks should disappear with increasing
temperature when $\xi $ becomes smaller than $\xi _{th}$, while the Hubbard
bands should exist as long as $T<U$.

While the location of the precursors of antiferromagnetic bands should be
accurate in our theory, the same will not be true for the location of the
upper and lower Hubbard bands. This is because our theory is tuned to the
low frequency behavior of the irreducible vertices and does not have the
right numerical coefficient in the high-frequency expansion of the
self-energy, as shown in Eq.(\ref{OG}) below. Nevertheless, our analytical
approach to date is the only one that agrees at least qualitatively with the
finding that precursors of antiferromagnetic bands as well as upper and
lower Hubbard bands can occur simultaneously. Note however that a four peak
structure at $n=1$ was also obtained in Ref.\cite{Matsumoto} but the
physical difference between Hubbard bands and precursors of
antiferromagnetic bands was not clearly spelled out. We comment on recent
findings of the FLEX approach in Sec.(\ref{SecFLEX})\cite{Schmalian}\cite
{Langer}\cite{Serene}.

\subsection{Can the precursors of antiferromagnetic bands exist in three
dimensions?}

\label{SubSecAbsIn3d}

In two dimensions, the finite-temperature phase is disordered, but the
zero-temperature one is ordered and has a finite gap, except at the quantum
critical point away from half-filling. Hence, precursors of
antiferromagnetic bands that appear in the paramagnetic state do so with a
finite pseudogap which appears consistent with the finite zero-temperature
gap towards which the system is evolving. By contrast, in higher dimensions
the gap opens-up with a zero value at the transition temperature. Based on
this simple argument, one does not expect precursors of antiferromagnetic
bands in dimensions larger than two (see, however, below). Here, we will
also show that there is no phase space reasons for the existence of
precursors of the antiferromagnetic bands when $d>2$.

We have already shown that in three dimensions the quasiparticle at the
Fermi level at half-filling will have an imaginary part of the self-energy
that grows like $T\ln \xi ,$ an effect that is much weaker than $T\xi $
found in two dimensions. Despite this small effect, in three dimensions the
classical fluctuations do not affect the self-energy for energies larger
than $v_F\xi ^{-1}$. Indeed, consider the contribution of classical thermal
fluctuations to the self-energy Eq.(\ref{Sigm}). In two dimensions, we have
for $\left| \omega \right| >v_F\xi ^{-1}$%
\begin{equation}
\text{Re}\left[ \Sigma _{cl}^{2d}\left( {\bf k}_F,\omega \right) \right]
\cong \frac{UT}{2\xi _0^2}\int \frac{d^2q}{(2\pi )^2}\frac 1{q^2+\xi ^{-2}}%
\frac 1\omega .  \label{Sigm2}
\end{equation}
which allows us to recover the approximate formula for the spectral weight
given in Eq.(\ref{SpectralWeight}) above. In three dimensions however, this
approximation cannot be done because the integral is {\it not} dominated by
small values of $q$. To see this explicitly in three dimensions, consider
the contribution of classical thermal fluctuations 
\begin{equation}
\Sigma _{cl}^{3d}\left( {\bf k}_F,\omega +i\eta \right) \cong \frac{UT}{2\xi
_0^2}\int \frac{dq_{\Vert }}{2\pi }\int \frac{d^2q_{\bot }}{(2\pi )^2}\frac 1%
{q_{\bot }^2+q_{\Vert }^2+\xi ^{-2}}\frac 1{\omega +i\eta +v_Fq_{\Vert }}
\end{equation}
\begin{equation}
\cong \frac{UT}{2\xi _0^2}\frac 1{4\pi }\int \frac{dq_{\Vert }}{2\pi }\ln
\left[ \frac{\Lambda _{\bot }^2+q_{\Vert }^2+\xi ^{-2}}{q_{\Vert }^2+\xi
^{-2}}\right] \frac 1{\omega +i\eta +v_Fq_{\Vert }}.
\end{equation}
As long as $\left| \omega \right| >v_F\xi ^{-1}$, the logarithmic
singularity that develops at $q_{\Vert }=0$ when $\xi ^{-1}\rightarrow 0$ is
integrable and gives no singular contribution to the self-energy. Hence,
unusual effects of classical thermal fluctuations are confined to the range
of frequencies $\left| \omega \right| <v_F\xi ^{-1}.$ At higher frequencies, 
$\left| \omega \right| >v_F\xi ^{-1}$, all bosonic Matsubara frequencies in
Eq.(\ref{param}) need to be taken into account and from phase space
considerations alone there is no reason for the existence of precursors of
antiferromagnetic bands in the $3D$ case. However, the existence of such
bands in $3D$ cannot be completely excluded based on dimensional arguments
alone because they occur at finite frequencies and strictly speaking they
are non-universal. In particular, as discussed in Ref.\cite{Dare}, one
expects to see precursors that look like $2D$ antiferromagnetic bands
(shadow bands) in the vicinity of the finite temperature phase transition in
strongly anisotropic quasi-two-dimensional material. On the other hand, such
bands do not generically exist in the almost isotropic $3D$ case, because
even in $2D$ the conditions for such bands are quite stringent. The
difference between shadow bands and Hubbard bands has been discussed in the
previous subsection and the discussion of non-analyticities sometimes
encountered in Fermi liquid theory can be found in Appendix (\ref
{AppRealFrequ}).

\subsection{Away from half-filling}

\label{SubSecAway}

Close to half-filling, in the nearest-neighbor hopping model, one can enter
a renormalized classical regime with large antiferromagnetic correlation
length, even though the zero-temperature Fermi surface properties may favor
incommensurate correlations. This renormalized-classical regime with large $%
\left( \pi ,\pi \right) $ correlations occurs when $T_{X}\gg \mu _{0}$. By
arguments similar to those above, one finds that in this regime one still
has precursors of antiferromagnetic bands. However, the chemical potential
is in or near the lower precursor band and the system remains metallic. The
high-frequency precursor appears only below $T_{X}$ at $\omega \approx 
\tilde{\varepsilon}_{{\bf k}+{\bf Q}}$.

With second-neighbor hopping, the points of the Fermi surface that intersect
the magnetic Brillouin zone (hot spots) behave as does the whole Fermi
surface of the nearest-neighbor (nested) case discussed above. These
questions were discussed in detail in Ref.\cite{Yury3}.

\subsection{The pairing pseudogap and precursors of superconducting bands in
two dimensions}

\label{Pairing}

As we have already pointed out above, the results for the single particle
spectra obtained for the half-filled nearest-neighbor hopping repulsive
Hubbard model can be directly applied to the corresponding attractive
Hubbard model, in which case the pseudogap opens up in the renormalized
classical regime of pairing and charge fluctuations. Away from half-filling,
the symmetry between charge and pair correlations is lost and pair
fluctuations dominate, becoming infinite at the Kosterlitz-Thouless
transition temperature. This temperature is below the temperature at which
the magnitude of the pair order parameter acquires rigidity despite the
randomness of its phase. One expects then that a pseudogap will also open in
this case when the correlation length for pairing fluctuations becomes
larger than the single-particle thermal de Broglie wavelength $\xi
_{pairing}>\xi _{th}=v_F/T$. This should occur below the crossover
temperature to the renormalized classical regime of pairing fluctuations but
above the Kosterlitz-Thouless transition temperature.

The quantitative microscopic theory for the negative $U$ Hubbard model will
be considered in a separate publication. By contrast with all other sections
of this paper, our considerations here will be more phenomenological.
Nevertheless, they will allow us to present some analytical results for the
self-energy obtained in the critical regime dominated by pairing
fluctuations. Details of the model should not be very important since we are
in a regime where everything is dominated by long wave length fluctuations.

The derivation of $\Sigma ({\bf k},\omega )$ in the pairing case is a
straightforward extension of what we did in the antiferromagnetic case (see
Subsecs.(\ref{SubSecEffectOnQP}), (\ref{SubSecPrecursors}) and Ref.\cite
{Yury3}). In particular, in complete analogy with the magnetic case, the
main contribution to the self-energy in the critical regime comes from the
classical thermal fluctuations $iq_n=0$. Assuming some effective coupling
constant $g^{\prime }$ between quasiparticles and pairing fluctuations,
which in general can be momentum dependent, one can write in the one loop
approximation 
\begin{equation}
\Sigma _{cl}\left( {\bf k},ik_n\right) \approx Tg^{\prime }({\bf k})\int 
\frac{d^2q}{(2\pi )^2}\frac 1{\xi _p^{-2}+q^2}\frac 1{ik_n+\tilde{\epsilon}%
_{-{\bf k+q}}}  \label{SigmPairing}
\end{equation}
Here $\tilde{\epsilon}_{{\bf k}}$ is the electron dispersion relative to the
chemical potential, and all factors in front of integral are reabsorbed into
the coupling constant $g^{\prime }$. This expression is similar to the
expression Eq.(\ref{Sigm}) in the magnetic case but there are two important
differences: {\it i)} instead of $\tilde{\epsilon}_{{\bf k}+{\bf Q+q}}$ we
have now $\tilde{\epsilon}_{-{\bf k+q}};$ {\it ii)} there is no minus sign
in front of $\tilde{\epsilon}_{-{\bf k+q}}$. The first difference is due to
the fact that superconductivity usually occurs with zero center of mass
momentum for the pair, and hence the pairing susceptibility in the normal
state $\chi _p\propto 1/(\xi _p^{-2}+q^2)$ must be peaked near ${\bf q}=0$,
(the integration variable ${\bf q}$ in Eq. (\ref{Sigm}) was measured
relative to ${\bf Q}=(\pi ,\pi )$). The second difference comes from the
fact that we are now considering the contribution to $\Sigma $ coming from
the particle-particle channel instead of the particle-hole channel. Taking
the integrals over ${\bf q}$ and using the fact that small ${\bf q}$ only
will contribute we neglect the ${\bf q}$ dependence of the coupling constant
and obtain for the imaginary part of $\Sigma _{cl}$ the following expression 
\begin{equation}
\Sigma ^{\prime \prime }({\bf k},\omega )=-\frac{g^{\prime }({\bf k})T}{4%
\sqrt{\left( \omega +\tilde{\varepsilon}_{-{\bf k}}\right) ^2+v_{-{\bf k}%
}^2\xi _p^{-2}}}  \label{SigI}
\end{equation}

In the renormalized classical regime the pairing correlation length $\xi _p$
increases faster with decreasing temperature than $\xi _{th}=v_F/T$.
Consequently, $\Sigma ^{\prime \prime }({\bf k}_F,0)$ tends to diverge with
decreasing temperature and a pairing pseudogap in the spectral function $A(%
{\bf k}_F,\omega )$ opens up over the complete Fermi surface, except maybe
at a few points where $g^{\prime }({\bf k})=0$. This is different from the
antiferromagnetic case, where the pseudogap in $A({\bf k}_F={\bf k}%
_{h.sp.},\omega )$ opens up only when, so called, ``hot spots'' ($\tilde{%
\varepsilon}({\bf k}_{h.sp.}+{\bf Q})=\tilde{\varepsilon}({\bf k}_{h.sp.})=0$%
) exist in a given model\cite{Yury3}. The antiferromagnetic pseudogap opens
everywhere on the Fermi surface only in the case of perfect nesting, where
all points on the Fermi surface are ``hot spots''.

The real part of the self-energy can be obtained from Eq.(\ref{SigI}) using
the Kramers-Kronig relation and has the form:

\begin{equation}
\Sigma ^{\prime }({\bf k},\omega )=\frac{g^{\prime }({\bf k})T}{4\pi \sqrt{%
\left( \omega +\tilde{\varepsilon}_{-{\bf k}}\right) ^2+v_{-{\bf k}}^2\xi
_p^{-2}}}\ln \left| \frac{\omega +\tilde{\varepsilon}_{-{\bf k}}+\sqrt{%
\left( \omega +\tilde{\varepsilon}_{-{\bf k}}\right) ^2+v_{-{\bf k}}^2\xi
_p^{-2}}}{\omega +\tilde{\varepsilon}_{-{\bf k}}-\sqrt{\left( \omega +\tilde{%
\varepsilon}_{-{\bf k}}\right) ^2+v_{-{\bf k}}^2\xi _p^{-2}}}\right|
\label{SigR}
\end{equation}
To understand how precursors of the superconducting bands develop, let us
look at $\Sigma ^{\prime }({\bf k},\omega )$ at frequencies $|\omega +\tilde{%
\varepsilon}_{-{\bf k}}|\gg v_{-{\bf k}}\xi _p^{-1}$. In this case, using
inversion symmetry $\tilde{\varepsilon}_{-{\bf k}}=\tilde{\varepsilon}_{{\bf %
k}}$, one can obtain from Eq.(\ref{SigR}) the following asymptotic form 
\begin{equation}
\Sigma ^{\prime }({\bf k},\omega )\approx \frac{g^{\prime }({\bf k})}{2\pi }%
\frac{T\ln \xi _p}{\omega +\tilde{\varepsilon}_{{\bf k}}}
\label{SigRlimitLargeW}
\end{equation}
When, $\xi _p\sim \exp (const/T)$ (see, more general case below) this form
of the self-energy leads to the usual BCS result $\Sigma ^{\prime }({\bf k}%
,\omega )\approx \Delta ^2\left( {\bf k}\right) /(\omega +\tilde{\varepsilon}%
_{{\bf k}})$ with the gap $\Delta ^2\left( {\bf k}\right) \approx (g^{\prime
}({\bf k})/2\pi )T\ln \xi _p$. On the other hand, the imaginary part $\Sigma
^{\prime \prime }({\bf k},\omega ),$ Eq.(\ref{SigI}), vanishes everywhere in
the $T=0$ limit, except when $\omega =-\tilde{\varepsilon}_{{\bf k}}$ where
it becomes infinite. The results for $\Sigma ^{\prime }$ and $\Sigma
^{\prime \prime }$ can thus be combined to write for the corresponding limit
of the retarded self-energy $\Sigma ^R=\Delta ^2\left( {\bf k}\right)
/(\omega +\tilde{\varepsilon}_{{\bf k}}+i\eta )$. This limit leads to the
standard BCS expression for the normal Green's function when substituted
back into the Dyson equation $G^R=1/\left( \omega +i\eta -\tilde{\varepsilon}%
_{{\bf k}}-\Sigma ^R({\bf k},\omega )\right) $. Above the transition
temperature, the anomalous Green's function remains zero since there is no
broken symmetry. The qualitative picture for the development of the pairing
pseudogap and of the precursors of superconducting bands at ${\bf k}={\bf k}%
_F$ is illustrated in Fig.(\ref{FigSpectralWeight}) and in the left part of
Fig.(\ref{Akom}). While in the case of magnetic critical fluctuations these
figures describe the precursor effect in $A({\bf k}_F,\omega )$ for
perfect-nesting or for the ``hot spots'' (when such points exist), in the
case of pairing fluctuations they describe the spectra for all ${\bf k}_F$
and for all fillings where the ground state is superconducting.

We need to comment on a subtle difference between the antiferromagnetic and
the pairing precursor effects in the single particle spectra. While the
magnetic order parameter has three components and can order only at zero
temperature in the two-dimensional repulsive model, away from half-filling
in the attractive model the pairing order parameter becomes the only
relevant order parameter at low temperature. Since it has only two
components, a finite temperature Kosterlitz-Thouless phase transition is
then allowed in two dimensions. The critical behavior in vicinity of this
transition is given by $\xi _{p}\propto \exp [const/(T-T_{KT})^{1/2}]$
instead of $\xi \propto \exp (const/T)$ as in the magnetic case. To take
this properly into account one would need a treatment of the problem that is
more sophisticated than that given above. In particular, one would have to
take into account corrections to the simple form that we used for the
pairing susceptibility $\chi _{p}(q,0)\propto 1/(\xi _{p}^{-2}+q^{2}).$ This
Lorentzian form of the susceptibility in the critical regime is strictly
valid only in the $n=\infty $ limit ( $n$ is the number of the components of
the order parameter) and is, clearly, a less accurate approximation in the
case of pairing fluctuations $\left( n=2\right) $ than in the case of the
antiferromagnetic fluctuations $\left( n=3\right) $. Nevertheless, we
believe that qualitatively the picture given above is correct for two
reasons. First, because in the Kosterlitz-Thouless picture the magnitude of
the order parameter is locally non-zero starting below a crossover
temperature $T_{X}$ that is larger than the transition temperature $T_{KT}$.
It is only the phase that is globally decorrelated above $T_{KT}$. This
means that locally the quasiparticles are basically in a superconducting
state even above $T_{KT}.$ A second reason to believe in the precursor
effects is that the superfluid density and the gap are {\em finite} as $%
T\rightarrow T_{K.T.}^{-}$ and, hence, the two peak structure in $A({\bf k}%
_{F},\omega )$ exists even as the phase transition point is approached from
the low-temperature side. By analogy with the antiferromagnetic case, this
two peak structure should not immediately disappear when one increases the
temperature slightly above $T_{KT}$.

Finally, we point out that the precursor phenomenon described above has to
be distinguished from, so-called, pre-formed pairs considered first by P.
Nozi\`{e}res and S. Schmitt-Rink\cite{NSchmitt} (see also \cite{Melo}).
These pre-formed pairs exist in any dimension when the coupling strength is
sufficiently large, while the precursor effect considered above can be
caused by arbitrarily small attractive interactions but only in two
dimensions. We think that recent Monte Carlo data\cite{RanderiaU<0} on the
negative $U=-W/2$ Hubbard model illustrates the opening of the
single-particle pseudogap due to critical fluctuations, rather than a
strong-coupling effect. In these simulations, the drop in the density of
states at the Fermi level should be accompanied by a simultaneous rapid
increase of the pairing structure factor $S_p({\bf q=}0,T)$ . The latter
must be exponential in the infinite 2D lattice and a size analysis of Monte
Carlo data similar to the one shown on Fig.(\ref{FigJoint}) would be
extremely helpful to clarify this issue.

\section{Absence of the Precursors of antiferromagnetic bands and upper and
lower Hubbard bands in Eliashberg-type self-consistent theories}

\label{SecFLEX}

In this section, we explain why the theories that use self-consistent
propagators but neglect the corresponding frequency-dependent vertex
corrections {\it fail} to see two important physical effects: namely upper
and lower Hubbard bands, as well as the precursors of antiferromagnetic
bands that we just discussed. The failure of this type of self-consistent
schemes to correctly predict upper and lower Hubbard bands has been realized
a long time ago in the context of calculations in infinite dimension\cite
{BandesHubbardDinf}\cite{Georges}. While one may brush aside this failure by
claiming that high-energy phenomena are not so relevant to low-energy
physics, we show that in fact these schemes also fail to reproduce the {\it %
low-energy} pseudogap and the precursors of antiferromagnetic bands for
essentially the same reasons that they fail to see Hubbard bands. It is thus
useful to start by a discussion of the better understood phenomenon of upper
and lower Hubbard bands and then to move to precursors of antiferromagnetic
bands.

\subsection{Why Eliashberg-type self-consistency for the electronic
self-energy kills Hubbard bands}

\label{HubbardBands}

We first note that ordinary perturbation theory satisfies the correct
high-frequency behavior Eq.(\ref{SelfHaut}) for the self-energy namely, for $%
k_n\gg W$%
\begin{equation}
\lim_{ik_n\rightarrow \infty }\Sigma _\sigma \left( {\bf k,}ik_n\right)
=Un_{-\sigma }+\frac{U^2n_{-\sigma }\left( 1-n_{-\sigma }\right) }{ik_n}%
+\ldots  \label{SelfHaut}
\end{equation}
It is the latter property that guarantees the existence of the Hubbard bands
for $U>W$. To see this, consider the half-filled case. In this case, $%
n_{-\sigma }=1/2$, $\mu =U/2$ and one finds for the spectral weight 
\begin{equation}
A\left( {\bf k},\omega \right) \sim \frac{-2\Sigma ^{\prime \prime }}{\left(
\omega -\frac{U^2}{4\omega }\right) ^2+\Sigma ^{\prime \prime 2}}
\label{HLHubbard}
\end{equation}
which has pronounced maxima at the upper and lower Hubbard bands, namely $%
\omega =\pm U/2,$has long as $\Sigma ^{\prime \prime }$is not too large.
Since these results are obtained using high-frequency asymptotics, they are
valid only when the asymptotic Eq.(\ref{SelfHaut}) has already set in when $%
\omega \sim U/2.$In the exact theory and in ordinary perturbation theory in
terms of {\em bare} Green functions $G^{\left( 0\right) }$, Eq.(\ref
{SelfHaut}) is valid for $|\omega |\gg W$and the Hubbard bands appears as
soon as $U$become larger than $W$.

The fact that this simple high-frequency behavior sets in at the energy
scale given by $W$rather than $U$, even when $W<U$, is a non-trivial
consequence of the Pauli principle. To see this we first recall the exact
result for the self-energy $\Sigma _\sigma \left( {\bf k,}ik_n\right) $in
the atomic limit\cite{Hubbard} 
\begin{equation}
\Sigma _\sigma ^{atomic}\left( {\bf k,}ik_n\right) =Un_{-\sigma }+\frac{%
U^2n_{-\sigma }\left( 1-n_{-\sigma }\right) }{ik_n+\mu -U\left( 1-n_{-\sigma
}\right) }  \label{Exact}
\end{equation}
Formally, the atomic limit means that hopping is the smallest of all energy
scales in the problem, including the temperature, $t\ll T$, which is not a
very interesting case. However, the same arguments that have been used to
derive the expression (\ref{Exact}) in the atomic limit can be used to show
that Eq.(\ref{Exact}) is valid at any $T/t$when $k_n\gg W$. Indeed, in the
equations of motion for two-particle correlators\cite{Hubbard} one can
neglect hopping terms when $k_n\gg W$. This is where the asymptotic behavior
(\ref{Exact}) sets in since the equations of motion then immediately lend
themselves to a solution without any additional approximation for the
interacting term. This solution is possible because the Pauli principle $%
n_{i\sigma }^2=n_{i\sigma }$allows us to collapse three-particle correlation
function which enters equation of motion to the two-particle one $%
U\left\langle T_\tau \left( n_{i-\sigma }\left( \tau _i\right) n_{i-\sigma
}\left( \tau _i\right) c_{i\sigma }\left( \tau _i\right) c_{j\sigma
}^{\dagger }\left( \tau _j\right) \right) \right\rangle =U\left\langle
T_\tau \left( n_{i-\sigma }\left( \tau _i\right) c_{i\sigma }\left( \tau
_i\right) c_{j\sigma }^{\dagger }\left( \tau _j\right) \right) \right\rangle$%
. Hence, the expression for atomic limit Eq.(\ref{Exact}) is also a general
result for the self-energy that is valid for $k_n\gg W.$At half-filling $%
n_{-\sigma }=1/2$, $\mu =U/2$and the asymptotic (\ref{SelfHaut}) sets in at $%
k_n\sim W$, as was pointed out above. Away from half-filling, as long as $%
\left| \mu -\Sigma \left( \infty \right) \right| $and $\left| \mu -U\left(
1-n_{-\sigma }\right) \right| $are both much smaller than $W,$(they both
vanish at half-filling), the asymptotic behavior will also start at $k_n\sim
W$.

The situation is qualitatively different when one uses dressed Green
functions, but does not take into account the frequency dependence of the
vertex, as it is done in FLEX (See Eq.\ref{BerkSFL})) or for second-order
perturbation theory with dressed $G$. For example, the second-order
expression for $\Sigma _\sigma \left( {\bf k,}ik_n\right) $in terms of full $%
G$does satisfy the asymptotics Eq.(\ref{SelfHaut}), but it sets in too late,
namely for $k_n\gg U$, instead of $k_n\gg W$. Indeed, when $k_n\gg W$, the
equation for the self-energy at half-filling in this type of theories
reduces to 
\begin{equation}
\Sigma \left( ik_n\right) =\frac{\Delta ^2}{ik_n-\Sigma \left( ik_n\right) }
\label{AsSigFLEX}
\end{equation}
where $\Delta ^2=cU^2/4$with $c$a constant of proportionality involving the
sum over all wave vectors and Matsubara frequencies of the self-consistent
dynamical susceptibilities. In a given theory the value of $c$ may differ
from its value $c=1$obtained from the exact result Eq. (\ref{Exact}), but
its always of order unity. The solution of Eq.(\ref{AsSigFLEX}) 
\begin{equation}
\Sigma \left( ik_n\right) =\frac 12ik_n-\frac 12\sqrt{\left( ik_n\right)
^2-4\Delta ^2}  \label{SigmaFLEX}
\end{equation}
has the analytically continued form 
\begin{equation}
\text{Re}\Sigma ^R\left( \omega \right) =\frac \omega 2-\frac \omega {%
2\left| \omega \right| }\theta \left( \left| \omega \right| -2\Delta \right) 
\sqrt{\omega ^2-4\Delta ^2}  \label{SigmaFLEXreal}
\end{equation}
\begin{equation}
\text{Im}\Sigma ^R\left( \omega \right) =-\frac 12\theta \left( 2\Delta
-\left| \omega \right| \right) \sqrt{4\Delta ^2-\omega ^2}
\end{equation}
From this one can immediately see that a $U^2/\omega $regime exists for Re$%
\Sigma ^R\left( \omega \right) $only when $\left| \omega \right| \gg U,$$%
\left( \text{with }2\Delta =U\right) $.

This means that such regime sets in too late to give the Hubbard bands
described by Eq.(\ref{HLHubbard}), because the Hubbard bands occur at $%
\omega =\pm U/2$ and for such $\omega $the asymptotic form $\Sigma ^R\propto
U^2/\omega $is not valid yet in FLEX and similar theories. Consequently,
instead of well defined peaks at $\omega =\pm U/2$ in the half-filled case,
one obtains only long tails in the spectral function $A_\sigma \left( {\bf k}%
,\omega \right) ,$ no matter how large $U$ is.\cite{BandesHubbardDinf}(see
also following subsection).

This explains why there is no Hubbard bands in any theory that uses
self-consistent Green's functions, but neglects the frequency dependence of
the vertex. This is an explicit example that illustrates what seems to be a
more general phenomenon when there is no Migdal theorem for vertex
corrections: a calculation with dressed Green's functions but no frequency
dependent vertex correction often gives worse results than a calculation
done with bare Green's functions and a frequency independent vertex.

\subsection{Why FLEX fails to see precursors of antiferromagnetic bands}

\label{SubSecWhyFLEX}

In this subsection we describe the qualitative differences between our
results and the results of FLEX approximations given by Eq.(\ref{BerkSFL})
with regards to the ``shadow bands'' and explain why we believe that the
failure of the FLEX to reproduce these bands is an artifact of that
approximation. To avoid any confusion, we first clarify the terminology,
because the term ``shadow bands'' has been used previously to describe
different physical effects. (see for details Ref. \cite{Yury3}). We note
that the so-called shadow features discussed in \cite{Yury3}, \cite{Langer}
as well as the pseudogap in the {\em total} density of states $N(\omega
)=(1/N)\sum_{{\bf k}}A\left( {\bf k},\omega \right) )$exist in both theories
and we will not discuss them here. Instead, we concentrate on the precursors
of antiferromagnetic bands in the spectral function $A\left( {\bf k}%
_F,\omega \right) $which correspond to two new solutions of the
quasi-particle equation 
\begin{equation}
\omega -\epsilon ({\bf k})+\mu -\Sigma ({\bf k},\omega )=0  \label{QPeq}
\end{equation}

We start by recalling a simple physical argument why the precursors of
antiferromagnetic bands must exist at finite temperatures in the vicinity of
the zero-temperature phase transition in two dimensions. This can be best
understood by contrasting this case with isotropic 3D case where such
precursor effect are highly unlikely. (For a discussion of the strongly
anisotropic case see Sec.(\ref{SubSecAbsIn3d})). Indeed, in three dimensions
there is a {\em finite} temperature phase transition and the gap is equal to
zero at this temperature $\Delta (T_N)=0$. Consequently at $T_N$there is
only one peak in the $A\left( {\bf k}_F,\omega \right) $at $\omega =0$which
starts to split into two peaks only below $T_N$. Based on this simple
physical picture, one would not expect to see precursors of
antiferromagnetic bands above $T_N$in this case. The situation is
qualitatively different in two-dimensions where classical thermal
fluctuations suppress long-range order at any finite temperature while at
the $T=0$ phase transition the system goes directly into the ordered state
with a {\em finite} gap. Clearly, the two peak structure in $A\left( {\bf k}%
_F,\omega \right) $ at $T=0$ cannot disappear as soon as we raise the
temperature.

\begin{figure}%
\centerline{\epsfxsize 6cm \epsffile{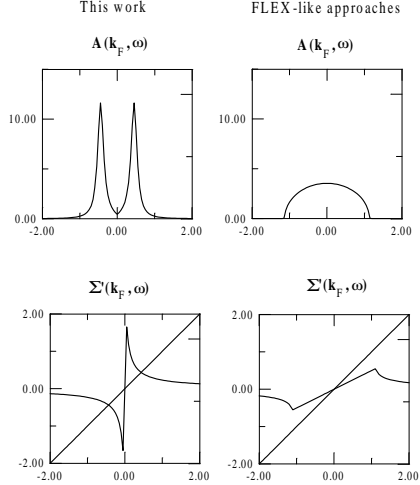}}%
\caption{Top two panels are qualitative sketches of the spectral weight at the 
Fermi wave vector at half-filling. The plots are obtained by retainig only the classical contribution to 
the self-energy for $T<T_X$ using parameters corresponding to the typical $U=4$, 
of Monte Carlo 
simulations. The two bottom panels are the corresponding plots of $Re \Sigma (\omega)$.
The left-hand side of this figure is obtained using our approximation while the
right-hand side is obtained from the FLEX-like approach. 
The intersection with the $45$ degree line $\omega$ in the bottom-left panel gives rise to the
precursors of antiferromagnetic bands seen right above it.}%
\label{Akom}%
\end{figure}

For simplicity we again consider half-filling. As we have seen in section (%
\ref{SubSecPrecursors}) two new quasi-particle peaks do appear in the
renormalized classical regime $T<T_X$in our theory. We have also found a
pseudogap with the minimum at $\omega =0$in this regime. In contrast, the
numerical solution of the FLEX equations\cite{Serene} found a spectral
function with a single maximum in $A\left( {\bf k}_F,\omega \right) $at $%
\omega =0$even when $\tilde{\chi}^{RPA}({\bf q},0)$becomes strongly peaked
at ${\bf q}={\bf Q}$. With decreasing temperature this central maximum
becomes anomalously broad, but the two peak structure does not appear. The
clear deviation from the Fermi liquid is signaled by the positive sign of $%
\partial \Sigma ^{\prime }\left( {\bf k}_F,\omega \right) /\partial \omega
>0 $. However the value of $\partial \Sigma ^{\prime }\left( {\bf k}%
_F,\omega \right) /\partial \omega $does not become larger than unity. The
latter would unavoidably lead to the existence of two new quasi-particle
peaks away from $\omega =0$as is clear from the graphical solution of the
quasiparticle equation Eq.(\ref{QPeq}) shown on the bottom left panel of
Figs.(\ref{Akom}).

We now explain analytically the origin of these qualitative differences in
the two theories. In our theory $\partial \Sigma ^{\prime }({\bf k}_F,\omega
)/\partial \omega |_{\omega =0}\propto T\xi ^2$and hence it quickly becomes
larger than unity in the renormalized classical regime $\xi \propto \exp
(const/T)$. In addition, for $\omega >v_F\xi ^{-1}$the real part of the
self-energy has the same behavior as in the ordered state $\Sigma ({\bf k}%
_F,\omega )\propto \Delta ^2/\omega $with $\Delta ^2\propto T\ln \xi =const$%
. The important point is that this asymptotic behavior $\Sigma ({\bf k}%
_F,\omega )\propto \Delta ^2/\omega $of the self-energy already sets in for $%
\omega \sim \Delta \gg v_F\xi ^{-1}$. It is this property that leads to the
appearance of the precursors of antiferromagnetic bands at $\omega =\pm
\Delta $in a manner analogous to the appearance of the Hubbard bands in the
strong coupling limit that is discussed in the previous subsection. Let's
now try to understand analytically what happens in the FLEX approximation.
As in our theory, the main contribution to the self-energy in the strongly
fluctuating regime comes from the zero-frequency term in the Matsubara sum
in the equation for the self-energy (Eq.(\ref{ClassicalContribution}) in our
theory and Eq.(\ref{BerkSFL}) in FLEX). An upper bound of the effect of the
critical spin fluctuations can be obtained by approximating $T\tilde{\chi}%
^{RPA}({\bf q},0)\propto \delta ({\bf q})$. Then one immediately obtains the
same expression for the self-energy as the one obtained in FLEX in the
context of Hubbard bands Eq.(\ref{AsSigFLEX}). (The only difference is that
the parameter $\Delta $ is now defined by the zero-frequency Matsubara
contribution of $\tilde{\chi}^{RPA}({\bf q},0)$, rather than by the sum over
all Matsubara frequencies.) As we have already discussed in the context of
Hubbard bands, such a form for $\Sigma $ does not lead to the appearance of
two new quasiparticle solution away from $\omega =0$ because the
characteristic behavior $\Sigma ({\bf k}_F,\omega )\propto \Delta ^2/\omega $
sets in too late, namely for $\omega \gg \Delta $. In addition, the slope of 
$\Sigma ^{\prime }\left( {\bf k}_F,\omega \right) $ at $\omega \rightarrow 0$
does not diverge with decreasing temperature as in our theory but instead
saturates to its value given by the analog of Eq.(\ref{SigmaFLEXreal}), {\it %
i.e.} $\partial \Sigma ^{\prime }\left( {\bf k}_F,\omega \right) /\partial
\omega <1/2$. As we mentioned above, a value larger than unity $\partial
\Sigma ^{\prime }\left( {\bf k}_F,\omega \right) /\partial \omega >1$ would
guarantee the existence of two new solutions of the quasiparticle equation
Eq.(\ref{QPeq}) away from $\omega =0$. The right-hand side of Fig.(\ref{Akom}%
) illustrates clearly what happens in a FLEX-like approach such as Eq.(\ref
{AsSigFLEX}). The contribution of classical fluctuations to the spectral
weight does not lead to a Fermi liquid since $A\left( {\bf k}_F,\omega
\right) $ saturates to a finite width as temperature decreases, but
nevertheless precursors of shadow bands do not occur because $\partial
\Sigma ^{\prime }\left( {\bf k}_F,\omega \right) /\partial \omega $ is
bounded below unity. (Note that the spectral weight would not vanish so
steeply at large frequencies if we had taken into account the quantum
contribution of the spin fluctuations, as in full FLEX calculations.)

We just saw that the self-consistency in the propagators without
corresponding self-consistency in the vertices inhibits the existence of the
shadow bands in essentially the same way as it inhibits the existence of the
Hubbard bands. It thus seems to us very likely that the absence of the
precursors of antiferromagnetic bands below $T_X$ in FLEX is an artifact.
This conclusion can be reliably verified by comparison with Monte Carlo data
despite the fact that the latter is done for finite lattices and in the
Matsubara formalism. This was discussed in more detail in Sec.(\ref
{SubSecPrecursors}). Here we just note that the temperature dependence of
Matsubara quantities such as $G({\bf k}_F,\tau =\beta /2)$ and $\Sigma ({\bf %
k}_F,ik_1)$ have a very characteristic form in the pseudogap regime. For
example, $\Sigma ({\bf k},ik_1)\propto 1/(i\pi T)$ in the pseudogap regime,
while in FLEX we would expect a much weaker temperature dependence of this
quantity (the upper bound being given by the analog of Eq.(\ref{SigmaFLEX})).

We also would like to comment on the 1D model\cite{McKenzie} which describes
the interaction of electrons with {\em static} spin fluctuations
characterized by the susceptibility $\chi _{sp}\propto \delta (\omega )[\xi
^{-1}/({\bf q}-{\bf Q})^2+\xi ^{-2}]$. The nice thing about this model is
that it has an exact solution which shows the development of shadow bands
and of the pseudogap in $A\left( {\bf k}_F,\omega \right) $. A treatment
similar to ours which uses non-interacting Green's functions in the one-loop
approximation also reproduces this feature\cite{McKenzie}. However, the
analogous approximation with dressed Green's functions leads to Eq.(\ref
{AsSigFLEX}) and hence inhibits the existence of the ``shadow bands'' and of
the pseudogap in $A\left( {\bf k}_F,\omega \right) $.

In closing we comment on semantics and on the physical interpretation of
some results obtained in the FLEX approximation. The expression ``conserving
approximation'' has been widely used to describe FLEX calculations of the
single particle properties and, in particular, in the context of the shadow
bands and of the failure of Luttinger's theorem \cite
{Serene,Langer,Schmalian}. The conserving aspect has been emphasized, but in
fact the only desirable feature in the calculation of the single-particle
properties is that the self-energy $\Sigma $ is obtained from a functional
derivative of the Luttinger-Ward functional $\Sigma =\delta \Phi /\delta G$
and hence it is guaranteed to satisfy Luttinger's theorem whenever
appropriate. Only on the next level does this scheme lead to a calculation
of the ``true'' susceptibilities\cite{FLEX} and of collective modes that
satisfy conservation laws (Ward identities). However, these ``true''
susceptibilities are never substituted back in the calculation of the
self-energy and the effect of ``true'' collective modes on the
single-particle spectrum is an open question in FLEX. In fact, the RPA
propagators $\tilde{\chi}_{RPA}$ appearing in the self-energy expression are
different from susceptibilities from which collective modes should be
computed and further they explicitly break conservation laws, as can be seen
from the fact that RPA-like expressions $\tilde{\chi}_{RPA}=\tilde{\chi}%
_0/(1-U\tilde{\chi}_0)$ with a {\em dressed} bubble $\tilde{\chi}_0$ have
the unphysical properties that are mentioned in Eqs.(\ref{WardTriste}) and (%
\ref{fTriste}) of Appendix (\ref{SecSumRules}). The fact that there are in
effect two susceptibilities in the FLEX approximation leads, in our opinion,
to some confusion and incorrect physical interpretation of the results in
the literature. In particular, it was argued that the non-Fermi-liquid
behavior and deviations from Luttinger theorem found in FLEX \cite
{Serene,Langer,Schmalian} are not due to critical thermal fluctuation in the
vicinity of the phase transition but are rather the result of large $U$. The
reasoning for such claim was that although the RPA susceptibilities $\tilde{%
\chi}_{RPA}$ is very strongly peaked at ${\bf q}={\bf Q}$, the ``true'' FLEX
susceptibility is not. In our opinion, such claim could be justified only if
one would substitute the ``true'' susceptibility back in the calculation of $%
\Sigma $ (for example using the exact Eq.(\ref{2})) and found that the
deviation from the Luttinger theorem and other qualitative changes in $A(%
\vec{k},\omega )$ increase with decreasing temperature without almost
divergent behavior of the {\em conserving} susceptibility $\chi _{sp}({\bf Q}%
,0)$ and of the static structure factor $S_{sp}\left( {\bf Q}\right) .$

The Monte Carlo data on Fig.(\ref{FigJoint}) are also instructive since they
clearly show that qualitative changes in the single-particle spectra occur
when the system enters the renormalized classical regime with rapidly
growing $S_{sp}\left( {\bf Q}\right) $. The fact that the FLEX ``true''
susceptibility does not show such behavior at half-filling\cite{Serene}
tells us that it even more drastically disagrees with the Monte Carlo data
than the RPA-like $\tilde{\chi}$ which enters the expression for
self-energy. Moreover, even away from half-filling the ``true''
susceptibility in FLEX at ${\bf q}={\bf Q}$ significantly underestimates the
strength of the spin fluctuations, as is clear from the comparison with
Monte Carlo data in Fig.(\ref{FigChiT}). In our opinion the, so-called,
``true susceptibility'' in FLEX is the key element in the confusion
surrounding the interpretation of FLEX results for the self-energy because
the ``true susceptibility'' never comes in the calculation of the
self-energy. For all practical purposes these calculations of the
self-energy should be considered as consistent with Luttinger's theorem at $%
T=0$ but based on a non-conserving susceptibility. Consistency with
conservation laws and consistency with Luttinger's theorem are not identical
requirements because to satisfy rigorously Luttinger's theorem one needs
that $\Sigma =\delta \Phi /\delta G$, while to have conserving
susceptibilities one needs that the irreducible vertices used in
Bethe-Salpeter equation Eq.(\ref{1a}) should be obtained from $\Gamma
=\delta ^2\Phi /\delta G\delta G$.

\section{Domain of validity of our approach}

\label{SecValidity}

Our approach is not valid beyond intermediate coupling. That is perhaps best
illustrated by Fig.(\ref{FigTXU}) that shows that the crossover temperature
first increases with $U$ and then saturates instead of decreasing. The
decrease is expected on general grounds from the fact that at strong
coupling the tendency to antiferromagnetism should decrease roughly as $%
J\sim t^2/U.$ The reason for this failure of our approach is clear. As we
know from studies in infinite dimension\cite{Georges}, to account for
strong-coupling effects it is necessary to include at least a frequency
dependence to the self-energy and to the corresponding irreducible vertices.

Our theory also fails at half-filling deep in the renormalized classical
regime, i.e. $T\ll T_X$ mainly for two reasons. First, the {\it ansatz} $%
U_{sp}=Ug_{\uparrow \downarrow }(0)$, Eq.(\ref{Usp}), fails in the sense
that $g_{\uparrow \downarrow }(0)$ eventually reaches zero at $T\rightarrow
0 $ because of the $\log ^2T$ divergence in the irreducible susceptibility $%
\chi _0\left( \pi ,\pi \right) $ due to perfect nesting. The physically
appropriate choice for $g_{\uparrow \downarrow }(0)$ in the renormalized
classical regime is to keep its value fixed to its crossover-temperature
value (See Fig.(\ref{FigGUpDown}) and Sec.(\ref{SecMonteCarlo})). The more
serious reason why our approach fails for $T\ll T_X$ is that, as we just
saw, critical fluctuations destroy completely the Fermi liquid
quasiparticles and lead to a pseudogap. This invalidates our starting point.
It is likely that in a more self-consistent theory, the logarithmic
divergence of the appropriate irreducible susceptibility will be cutoff by
the pseudogap. However, just a simple dressing of the Green's function is
not the correct solution to the problem because it would make the theory
non-conserving, as we discussed in Sec.(\ref{SecConstTwoPart}). One needs to
take into account wave vector and energy dependent vertex corrections
similar to those discussed by Schrieffer\cite{SchriefferVertex}\cite{Blaizot}%
.

\section{Comparisons with other approaches}

\label{SecComparisons}

In Appendix (\ref{DiscussionOtherApproaches}), we discuss in detail various
theories, pointing out limitations and advantages based on the criteria
established in Appendix(\ref{SubSecConstSingle}) and (\ref{SecConstTwoPart}%
). More specifically, we include in our list of desirable properties, the
local Pauli principle $\left\langle n_{\uparrow }^2\right\rangle
=\left\langle n_{\uparrow }\right\rangle ,$ the Mermin-Wagner theorem Eq.(%
\ref{SusSpinSum}), the Ward identities Eq.(\ref{Ward}), and $f-$sum rule Eq.(%
\ref{f2}), one-particle {\it vs} two-particle consistency $\Sigma _\sigma
\left( 1,\overline{1}\right) G_\sigma \left( \overline{1},1^{+}\right)
=U\left\langle n_{\uparrow }n_{\downarrow }\right\rangle $ Eq.(\ref
{Consistency}), Luttinger's theorem, and the large frequency asymptotic for
the self-energy Eq.(\ref{SelfHaut}), which is important for the existence of
the Hubbard bands. Comments on the forward scattering sum-rule Eq.(\ref
{Forward}) appear in Appendix (\ref{Lat}). In the present section, we only
state without proof where each theory has strengths and weaknesses.

In standard paramagnon theories,\cite{Stamp}\cite{Enz} the spin and charge
fluctuations are computed by RPA, using either bare or dressed Green's
functions. Then the fluctuations are feedback in the self-energy. When RPA
with bare Green's functions are used for the collective modes, these satisfy
the $f-$sum rule, but that is the only one of our requirements that is
satisfied by such theories.

In conserving approximation schemes\cite{Baym},\cite{FLEX} the Mermin-Wagner
theorem, the Luttinger theorem and conservation laws are satisfied, but none
of the other above requirements are fulfilled.

In the parquet approach,\cite{parquet}\cite{FLEX-parquet} one enforces
complete antisymmetry of the four point function by writing down fully
crossing-symmetric equations for these. However, in actual calculations, the
local Pauli principle, the Mermin Wagner theorem, and the consistency
between one and two particle properties are only approximately satisfied,
while nothing enforces the other requirements.

In our approach, the high-frequency asymptotics and Luttinger's theorem are
satisfied to a very good degree of approximation while all other properties
in our list are exactly enforced. Let us specify the level of approximation.
Luttinger's theorem is trivially satisfied with our initial approximation
for the self-energy $\Sigma _\sigma ^{\left( 0\right) },$ but at the next
level of approximation, $\Sigma _\sigma ^{\left( 1\right) }$, one needs a
new chemical potential to keep the electron density $Tr\left[ G_\sigma
^{\left( 1\right) }\left( 1,1^{+}\right) \right] $ fixed. With this new
chemical potential the Fermi surface volume is preserved to a very high
accuracy. Finally, consider the high-frequency asymptotics. Since we use
bare propagators, the high-frequency asymptotics comes in at the appropriate
frequency scale, namely $ik_n\sim W$, which is crucial for the existence of
the Hubbard bands. However, the coefficient of the $1/ik_n$ term in the
high-frequency expansion Eq.(\ref{SelfHaut}) is incorrect because our
irreducible vertices $U_{sp}$ and $U_{ch}$ are tuned to the low frequencies.
If one would take into account the frequency dependence of $U_{sp}$ and $%
U_{ch}$ and assume that at high frequency they become equal to the bare
interaction $U$, then one would recover the exact result, provided the Pauli
principle in the form of Eq.(\ref{SpinChargePauli0}) is satisfied. The
difficulty with such a procedure is that frequency dependent irreducible
vertices requires frequency dependent self-energy in the calculation of
collective modes and that would make the theory much more complicated. Yet
it is, probably, the only way to extend the theory to strong coupling.

\section{Conclusion}

We have presented a new simple approach\cite{Vilk}\cite{Vilk2} to the
repulsive single-band Hubbard model. We have also critically compared
competing approaches, such as paramagnon, fluctuation exchange
approximation, and pseudo-potential parquet approaches. Our approach is
applicable for arbitrary band structure\cite{Veilleux} and gives us not only
a {\it quantitative} description of the Hubbard model, but also provide us
with some {\it qualitatively} new results. Let us summarize our theory
again. We first obtain spin and charge fluctuations by a self-consistent
parameterization of the two-particle effective interactions (irreducible
vertices) that satisfies a number of exact constraints usually not fulfilled
by standard diagrammatic approaches to the many-body problem. Then the
influence of collective modes on single-particle properties is taken into
account in such a way that single-particle properties are consistent with
two-particle correlators, which describe these collective modes. More
specifically, our approach satisfies the following constraints:

\begin{enumerate}
\item  Spin and charge susceptibilities, through the fluctuation-dissipation
theorem, satisfy the Pauli principle in the form $\left\langle n_{\uparrow
}^2\right\rangle =\left\langle n_{\uparrow }\right\rangle $ as well as the
local moment sum-rule, conservations laws and consistency with the equations
of motion in a local-field-like approximation.

\item  In two dimensions, the spin fluctuations satisfy the Mermin-Wagner
theorem.

\item  The effect of collective modes on single-particle properties is
obtained by a paramagnon-like formula that is consistent with the
two-particle properties in the sense that the potential energy obtained from 
$Tr\left[ \Sigma G\right] $ is identical to that obtained from applying the
fluctuation-dissipation theorem to spin and charge susceptibilities.

\item  Vertex corrections are included not only in spin and charge
susceptibilities $\left( U_{sp}\neq U_{ch}\neq U\right) $ but also in the
self-energy formula. In the latter case, this takes into account the fact
that there is no Migdal theorem controlling the effect of spin and charge
fluctuations on the self-energy.
\end{enumerate}

The results for both single-particle and two-particle properties are in {\it %
quantitative} agreement with Monte Carlo simulations for all fillings, as
long as $U$ is less than the bandwidth and $T$ is not much smaller than the
crossover temperature $T_X$ where renormalized-classical behavior sets in.
Both quantum-critical and renormalized-classical behavior can occur in
certain parameter ranges but the critical behavior of our approach is that
of the $O\left( n\right) $ model with $n\rightarrow \infty $.\cite{Dare}

The main predictions of physical significance are as follows:

\begin{enumerate}
\item  The theory predicts a magnetic phase diagram where magnetic order
persists away from half-filling but with completely suppressed
ferromagnetism.

\item  In the renormalized classical regime above the zero-temperature phase
transition, precursors of antiferromagnetic bands (shadow bands) appear in $%
A\left( {\bf k}_F,\omega \right) $. These precursors occur when $\xi >\xi
_{th}$ (or $\omega _{SF}<T$)$.$ Between these precursors of
antiferromagnetic bands a pseudogap appears at half-filling, so that the
Fermi liquid quasiparticles are completely destroyed in a wide temperature
range above the zero-temperature phase transition $0<T<T_X$. The upper
critical dimension for this phenomenon is three. We stress the qualitative
difference between the Hubbard bands and the precursors of antiferromagnetic
bands and we predict that in two dimensions one may see both sets of bands
simultaneously in certain parameter ranges. This prediction is consistent
with the results of numerical simulations.\cite{Dagotto}\cite{Hanke}. We
know of only one other analytic approach\cite{Matsumoto} which leads to
similar four peak structure in the spectral function.
\end{enumerate}

The zero temperature magnetic phase diagram is partly an open question
because, despite the qualitative agreement with other analytical approaches,
there is still an apparent contradiction with Monte Carlo simulations\cite
{ImadaDiagram}. Our prediction of precursors of antiferromagnetic bands on
the other hand is in agreement with Monte Carlo simulations. Neither this
effect nor upper and lower Hubbard bands are observed in self-consistent
schemes such as FLEX. This is because of inconsistent treatment of the
vertex and self-energy corrections in this approximation, as we have
explained in Sec.(\ref{SecFLEX}). However, if there was a Migdal theorem for
spin fluctuations, it would be justifiable to neglect the vertex corrections
and keep only the self-energy effects as is done in the FLEX approximation.
The presence of precursors of antiferromagnetic bands in two-dimensions is
then a clear case of qualitatively new Physics that would {\it not} appear
if there was a Migdal theorem for spin fluctuations. The same is true for
the Hubbard bands for large $U>W$ in any dimension.

We would like to state again clearly the nature of our critique of
approximation schemes which are based on using Migdal's theorem for systems
with electron-electron interactions. We do {\it not} imply that one does not
need at all to take into account the feedback of the single-particle spectra
on collective modes. The only point that we want to make here is that, based
on sum rules and comparison with Monte Carlo data, we see that frequency and
momentum dependent corrections to the self-energy and to the vertex often
tend to cancel one another and that ignoring this leads to qualitatively
incorrect results, in particular, with regards to the pseudogap. In this
paper we were able to look only at the beginning of the renormalized
classical regime when the pseudogap starts to form. The truly
self-consistent treatment of the one-particle and two-particle properties in
the pseudogap regime remains an open and very challenging problem. We hope
that by extending our approach to the ordered state and looking at how the
pseudogap starts to disappear as the temperature is raised, one can better
understand how to develop a more self-consistent theory in the pseudogap
regime. We now point out how our approach can be extended in other
directions.

As we mentioned in Sec.(\ref{SecDestruction}), the pseudogap and precursors
of antiferromagnetic bands in the two-dimensional repulsive Hubbard model
have interesting analogs in the attractive Hubbard model. In that model, one
expects a pairing pseudogap and precursors of superconducting quasiparticle
bands above $T_c$. At half-filling the negative and positive Hubbard models
are mapped onto one another by a canonical transformation and the present
theory is directly applicable to the attractive case. However, away from
half-filling the mapping between the two models is more complicated and the
microscopic theory requires additional sum-rule for pairing susceptibilities
to find self-consistently the effective pairing interaction. This work is
now in progress.

The present approach can be also extended to stronger coupling $U>W$. Again
the key idea would be to parameterize the irreducible vertices, which have
now to be frequency dependent, and then use the most important sum-rules to
find the parameterization coefficients. This will, of course, require
solving much more complicated self-consistent equations than in the present
approach, but we believe that the problem still can be made tractable.

Finally, we would like to make two comments about the magnetic and the
pairing pseudogap in the context of high-$T_c$ superconductors, based on the
results of our studies. First, as was stressed in Ref.\cite{Yury3}, to
understand clearly the physics of the single-particle pseudogap phenomena it
is important to distinguish {\it static} short-range order from {\it %
dynamical} short-range order. The former is defined by a nearly Lorentzian
form of the corresponding static structure factor $S({\bf q})\propto 1/((%
{\bf q}-{\bf Q})^2+\xi ^{-2})$ (${\bf Q}=(\pi ,\pi )$ in magnetic case, $%
{\bf Q}=0$ in the case of pairing), while the latter means only that the
corresponding susceptibility $\chi ({\bf q},0)$ has such a Lorentzian form.
A condition for the existence of the single particle pseudogap in the
vicinity of a given phase transition is that the corresponding short-range
order is quasi-static ({\it i.e.}$~\omega _{SF}\ll T$)\cite{Yury3}.
Experimentally, one can measure directly the dynamical spin structure factor 
$S({\bf q},\omega )$, and then obtain the static structure factor through
the integral $S({\bf q})=\int S({\bf q},\omega )d\omega /(2\pi )$. Even if
the zero-frequency dynamical structure factor $S_{sp}({\bf q},0)$ is very
strongly peaked at ${\bf q}\sim {\bf Q}$ it is possible that the static
structure factor $S_{sp}({\bf q)}$ is only weakly momentum dependent\cite
{Yury3}. Thus in order to know whether one should expect to see the
precursors of the antiferromagnetic bands and the corresponding pseudogap at
a given doping and temperature it is necessary to obtain the {\em static}
spin structure factor from the experimentally determined dynamical structure
factor and then analyze its momentum dependence to see both if it is peaked
and if it is quasi-two-dimensional.

The second comment that we would like to make is that both the pairing and
the magnetic single-particle pseudogap discussed above are an effect of low
dimensionality and hence they exist as long as there is a large
two-dimensional fluctuating regime before the real three-dimensional phase
transition. In this context, a pairing pseudogap could exist on either side
of optimal doping.\cite{Timusk} The much larger temperature range over which
a pseudogap appears in the underdoped compounds suggests that, in addition
to pairing fluctuations, other thermal fluctuations (charge, spin...)
prohibit finite-temperature ordering.\cite{EmeryKivelson} An example of this
occurs in the attractive Hubbard model where charge fluctuations push the
Kosterlitz-Thouless temperature to zero at half-filling, precisely where the
crossover temperature to the pseudogap regime is largest.

We thank A.E. Ruckenstein, Liang Chen and R. C\^{o}t\'{e} for helpful
discussions and A. Georges for pointing out useful references. This work was
partially supported by the Natural Sciences and Engineering Research Council
of Canada (NSERC), the Fonds pour la formation de chercheurs et l'aide \`{a}
la recherche from the Government of Qu\'{e}bec (FCAR), and (for Y.M.V.) the
National Science Foundation (Grant No. NSF-DMR-91-20000) through the Science
and Technology Center for Superconductivity and (for A.-M.S.T.) the Killam
Foundation as well as the Canadian Institute for Advanced Research (CIAR).
A.-M.S.T. would like to thank the Institute for Theoretical Physics, Santa
Barbara for hospitality during the writing of this work. Partial support
there was provided by the National Science Foundation under Grant No.
PHY94-07194.

\appendix 

\section{Sum rules, Ward identities and Consistency requirements}

\label{SecSumRules}

In this Appendix, we recall well known constraints on many-body theory that
follow from sum-rules and conservation laws and comment, wherever possible,
on their physical meaning and on where commonly used approaches fail to
satisfy these constraints. Although we come back on a detailed discussion of
various theories in a later appendix, we find it useful to include some of
this discussion here to motivate our approach. We consider in turn various
results that would be satisfied by any exact solution of the many-body
problem. They are all consequences of either anticommutation relations alone
(Pauli principle) or of anticommutation relations and the Heisenberg
equations of motion. We describe in turn: 1. The relation between
self-energy and two-body correlation functions that embodies the details of
the Hamiltonian; 2. Sum rules for one-particle properties; 3. Sum rules and
constraints on two-particle properties, in particular {\it f}-sum rule and
Ward identities that express conservation laws; 4. A few relations that are
crucial in Fermi liquid theory, namely Luttinger's theorem and the forward
scattering sum rule.

\subsection{Equations of motion and the relation between the self-energy $%
\Sigma $ and two-particle properties.}

The self-energy (we always mean one-particle irreducible self-energy) is
related to the potential energy, and hence to two-particle correlations
through the expression Eq.(\ref{Consistency}), which in the Kadanoff and
Baym notation can be written as 
\begin{equation}
\Sigma _\sigma \left( 1,\overline{1}\right) G_\sigma \left( \overline{1}%
,1^{+}\right) =U\left\langle n_{\uparrow }n_{\downarrow }\right\rangle .
\label{Consistency1}
\end{equation}
Here, the index with an overbar, $\overline{1},$ means that there is a sum
over corresponding lattice positions and an integral over imaginary time.
The notation $1^{+}$ means that the imaginary time implicit in $1$ is $\tau
_1+\eta $ where $\eta $ is a positive infinitesimal number. Eq.(\ref
{Consistency1}) is an important consistency requirement between self-energy
and double occupancy in the Hubbard model that can easily be proven as
follows. From the equations of motion for the single-particle Green's
function Eq.(\ref{DefG}) one finds, 
\[
\left[ \left( -\frac \partial {\partial \tau _i}+\mu \right) \delta _{i,\ell
}+t_{i\ell }\right] G_\sigma \left( {\bf r}_\ell -{\bf r}_j,\tau _i-\tau
_j\right) 
\]
\begin{equation}
=\delta _{i,j}\delta \left( \tau _i-\tau _j\right) -U\left\langle T_\tau
\left( c_{i-\sigma }^{\dagger }\left( \tau _i\right) c_{i-\sigma }\left(
\tau _i\right) c_{i\sigma }\left( \tau _i\right) c_{j\sigma }^{\dagger
}\left( \tau _j\right) \right) \right\rangle .  \label{Dyson1}
\end{equation}
Using the short-hand notation in Eqs.(\ref{DefG}) and (\ref{Gfur}) and the
definition of self-energy (Dyson's equation) the above equation is also
written in the form, 
\begin{equation}
G_0^{-1}\left( 1,\overline{1}\right) G_\sigma \left( \overline{1},2\right)
=\delta \left( 1-2\right) +\Sigma _\sigma \left( 1,\overline{1}\right)
G_\sigma \left( \overline{1},2\right) .  \label{Dyson2}
\end{equation}
Comparing the last two equations, the well known relation Eq.(\ref
{Consistency1}) (or Eq.(\ref{Consistency})) between self-energy, Green's
function and potential energy follows.

So-called conserving\cite{Baym} approaches to the many-body problem violate
the above consistency requirement Eq.(\ref{Consistency}) in the following
sense. The right-hand side can be computed from the collective modes using
the fluctuation-dissipation theorem. In conserving approximations, this
gives a result that is different from what is computed directly from the
left-hand side of the equation, namely from the self-energy and from the
Green's function. In fact, all many-body approaches satisfy the above
consistency requirement at best in an approximate way. However, it is a very
important requirement and Eq.(\ref{Consistency}) plays a key role in our
discussion. Seen in Matsubara frequency, it is a sum rule, or an integral
constraint that involves all frequencies, large and small.

\subsection{Constraints on single-particle properties:}

\label{SubSecConstSingle}The spectral weight $A_\sigma \left( {\bf k},\omega
\right) $ can be interpreted as a probability of having an electron in a
state $\left( \sigma ,{\bf k},\omega \right) $ and it satisfies the
normalization sum rule 
\begin{equation}
\int_{-\infty }^\infty \frac{d\omega }{2\pi }A_\sigma \left( {\bf k},\omega
\right) =\left\langle \left\{ c_{{\bf k}\sigma },c_{{\bf k}\sigma }^{\dagger
}\right\} \right\rangle =1.  \label{A1}
\end{equation}
Formally this is a consequence of the jump in the Green's function at $\tau
=0$, as can be seen from calculating 
\begin{equation}
G_\sigma \left( {\bf k},0^{-}\right) -G_\sigma \left( {\bf k},0^{+}\right)
=1=T\sum_{ik_n}\left( e^{ik_n\eta }-e^{-ik_n\eta }\right) \int_{-\infty
}^\infty \frac{d\omega }{2\pi }\frac{A_\sigma \left( {\bf k},\omega \right) 
}{ik_n-\omega }=\int_{-\infty }^\infty \frac{d\omega }{2\pi }A_\sigma \left( 
{\bf k},\omega \right) .
\end{equation}

To do perturbation theory directly for the Green's function to any {\em %
finite} order would require that the interaction $U$ be small not only in
comparison with the bandwidth $W$ but also in comparison with the smallest
Matsubara frequency $ik_1=2\pi T$. Also, the direct perturbation series for
the Green's function gives, after analytical continuation, poles of
arbitrary high order located at the unperturbed energies. These high-order
poles are inconsistent with the simple pole (or branch cut) structure of the
Green's function predicted by the spectral representation. Furthermore, the
high-order poles lead to a spectral weight that can be negative.\cite
{SenechalPairault} The common way to get around these difficulties is to
make approximations for the self-energy $\Sigma $ instead and then calculate
the Green's function using Dyson's equation Eq.(\ref{Gself}).

It is interesting to note that to satisfy the constraint Eq.(\ref{A1}), it
suffices that $\Sigma ({\bf k},ik_n)$, defined by Eq.(\ref{Gself}), has a
finite limit as $ik_n\rightarrow \infty .$ More constraints on
approximations for the self-energy may be found by continuing this line of
thought. A systematic way of doing this is to do a high-frequency expansion
for both the Matsubara Green's function and the self-energy and to find
coefficients using sum-rules. The sum-rules that we need then are\cite
{White91} 
\begin{equation}
\int_{-\infty }^\infty \frac{d\omega }{2\pi }\omega A_\sigma \left( {\bf k}%
,\omega \right) =\left\langle \left\{ \left[ c_{{\bf k}\sigma },\left( H-\mu
N\right) \right] ,c_{{\bf k}\sigma }^{\dagger }\right\} \right\rangle
=\epsilon _{{\bf k}}-\mu +Un_{-\sigma }  \label{A2}
\end{equation}
\begin{equation}
\int_{-\infty }^\infty \frac{d\omega }{2\pi }\omega ^2A_\sigma \left( {\bf k}%
,\omega \right) =(\epsilon _{{\bf k}}-\mu )^2+2U(\epsilon _{{\bf k}}-\mu
)n_{-\sigma }+U^2n_{-\sigma }  \label{A3}
\end{equation}
where $n_\sigma =n/2$ since we are in the paramagnetic state.

Using the spectral representation Eq.(\ref{SpectralG}) one can easily see
that the above sum rules give the coefficients of the high-frequency
expansion of the Matsubara Green's function 
\begin{equation}
\lim_{ik_n\rightarrow \infty }G_\sigma \left( {\bf k,}ik_n\right) =\frac 1{%
ik_n}+\left( \frac 1{ik_n}\right) ^2\int \frac{d\omega }{2\pi }\omega
A_\sigma \left( {\bf k},\omega \right) +\left( \frac 1{ik_n}\right) ^3\int 
\frac{d\omega }{2\pi }\omega ^2A_\sigma \left( {\bf k},\omega \right) +\ldots
\label{HighG}
\end{equation}
The self-energy has the same analytic properties as the Green's function.
Using its high frequency expansion in the expression for the Green's
function Eq.(\ref{Gself}), one finds that the first term in Eq.(\ref{HighG}%
), leads to the requirement that the self-energy has a finite limit at $%
ik_n\rightarrow \infty $. The second term fixes the value of this constant
to the Hartree-Fock result, and the last and second-term combine to give the
leading term in $1/ik_n\ $of the self-energy high-frequency expansion. In
short, we find the result quoted in Eq.(\ref{SelfHaut}), namely 
\begin{equation}
\lim_{ik_n\rightarrow \infty }\Sigma _\sigma \left( {\bf k,}ik_n\right)
=Un_{-\sigma }+\frac{U^2n_{-\sigma }\left( 1-n_{-\sigma }\right) }{ik_n}%
+\ldots  \label{SelfHaut2}
\end{equation}
The Kramers-Kronig relation for the self-energy 
\begin{equation}
\mathop{\rm Re}%
\left[ \Sigma _\sigma ^R\left( {\bf k,}\omega \right) -\Sigma _\sigma
^R\left( {\bf k,}\infty \right) \right] ={\cal P}\int \frac{d\omega ^{\prime
}}\pi \frac{%
\mathop{\rm Im}%
\left[ \Sigma _\sigma ^R\left( {\bf k,}\omega ^{\prime }\right) \right] }{%
\omega ^{\prime }-\omega }.  \label{KK-Re_Self}
\end{equation}
and the high-frequency result Eq.(\ref{SelfHaut2}) imply the following
sum-rule for the imaginary part of the self-energy 
\begin{equation}
-\int \frac{d\omega ^{\prime }}\pi 
\mathop{\rm Im}%
\left[ \Sigma _\sigma ^R\left( {\bf k,}\omega ^{\prime }\right) \right]
=U^2n_{-\sigma }\left( 1-n_{-\sigma }\right)
\end{equation}
Important consequences of this equation are that for a given $U$ the
integrated imaginary part of the self-energy is independent of temperature
and is increasing towards half-filling. The right-hand side of this equation
is also a measure the width of the single-particle excitation spectrum, as
can be seen from the spectral weight moments Eqs.(\ref{A2}) and (\ref{A3}), 
\begin{equation}
\overline{\omega ^2}-\overline{\omega }^2\equiv \int_{-\infty }^\infty \frac{%
d\omega }{2\pi }\omega ^2A_\sigma \left( {\bf k},\omega \right) -\left[
\int_{-\infty }^\infty \frac{d\omega }{2\pi }\omega A_\sigma \left( {\bf k}%
,\omega \right) \right] ^2=U^2n_{-\sigma }\left( 1-n_{-\sigma }\right)
\end{equation}

An important physical point is that the asymptotic behavior Eq.(\ref
{SelfHaut2}) is a necessary condition for the existence of upper and lower
Hubbard bands, as has been explained in Sec.\ref{HubbardBands}. However, it
is important to realize that it is not a sufficient condition. Indeed, the
following paradox has been noticed in explicit calculations in infinite
dimensions\cite{BandesHubbardDinf},\cite{Georges}. While ordinary
second-order perturbation theory with {\em bare} Green functions $G_{0\text{ 
}}$reproduces correctly the appearance of the Hubbard bands with increasing $%
U$, the perturbation theory with dressed Green function $G=[G_{0\text{ }%
}^{-1}-\Sigma ]^{-1}$ does not. The reason for this is that although the
second-order expression for $\Sigma _\sigma \left( {\bf k,}ik_n\right) $ in
terms of full $G$ does satisfy the asymptotics Eq.(\ref{SelfHaut2}), it sets
in too late, namely for $k_n\gg U$, instead of $k_n\gg W$. The fact that the
asymptotics should start at $k_n\sim W$ even when $U>W$ is a non-trivial
consequence of the Pauli principle, as explained in Sec.(\ref{HubbardBands}%
). Thus there are no Hubbard bands in any theory that uses self-consistent
Green functions but neglects the frequency dependence of the vertex. This is
an explicit example that illustrates what seems to be a more general
phenomena: a calculation with dressed Green's functions but no frequency
dependent vertex correction often gives worse results that the one done with
bare Green's functions and a frequency independent vertex. We will see in
the next subsection that this also happens in the calculation of the
two-particle properties. Also, as we have argued in Sec.(\ref{SecFLEX}), a
similar situation occurs with the precursors of antiferromagnetic bands in
the renormalized classical regime in two-dimensions.

Finally, we quote two more well known sum-rules that we will need. They
involve the Fermi function $f\left( \omega \right) $ and the spectral
weight. The first one follows from definition of $G_\sigma \left( {\bf k,}%
\tau \right) $ and the spectral representation 
\begin{equation}
\lim_{\tau \rightarrow 0^{-}}G_\sigma \left( {\bf k,}\tau \right) =\int 
\frac{d\omega }{2\pi }f\left( \omega \right) A_\sigma \left( {\bf k,}\omega
\right) =\left\langle c_{{\bf k}\sigma }^{\dagger }c_{{\bf k}\sigma
}\right\rangle \equiv n_{{\bf k}\sigma }.  \label{Occupation}
\end{equation}
The quantity $n_{{\bf k}\sigma }$ is the distribution function. It is equal
to the Fermi function only when the self-energy is frequency independent.
The next result, that follows simply from the equations of motion, 
\begin{equation}
\lim_{\tau \rightarrow 0^{-}}-\frac 1N\sum_{{\bf k}}\frac{\partial G_\sigma
\left( {\bf k,}\tau \right) }{\partial \tau }=\frac 1N\sum_{{\bf k}}\int 
\frac{d\omega }{2\pi }\omega f\left( \omega \right) A_\sigma \left( {\bf k,}%
\omega \right) =\frac 1N\sum_{{\bf k}}\left( \epsilon _{{\bf k}}-\mu \right)
n_{{\bf k}\sigma }+U\left\langle n_{\uparrow }n_{\downarrow }\right\rangle
\label{Energie}
\end{equation}
is useful to show to what extent certain dressed-propagator approaches fail
to satisfy the $f-$sum rule.

\subsection{Constraints on two-particle properties}

\label{SecConstTwoPart}

For any one-band model, independently of the Hamiltonian, the Pauli
principle (anticommutation relations) 
\begin{equation}
\left\langle n_{i\sigma }^2\right\rangle =\left\langle n_{i\sigma
}\right\rangle  \label{Pauli}
\end{equation}
implies the following two simple identities: 
\begin{equation}
\left\langle \left( n_{i\uparrow }\pm n_{i\downarrow }\right)
^2\right\rangle =n\pm 2\left\langle n_{i\uparrow }n_{i\downarrow
}\right\rangle .  \label{Simple}
\end{equation}
The correlation functions on the left-hand side are equal-time and
equal-position spin and charge correlation functions. The susceptibilities $%
\chi _{ch}\left( {\bf r}_i-{\bf r}_j,\tau \right) ,$ $\chi _{sp}\left( {\bf r%
}_i-{\bf r}_j,\tau \right) $ in Eqs.(\ref{SusCharge}) and (\ref{SusSpin})
are response functions for arbitrary $\left( {\bf r}_i-{\bf r}_j,\tau
\right) $ so they must reduce to the above equal-time equal-position
correlation functions when ${\bf r}_i={\bf r}_j$ and $\tau =0$. This is one
special case of the imaginary-time version of the fluctuation-dissipation
theorem Eqs.(\ref{SusSpin})(\ref{SusCharge}). This translates into
local-moment and local-charge sum-rules for the susceptibilities 
\begin{equation}
\frac TN\sum_{{\bf q}}\sum_{iq_n}\chi _{sp}\left( {\bf q},iq_n\right)
=2\left\langle n_{\uparrow }n_{\uparrow }\right\rangle -2\left\langle
n_{\uparrow }n_{\downarrow }\right\rangle =n-2\left\langle n_{\uparrow
}n_{\downarrow }\right\rangle  \label{SusSpinSum}
\end{equation}
\begin{equation}
\frac TN\sum_{{\bf q}}\sum_{iq_n}\chi _{ch}\left( {\bf q},iq_n\right)
=2\left\langle n_{\uparrow }n_{\uparrow }\right\rangle +2\left\langle
n_{\uparrow }n_{\downarrow }\right\rangle -n^2=n+2\left\langle n_{\uparrow
}n_{\downarrow }\right\rangle -n^2  \label{SusChargeSum}
\end{equation}
where we have removed the $i$ dependence of $\left\langle n_{i\uparrow
}n_{i\downarrow }\right\rangle $ using translational invariance. The
right-hand side of the local-moment sum-rule is equal to $\left\langle
\left( S^z\right) ^2\right\rangle ,$while that of the local-charge sum rule
is equal to $\left\langle \rho ^2\right\rangle -n^2$

If arbitrary sets of diagrams are summed, nothing can prevent the right-hand
side from taking unphysical values. For example, the Pauli principle may be
violated, {\it i.e. }$\left\langle n_{\uparrow }n_{\uparrow }\right\rangle
\neq \left\langle n_{\uparrow }\right\rangle .$ To see this, notice that
when the Pauli principle is satisfied, our two sum rules Eqs.(\ref
{SusSpinSum}) and (\ref{SusChargeSum}) lead to 
\begin{equation}
\frac TN\sum_{{\bf q}}\sum_{iq_n}\left[ \chi _{sp}\left( {\bf q},iq_n\right)
+\chi _{ch}\left( {\bf q},iq_n\right) \right] =2n-n^2.
\label{SpinChargePauli}
\end{equation}
It is easy to check that well known approaches to the many-body problem,
such as RPA, violate this basic requirement. Indeed, the ordinary RPA
expressions for spin and charge are 
\begin{equation}
\chi _{sp}^{RPA}(q)\equiv \frac{\chi _0}{1-\frac U2\chi _0}  \label{RPASp}
\end{equation}
\begin{equation}
\chi _{ch}^{RPA}(q)\equiv \frac{\chi _0}{1+\frac U2\chi _0}  \label{RPACh}
\end{equation}
where 
\begin{equation}
\chi _0\left( q\right) =-2\frac TN\sum_kG^{\left( 0\right) }\left( k\right)
G^{\left( 0\right) }\left( k+q\right) .  \label{Chi0}
\end{equation}
That RPA does not satisfy the sum rule Eq.(\ref{SpinChargePauli}) already to
second order in $U$ can be easily seen by expanding the denominators.

To satisfy the Mermin-Wagner theorem, approximate theories must also prevent 
$\left\langle n_{\uparrow }n_{\downarrow }\right\rangle $ from taking
unphysical values. This quantity is positive and bounded by its value for $%
U=\infty $ and its value for non-interacting systems, namely $0\leq
\left\langle n_{\uparrow }n_{\downarrow }\right\rangle \leq n^2/4$. Hence,
the right-hand side of the local-moment sum-rule Eq.(\ref{SusSpinSum}) is
contained in the interval $\left[ n,n-\frac 12n^2\right] .$ Any theory that
prevents the right-hand side of the local-moment sum rule from taking
infinite values satisfies the Mermin-Wagner theorem.

\begin{description}
\item[Proof:]  Near a magnetic phase transition, the zero
Matsubara-frequency component of the spin susceptibility takes the
Ornstein-Zernicke form 
\begin{equation}
\chi _{sp}\left( {\bf q+Q},0\right) \sim \frac 1{q^2+\xi ^{-2}}
\end{equation}
where $q$ is measured with respect to the ordering wave vector ${\bf Q}$ and
where $\xi ^2$ is the square of the correlation length. Near its maximum,
the above susceptibility is of order $\xi ^2$ while all finite
Matsubara-frequency components at the ordering wave vector are at most of
order $1/\left( 2\pi T\right) ^2$ which is much smaller than $\xi ^2$.
Hence, one can keep only the zero-Matsubara frequency contribution on the
left-hand side of the local-moment sum rule Eq.(\ref{SusSpinSum}) obtaining 
\begin{equation}
T\int \frac{d^d{\bf q}}{\left( 2\pi \right) ^d}\frac 1{q^2+\xi ^{-2}}=%
\widetilde{C}  \label{IntegLorentz=C}
\end{equation}
where $\widetilde{C}$ contains non-zero Matsubara frequency contributions as
well as $n-2\left\langle n_{\uparrow }n_{\downarrow }\right\rangle .$ Since $%
\widetilde{C}$ is finite, this means that in two dimensions $\left(
d=2\right) $, it is impossible to have $\xi ^{-2}=0$ on the left-hand side
otherwise the integral would diverge logarithmically.
\end{description}

Finally, the $f$-sum rule on spin and charge susceptibilities follows as
usual from the fact that the Hamiltonian conserves particle number.
Computing $\left. \left\langle \left[ \rho _{{\bf q}},\frac{\partial \rho _{-%
{\bf q}}}{\partial \tau }\right] \right\rangle \right| _{\tau =0}$ and $%
\left. \left\langle \left[ S_{{\bf q}},\frac{\partial S_{-{\bf q}}}{\partial
\tau }\right] \right\rangle \right| _{\tau =0}$one obtains for either charge
or spin 
\begin{equation}
\int \frac{d\omega }\pi \omega \chi _{ch,sp}^{\prime \prime }\left( {\bf q,}%
\omega \right) =\lim_{\eta \rightarrow 0}T\sum_{iq_n}\left( e^{-iq_n\eta
}-e^{iq_n\eta }\right) iq_n\chi _{ch,sp}\left( {\bf q},iq_n\right) =\frac 1N%
\sum_{{\bf k}\sigma }\left( \epsilon _{{\bf k+q}}+\epsilon _{{\bf k-q}%
}-2\epsilon _{{\bf k}}\right) n_{{\bf k}\sigma }.  \label{f2}
\end{equation}
As can be seen from the spectral representations of spin and charge
susceptibilities, Eq.(\ref{SpectCh}), the quantity that obeys the $f-$sum
rule is the coefficient of the leading term in the $1/q_n^2$ high-frequency
expansion of the susceptibilities.

The single-particle energies $\epsilon _{{\bf k}}$ entering explicitly the
right-hand side of the $f-$sum rule are independent of interactions, so
interactions influence the $f-$sum rule only very weakly through the $n_{%
{\bf k}\sigma }$. In fact, in a continuum $\epsilon _{{\bf k}}\propto {\bf k}%
^2$ so $n_{{\bf k}\sigma }$ enters only in the form $\sum_{{\bf k}\sigma }n_{%
{\bf k}\sigma }=n$ . In this case, the right-hand side of the $f-$sum rule
is proportional to $q^2n$ and hence is independent of interactions. On a
lattice however, the energies cannot in general be taken out of the sum and
interactions influence the value of the right-hand side, but only through
the fact that $n_{{\bf k}\sigma }$ differs from the non-interacting Fermi
function $f_{{\bf k}\sigma }$. At strong-coupling, where the self-energy is
strongly frequency dependent, this difference between $n_{{\bf k}\sigma }$
and $f_{{\bf k}\sigma }$ becomes important. But from weak to intermediate
coupling, calculations where $f_{{\bf k}\sigma }$ appears on the right-hand
side should be good approximations. In the explicit examples that we have
treated, the $U$ dependence of the $f-$sum rule becomes important only close
to half-filling and for $U>4$, signaling the breakdown of approximations
based on frequency-independent self-energies.

While RPA-like theories that use $f_{{\bf k}\sigma }$ instead of $n_{{\bf k}%
\sigma }$ violate only weakly the $f-$ sum rule in the weak to intermediate
coupling regime, self-consistent theories that use frequency-dependent
self-energies but no frequency-dependent vertices violate conservations laws
in general, and the $f-$ sum rule in particular, in a much more dramatic
way. The point is that susceptibilities with a dressed bubble, $\tilde{\chi}%
_{RPA}=\tilde{\chi}_0/(1-\frac 12U\tilde{\chi}_0),$ are bad approximations
because they have the following properties, for any value of $U$ 
\begin{equation}
\tilde{\chi}_{RPA}({\bf q}=0,iq_n\neq 0)\neq 0  \label{WardTriste}
\end{equation}
\begin{equation}
\int \frac{d\omega }{2\pi }\omega \tilde{\chi}_{RPA}^{\prime \prime }\left( 
{\bf q,}\omega \right) =\frac 1N\sum_{{\bf k,}\sigma }\left( \epsilon _{{\bf %
k+q}}+\epsilon _{{\bf k-q}}-2\epsilon _{{\bf k}}\right) n_{{\bf k}\sigma
}+4U\left( \left\langle n_{\uparrow }\right\rangle \left\langle
n_{\downarrow }\right\rangle -\left\langle n_{\uparrow }n_{\downarrow
}\right\rangle \right) .  \label{fTriste}
\end{equation}
The first of these equations explicitly violates the Ward identity, Eq.(\ref
{Ward}) below, at all frequencies, including small non-zero ones, since at
zero wave vector we should have $\chi ({\bf q}=0,iq_n\neq 0)=0$ for {\it all 
}frequencies except zero. The second equation, Eq.(\ref{fTriste}) violates
the $f$-sum rule Eq.(\ref{f2}) at {\it all wave }vectors{\it , }by a
constant term $4U\left( \left\langle n_{\uparrow }\right\rangle \left\langle
n_{\downarrow }\right\rangle -\left\langle n_{\uparrow }n_{\downarrow
}\right\rangle \right) $ which in practical calculations, say at $U=4$, is
of the same order as the first term, which is the only one that should be
there according to the $f$-sum rule.

\begin{description}
\item[Proof:]  Eqs.(\ref{WardTriste}) and (\ref{fTriste}) are proven as
follows. Consider the standard RPA expression but with dressed bubbles $%
\tilde{\chi}_0$ 
\begin{equation}
\tilde{\chi}_{RPA}=\tilde{\chi}_0/(1-\frac U2\tilde{\chi}_0).
\label{RPApseudo}
\end{equation}
Using the spectral representation for the Green's function and inversion
symmetry in the Brillouin zone one finds, 
\begin{equation}
\tilde{\chi}_0\left( {\bf q},iq_n\right) =\frac 2N\sum_{{\bf k}}\int \frac{%
d\omega }{2\pi }\int \frac{d\omega ^{\prime }}{2\pi }A\left( {\bf k,}\omega
\right) A\left( {\bf k+q,}\omega ^{\prime }\right) \frac{\left( \omega
-\omega ^{\prime }\right) \left( f\left( \omega ^{\prime }\right) -f\left(
\omega \right) \right) }{\left( \omega -\omega ^{\prime }\right) ^2+q_n^2}.
\end{equation}
When the bubble is not dressed, the spectral weights are delta functions so
that at ${\bf q}=0$ the susceptibility would vanish for all non-zero values
of $q_n$, as required by the Ward identity. However, here because the
spectral weight has a width and because the integrand is even and positive,
then the integral will not vanish, resulting in the first anomaly Eq.(\ref
{WardTriste}) we mention. To prove the second equation, Eq.(\ref{fTriste}),
it suffices to remember from the spectral representation of the
susceptibility Eq.(\ref{SpectCh}) and the derivation of the $f-$sum rule
Eqs.(\ref{f2}) that we are looking for the coefficient of the $1/q_n^2$ term
in the high-frequency expansion. Given the RPA form, Eq.(\ref{RPApseudo}),
only the numerator contributes to this limit. One obtains, for the
coefficient of the $1/q_n^2$ term, 
\begin{equation}
\frac 2N\sum_{{\bf k}}\int \frac{d\omega }{2\pi }\int \frac{d\omega ^{\prime
}}{2\pi }A\left( {\bf k,}\omega \right) A\left( {\bf k+q,}\omega ^{\prime
}\right) \left( \omega -\omega ^{\prime }\right) \left( f\left( \omega
^{\prime }\right) -f\left( \omega \right) \right)
\end{equation}
from which Eq.(\ref{fTriste}) follows using the sum rules for occupation
number Eq.(\ref{Occupation}) and for energy (\ref{Energie}).
\end{description}

Conservation laws have general consequences not only on equal-time
correlation functions, as in the $f-$sum rule above, but also on
time-dependent correlation functions. For example, from the Heisenberg
equations of motion and anti-commutation relations, follow the Ward
identities\cite{Dare1} 
\[
\sum_{{\bf k}}\sum_{\sigma =\pm 1}\sum_{\sigma ^{\prime }=\pm 1}\left( \frac %
\partial {\partial \tau }+\left( \epsilon _{{\bf k+q}}-\epsilon _{{\bf k}%
}\right) \right) \left\langle T_\tau c_{{\bf k}\sigma }^{\dagger }\left(
\tau \right) \sigma ^\ell c_{{\bf k+q}\sigma }\left( \tau \right) c_{{\bf k}%
^{\prime }+{\bf q}\sigma ^{\prime }}^{\dagger }\left( \tau _1\right) \sigma
^{\prime \ell }c_{{\bf k}^{\prime }\sigma ^{\prime }}\left( \tau _2\right)
\right\rangle 
\]
\begin{equation}
=\delta \left( \tau -\tau _1\right) \sum_{\sigma ^{\prime }=\pm 1}\sigma
^{\prime \ell }G_{\sigma ^{\prime }}\left( {\bf k}^{\prime },\tau _2-\tau
\right) -\delta \left( \tau -\tau _2\right) \sum_{\sigma ^{\prime }=\pm
1}\sigma ^{\prime \ell }G_{\sigma ^{\prime }}\left( {\bf k}^{\prime }+{\bf q}%
,\tau -\tau _1\right)  \label{Ward}
\end{equation}
where $\ell =0$ for charge, and $\ell =1$ for spin. The $f-$sum rule above,
Eq.(\ref{f2}), follows from the above identity by simply taking $\tau
_1=\tau _2^{+}$, summing over ${\bf k}^{\prime }$ and subtracting the two
results for $\tau \rightarrow \tau _1^{+}$ and $\tau \rightarrow \tau _1^{-}$%
.

We have seen in this section that there are strong cancelations for
two-particle properties between the frequency dependence of self-energy and
that of the vertex corrections, so that putting a frequency dependence in
only one of them is a bad approximation. We have adopted the Kadanoff-Baym
formalism in the main text since it can be used as a guide to make
approximations that satisfy conservation laws.

\subsection{When there is a Fermi surface}

\label{Lat}

When perturbation theory converges (no phase transition) then at zero
temperature $T=0$ the imaginary part of the self-energy vanishes, $\Sigma
_{\sigma }^{\prime \prime }\left( {\bf k,}\omega =0\right) =0$, for all $%
{\bf k}$ values and the Fermi surface defined by 
\begin{equation}
\epsilon _{{\bf k}}-\mu -\Sigma _{\sigma }^{\prime }\left( {\bf k,}\omega
=0\right) =0  \label{FS}
\end{equation}
encloses a volume that is equal to the volume enclosed by non-interacting
particles

\begin{equation}
\frac 1N\sum_{{\bf k}}\theta \left( \mu -\epsilon _{{\bf k}}-\Sigma _\sigma
^{\prime }\left( {\bf k,}0\right) \right) =\frac 1N\sum_{{\bf k}}\theta
\left( \mu _0-\epsilon _{{\bf k}}\right) =n_\sigma .  \label{Lat1}
\end{equation}

This is the content of Luttinger's theorem.\cite{Luttinger}\cite{AGD}. It
implies that there is a strong cancelation between the change of the
chemical potential and the change of the self-energy on the Fermi surface.
In particular, when $\Sigma _\sigma ^{\prime }\left( {\bf k}_F{\bf ,}%
0\right) $ does not depend on ${\bf k}$ or on the direction of ${\bf k}_F$
(infinite $D$ Hubbard model, electron gas) the change in $(\mu -\mu _0)$ is
exactly canceled by $\Sigma _\sigma ^{\prime }\left( {\bf k}_F{\bf ,}%
0\right) $

\begin{equation}
\mu -\mu _0=\Sigma _\sigma ^{\prime }\left( {\bf k}_F{\bf ,}0\right) .
\label{Lat2}
\end{equation}

Luttinger's theorem is satisfied when

\begin{equation}
\lim_{T\rightarrow 0}\int \frac{\partial \Sigma _\sigma \left( {\bf k,}i\nu
\right) }{\partial (i\nu )}G_\sigma \left( {\bf k,}i\nu \right) d\nu d{\bf k}%
=0.
\end{equation}
Any theory that calculates its self-energy from a functional derivative of
the Luttinger-Ward functional $\Sigma =\delta \Phi [G]/\delta G$ will
satisfy Luttinger's theorem\cite{Luttinger}\cite{AGD}. The latter procedure
requires self-consistent determination of the self-energy as a function of
momentum and frequency $\Sigma _\sigma \left( {\bf k,}ik_n\right) $ and is
usually quite computationally involved. However, even when this procedure to
calculate the self-energy is not followed, it turns out to be rather easy to
satisfy this theorem to an excellent degree of approximation in the weak to
intermediate coupling regime. The reason for this is that any
frequency-independent self-energy will preserve Luttinger's theorem and weak
frequency dependence will not cause great harm. For the electron gas,
Luttinger\cite{Luttinger} suggests a way to build a perturbation theory in
terms of non-interacting Green's functions which allows to satisfy
Luttinger's theorem to very good accuracy. The trick is that the chemical
potential for the interacting electrons $\mu $ should always enter the
calculations in the form of the difference with the shift of the self-energy
on the Fermi surface $\tilde{G}_0=1/[ik_n-\epsilon _{{\bf k}}+(\mu -\Sigma
_\sigma ^{\prime }\left( {\bf k}_F{\bf ,}0\right) )]$. The
``non-interacting'' Green's function $\tilde{G}_0$ in this formalism is the
Green's function of some effective non-interacting system and, in general,
it is different from both $1/\left( ik_n-\epsilon _{{\bf k}}+\mu \right) $
and $1/\left( ik_n-\epsilon _{{\bf k}}+\mu _0\right) $. However, when $%
T\rightarrow 0$ Luttinger's theorem requires that $(\mu -\Sigma _\sigma
^{\prime }\left( {\bf k}_F{\bf ,}0\right) )\rightarrow \mu _0$ and one can
approximate $\tilde{G}_0$ by the Green's function for a non-interacting
system of the same density $G_0=1/\left( ik_n-\epsilon _{{\bf k}}+\mu
_0\right) $. In practice, one can also have a phase transition (or
crossover) at a finite temperature $T_c$ ($T_X$). In these cases Luttinger's
theorem is satisfied only approximately since the zero-temperature limit
cannot be reached without a breakdown of perturbation theory. Then the
relevant question is how well it is satisfied at $T_c$ ($T_X$). (See also
Sec.(\ref{Single-particle}) for a discussion of Luttinger's theorem in our
approach)

When Luttinger's theorem holds, one can usually develop a Landau Fermi
liquid theory. In this approach, the Pauli principle is implemented only for
momentum states near the Fermi surface by imposing the forward scattering
sum rule. This sum rule, {\it in two dimensions}, reads 
\begin{equation}
\sum_\ell \left[ \frac{F_\ell ^s}{1+F_\ell ^s}+\frac{F_\ell ^a}{1+F_\ell ^a}%
\right] =0  \label{Forward}
\end{equation}
where $F_\ell ^s$ and $F_\ell ^a$ are the symmetric and antisymmetric Landau
parameters expanded on the $e^{-i\theta \ell }$ basis instead of the
Legendre polynomial basis. Recent renormalization group analysis has however
claimed\cite{Chitov} that the forward scattering sum rule comes from an
inaccurate use of crossing symmetry and is not the proper way to enforce the
Pauli principle. Most approaches to the many-body problem disregard this sum
rule anyway, in the same way that they disregard the local Pauli principle.

\section{Proofs of various formal results}

\label{Proofs}

In this appendix, we give the proofs of various relations mentioned in Secs.(%
\ref{SecFormal}) and (\ref{SubSecAccuracy}).

\begin{enumerate}
\item  The general expression for the self-energy Eq.(\ref{2a}) can be
obtained as follows. Use the equations of motion and the definition of the
self-energy Eqs.(\ref{Dyson1})(\ref{Dyson2}) which in the present notation
give 
\begin{equation}
\Sigma _\sigma \left( 1,\overline{1}\right) G_\sigma \left( \overline{1}%
,2\right) =-U\left\langle T_\tau \left[ \psi _{-\sigma }^{+}\left(
1^{++}\right) \psi _{-\sigma }\left( 1^{+}\right) \psi _\sigma \left(
1\right) \psi _\sigma ^{+}\left( 2\right) \right] \right\rangle
\end{equation}
\begin{equation}
=-U\left[ \frac{\delta G_\sigma \left( 1,2\right) }{\delta \phi _{-\sigma
}\left( 1,1^{+}\right) }-G_{-\sigma }\left( 1,1^{+}\right) G_\sigma \left(
1,2\right) \right]
\end{equation}
Substituting the equation for the three-point susceptibility (collective
modes) Eq.(\ref{1a}) in this last equation and multiplying on both sides by $%
G^{-1}$ proves\cite{BaymKadanoff} the expression Eq.(\ref{2a}) for the
self-energy.

\item  We now show that our approach satisfies the consistency requirement
between single-particle properties and collective modes in the form of Eq.(%
\ref{sumS}). Using our expression Eq.(\ref{param}) for $\Sigma ^{\left(
1\right) }$ and the definition of $\chi _0$ Eq.(\ref{Chi0}) we obtain 
\begin{equation}
\lim_{\tau \rightarrow 0^{-}}\frac TN\sum_k\Sigma _\sigma ^{\left( 1\right)
}\left( k\right) G_\sigma ^{\left( 0\right) }\left( k\right) e^{-ik_n\tau
}=Un_{-\sigma }^2-\frac U4\frac TN\sum_q\left[ U_{sp}\chi
_{sp}(q)+U_{ch}\chi _{ch}(q)\right] \frac{\chi _0\left( q\right) }2
\end{equation}
Using 
\begin{equation}
\chi _{sp}(q)-\chi _0\left( q\right) =\frac{U_{sp}}2\chi _0\left( q\right)
\chi _{sp}(q)
\end{equation}
\begin{equation}
\chi _0\left( q\right) -\chi _{ch}(q)=\frac{U_{ch}}2\chi _0\left( q\right)
\chi _{ch}(q)
\end{equation}
and the local moment Eq.(\ref{Spin}) and local charge Eq.(\ref{Charge}) sum
rules proves the result. The result is also obvious if we follow the steps
in the first part of this Appendix to deduce the self-energy expression Eq.(%
\ref{2}) using the collective mode equation Eq.(\ref{1}) adapted to our
approximation.
\end{enumerate}

\section{Ansatz for relation between $U_{sp}$ and $\langle n_{\uparrow
}n_{\downarrow }\rangle .$}

\label{Ansatz}

Using the present notation and formalism, we now give a physical derivation
of Eq.(\ref{Usp}) that is equivalent to the one already given using the
equations of motion approach\cite{Vilk}. (The latter derivation was inspired
by the local field approximation of Singwi {\it et al.}\cite{singwi}). Since
our considerations on collective modes are independent of the precise value
of the interaction $U$, we do have to use the equations of motion, or the
equivalent, to feed that information back in the definition of irreducible
vertices. The two irreducible vertices that we need are in principle
calculable from 
\begin{equation}
\Gamma _{\sigma \sigma ^{\prime }}\delta \left( 1-3\right) \delta \left(
2-4\right) \delta \left( 2-1^{+}\right) =\frac{\delta \Sigma _\sigma \left(
1,2\right) }{\delta G_{\sigma ^{\prime }}\left( 3,4\right) }=\frac{\delta
\left[ \Sigma _\sigma \left( 1,\overline{1}\right) G_\sigma \left( \overline{%
1},\overline{2}\right) G_\sigma ^{-1}\left( \overline{2},2\right) \right] }{%
\delta G_{\sigma ^{\prime }}\left( 3,4\right) }
\end{equation}
The rewriting on the right-hand side has been done to take advantage of the
fact that in the Hubbard model, the equations of motion (see Eqs.(\ref
{Dyson1})(\ref{Dyson2})) give us the product $\Sigma _\sigma \left( 1,%
\overline{1}\right) G_\sigma \left( \overline{1},\overline{2}\right) $ as
the highly local four field correlation function $-U\left\langle T_\tau
\left[ \psi _{-\sigma }^{+}\left( 1^{++}\right) \psi _{-\sigma }\left(
1^{+}\right) \psi _\sigma \left( 1\right) \psi _\sigma ^{+}\left( \overline{2%
}\right) \right] \right\rangle .$ Ordinary RPA amounts to a Hartree-Fock
factoring of this correlation function. Pursuing the philosophy that the
minimum number of approximations should be done on local correlation
functions, we do this factoring in such a way that it becomes exact when all
points are identical, namely when $\overline{2}=1^{+}$. In other words, we
write 
\begin{equation}
-U\left\langle T_\tau \left[ \psi _{-\sigma }^{+}\left( 1^{++}\right) \psi
_{-\sigma }\left( 1^{+}\right) \psi _\sigma \left( 1\right) \psi _\sigma
^{+}\left( \overline{2}\right) \right] \right\rangle \sim U\frac{\langle
n_{\uparrow }\left( 1\right) n_{\downarrow }\left( 1\right) \rangle }{%
\langle n_{\uparrow }\left( 1\right) \rangle \left\langle n_{\downarrow
}\left( 1\right) \right\rangle }G_{-\sigma }\left( 1,1^{+}\right) G_\sigma
\left( 1,\overline{2}\right)
\end{equation}
All quantities are evaluated as functionals of $G$ up to this point. We can
now evaluate the functional derivative 
\begin{equation}
\frac{\delta \Sigma _\sigma \left( 1,2\right) }{\delta G_{\sigma ^{\prime
}}\left( 3,4\right) }=\frac{\delta \left[ U\frac{\langle n_{\uparrow }\left(
1\right) n_{\downarrow }\left( 1\right) \rangle }{\langle n_{\uparrow
}\left( 1\right) \rangle \left\langle n_{\downarrow }\left( 1\right)
\right\rangle }G_{-\sigma }\left( 1,1^{+}\right) \delta \left( 1-2\right)
\right] }{\delta G_{\sigma ^{\prime }}\left( 3,4\right) }
\end{equation}
\begin{equation}
=\frac{\delta \left[ U\frac{\langle n_{\uparrow }\left( 1\right)
n_{\downarrow }\left( 1\right) \rangle }{\langle n_{\uparrow }\left(
1\right) \rangle \left\langle n_{\downarrow }\left( 1\right) \right\rangle }%
\right] }{\delta G_{\sigma ^{\prime }}\left( 3,4\right) }G_{-\sigma }\left(
1,1^{+}\right) \delta \left( 1-2\right) +U\frac{\langle n_{\uparrow }\left(
1\right) n_{\downarrow }\left( 1\right) \rangle }{\langle n_{\uparrow
}\left( 1\right) \rangle \left\langle n_{\downarrow }\left( 1\right)
\right\rangle }\frac{\delta G_{-\sigma }\left( 1,1^{+}\right) }{\delta
G_{\sigma ^{\prime }}\left( 3,4\right) }\delta \left( 1-2\right)
\end{equation}
The functional derivatives are now evaluated for the actual equilibrium
value of $G.$ Hence, we can use rotational invariance to argue that the
first term is independent of $\sigma $ and $\sigma ^{\prime }$ whereas the
last one is proportional to $\delta _{-\sigma ,\sigma ^{\prime }}.$ Since $%
U_{sp}=\Gamma _{\uparrow \downarrow }-\Gamma _{\uparrow \uparrow }$, only
this last term proportional to $\delta _{-\sigma ,\sigma ^{\prime }}$
contributes to $U_{sp}$. To obtain this term, it suffices to note that 
\begin{equation}
\frac{\delta G_{-\sigma }\left( 1,1^{+}\right) }{\delta G_{\sigma ^{\prime
}}\left( 3,4\right) }=\delta _{-\sigma ,\sigma ^{\prime }}\delta \left(
1-3\right) \delta \left( 4-1^{+}\right)
\end{equation}
and we obtain the desired result Eq.(\ref{Usp}) for $U_{sp}$.

\section{Real-frequency analysis of the self-energy and Fermi liquid limit}

\label{AppRealFrequ}

It is instructive to recover the two-dimensional result for precursors of
antiferromagnetic bands using the real-frequency formalism since it also
clarifies the limit in which the Fermi liquid result is recovered. Again we
neglect the contribution of charge fluctuations. Starting from our
expression for the self-energy Eq.(\ref{param}), one uses the spectral
representation for the susceptibility and for $G^{\left( 0\right) }$. The
Matsubara frequency sums can be then done and the result is trivially
continued to real frequencies.\cite{AGD2} One obtains, for the contribution
of classical and quantum spin fluctuations to the self-energy in $d$
dimensions 
\begin{equation}
\Sigma ^R\left( {\bf k},\omega \right) =\frac{UU_{sp}}4\int \frac{d^dq}{%
\left( 2\pi \right) ^d}\int \frac{d\omega ^{\prime }}\pi \left[ n\left(
\omega ^{\prime }\right) +f\left( \varepsilon _{{\bf k+q}}\right) \right] 
\frac{\chi _{sp}^{\prime \prime }\left( {\bf q},\omega ^{\prime }\right) }{%
\omega +i\eta +\omega ^{\prime }-\left( \varepsilon _{{\bf k}+{\bf q}}-\mu
_0\right) }  \label{SigmaReel}
\end{equation}
where $\mu _0=0$ at half-filling in the nearest-neighbor model and where $f$
is, as usual, the Fermi function, while $n\left( \omega \right) =\left(
e^{\beta \omega }-1\right) ^{-1}$ is the Bose-Einstein distribution. To
analyze this result in various limiting cases we need to know more about the
frequency dependence of the spin susceptibility. When the antiferromagnetic
correlation length is large, the zero-frequency result Eq.(\ref{ChiAs})
mentioned above can be generalized to 
\begin{equation}
\chi _{sp}^R({\bf q+Q}_d{\bf ,}\omega )\approx \xi ^2\frac 2{U_{sp}\xi _0^2}%
\left[ \frac 1{1+{\bf q}^2\xi ^2-i\omega /\omega _{SF}}\right]
\label{chiRPA}
\end{equation}
where, $\omega _{SF}=D/\xi ^2$ is the characteristic spin relaxation
frequency. In the notation of Ref.\cite{Dare}, the microscopic diffusion
constant $D$ is defined by 
\begin{equation}
\frac 1D\equiv \frac{\tau _0}{\xi _0^2}
\end{equation}
with the microscopic relaxation time, 
\begin{equation}
\tau _0=\frac 1{\chi _0\left( {\bf Q}_d\right) }\left. \frac{\partial \chi
_0^{^R}\left( {\bf Q}_d{\bf ,}\omega \right) }{\partial i\omega }\right|
_{\omega =0}.  \label{Gamma0}
\end{equation}
This relaxation-time is non-zero in models where the Fermi surface
intersects the magnetic Brillouin zone. Clearly, the frequency dependence of 
$\chi _{sp}^R({\bf q+Q}_d{\bf ,}\omega )$ is on a scale $\omega _{SF}=D/\xi
^2.$ The $1/\omega $ decrease of $\chi _{sp}^{\prime \prime }$ at
high-frequency is not enough to ensure that the real frequency version of
the local-moment sum rule is satisfied and the simplest way to cure this
problem is to introduce\cite{Millis2} a high-frequency cutoff $\Omega _{cut}$%
. The large correlation length makes the characteristic energy of the spin
fluctuations $\omega _{SF}$ a small number (critical slowing down). We
consider in turn two limiting cases.\cite{Crisan} The Fermi-liquid regime
appears for $\omega _{SF}\gg T$ and the non-Fermi liquid regime in the
opposite (renormalized classical) regime $\omega _{SF}\ll T.$

\subsection{Fermi liquid and nested Fermi liquid regime $\omega _{SF}\gg T$}

Perhaps the best known characteristic of a Fermi liquid is that $\Sigma
^{\prime \prime R}\left( {\bf k}_F,\omega ;T=0\right) \propto \omega ^2$ and 
$\Sigma ^{\prime \prime R}\left( {\bf k}_F,\omega =0;T\right) \propto T^2$.
To recover this result in the regime $\omega _{SF}\gg T$ far from phase
transitions, we start from the above expression Eq.(\ref{SigmaReel}) for the
self-energy to obtain 
\begin{equation}
\Sigma ^{\prime \prime R}\left( {\bf k}_F,\omega \right) =-\frac{UU_{sp}}4%
\frac 1{2v_F}\int \frac{d^{d-1}q_{\perp }}{\left( 2\pi \right) ^{d-1}}\int 
\frac{d\omega ^{\prime }}\pi \left[ n\left( \omega ^{\prime }\right)
+f\left( \omega +\omega ^{\prime }\right) \right] \chi _{sp}^{\prime \prime
}\left( q_{\perp },q_{\Vert }\left( q_{\perp },{\bf k}_F,\omega ,\omega
^{\prime }\right) ;\omega ^{\prime }\right)  \label{ImSigmaReel}
\end{equation}
where $q_{\Vert }$, the component of ${\bf q}$ parallel to the Fermi
momentum ${\bf k}_F$, is obtained from the solution of the equation 
\begin{equation}
\varepsilon _{{\bf k}+{\bf q}}-\mu _0=\omega +\omega ^{\prime }
\label{Condition_q_par}
\end{equation}
The key to understanding the Fermi liquid {\it vs} non-Fermi liquid regime
is in the relative width in frequency of $\chi _{sp}^{\prime \prime }\left( 
{\bf q,}\omega ^{\prime }\right) /\omega ^{\prime }$ {\it vs} the width of
the combined Bose and Fermi functions. In general, the function $n\left(
\omega ^{\prime }\right) +f\left( \omega +\omega ^{\prime }\right) $ depends
on $\omega ^{\prime }$ on a scale $Max\left( \omega ,T\right) $ while far
from a phase transition, the {\it explicit} frequency dependence of $\chi
_{sp}^{\prime \prime }\left( {\bf q,}\omega ^{\prime }\right) /\omega
^{\prime }$ is on a scale $\omega _{SF}\sim E_F\gg T$. Hence, in this case
we can assume that $\chi _{sp}^{\prime \prime }\left( {\bf q,}\omega
^{\prime }\right) /\omega ^{\prime }$ is a constant in the frequency range
over which $n\left( \omega ^{\prime }\right) +f\left( \omega +\omega
^{\prime }\right) $ differs from zero. Also, since $\chi _{sp}^{\prime
\prime }\left( {\bf q,}\omega ^{\prime }\right) /\omega ^{\prime }$ depends
on wave vector ${\bf q}$ over a scale of order $q_F,$ one can neglect the $%
\omega +\omega ^{\prime }$ dependence of $q_{\Vert }$ obtained from Eq.(\ref
{Condition_q_par}). Hence, we can approximate our expression Eq.(\ref
{ImSigmaReel}) for $\Sigma ^{\prime \prime R}$ by 
\begin{equation}
\Sigma ^{\prime \prime R}\left( {\bf k}_F,\omega \right) \simeq -\frac{%
UU_{sp}}4\frac{A\left( {\bf k}_F\right) }{2v_F}\int \frac{d\omega ^{\prime }}%
\pi \left[ n\left( \omega ^{\prime }\right) +f\left( \omega +\omega ^{\prime
}\right) \right] \omega ^{\prime }=-\frac{UU_{sp}}4\frac{A\left( {\bf k}%
_F\right) }{4v_F}\left[ \omega ^2+\left( \pi T\right) ^2\right]
\label{ImSigmaFermiLiquid}
\end{equation}
where the substitution $x=e^{\beta \omega \text{ }}$allowed the integral to
be done exactly and where 
\begin{equation}
A\left( {\bf k}_F\right) \equiv \int \frac{d^{d-1}q_{\perp }}{\left( 2\pi
\right) ^{d-1}}\lim_{\omega \rightarrow 0}\frac{\chi _{sp}^{\prime \prime
}\left( q_{\perp },q_{\Vert }\left( q_{\perp },{\bf k}_F,0,0\right) ;\omega
^{\prime }\right) }{\omega ^{\prime }}
\end{equation}

In general, $A$ depends on the orientation of the Fermi wave vector, $%
\widehat{{\bf k}}_F$, because it determines the choice of parallel and
perpendicular axis $q_{\perp },q_{\Vert }$. The above result Eq.(\ref
{ImSigmaFermiLiquid}) for $\Sigma ^{\prime \prime R}$ is the well known
Fermi liquid result.

There are known corrections to the Fermi liquid self-energy that come from
the non-analytic $\omega ^{\prime }/v_Fq$ behavior of $\chi _{sp}^{\prime
\prime }\left( {\bf q,}\omega ^{\prime }\right) /\omega ^{\prime }$ near the
ferromagnetic (zone center) wave vector. In three dimensions\cite
{BaymPethick} this non-analyticity leads to subdominant $\omega ^3\ln \omega 
$ corrections, while in two dimensions it leads to the dominant $\omega
^2\ln \omega $ behavior.\cite{StampSingular}\cite{Wilkins} In the case under
consideration, the antiferromagnetic contribution has a larger prefactor.
Even when it dominates however, it can also lead to non-analyticities in the
case of a nested Fermi surface. Indeed, we note that 
\begin{equation}
\text{Im}\chi _0^R({\bf Q}_d,\omega )=\pi {N}_d(\frac \omega 2)\text{tanh}%
\left( \frac \omega {4T}\right) .
\end{equation}
In two dimensions, the logarithmic divergence of the density of states ${N}%
_d(\frac \omega 2)$ at the van Hove singularity makes the zero-frequency
limit of the microscopic relaxation time Eq.(\ref{Gamma0}) ill-defined,
because of the logarithmic divergence at $\omega =0$. However, this leads
only to logarithmic corrections. If we drop logarithmic dependencies then
for $\omega <T$ one has $\left. \partial \chi _0^R\left( {\bf Q}_d{\bf ,}%
\omega \right) /di\omega \right| _{\omega \sim T}\sim 1/T$ and this $1/T$
dependence of $\left. \partial \chi _0^R\left( {\bf Q}_d{\bf ,}\omega
\right) /di\omega \right| _{\omega =0}$ changes the temperature dependence
of $\Sigma ^{\prime \prime R}\left( {\bf k}_F,0\right) $ from $T^2$ to $T$
as discussed in the ``Nested Fermi Liquid'' approach.\cite{Ruvalds}

\subsection{Non-Fermi liquid regime $\omega _{SF}\ll T$}

Near an antiferromagnetic phase transition, the spin-fluctuation energy
becomes much smaller than temperature. This is the renormalized classical
regime. The condition $\omega _{SF}\ll T$ means that $\chi _{sp}^{\prime
\prime }\left( q_{\perp },q_{\Vert };\omega ^{\prime }\right) $ is peaked
over a frequency interval $\omega ^{\prime }\ll T$ much narrower than the
interval $\omega ^{\prime }\sim T$ over which $n\left( \omega ^{\prime
}\right) +f\left( \omega +\omega ^{\prime }\right) $ changes. This situation
is the opposite of that encountered in the Fermi liquid regime. To evaluate $%
\Sigma ^{\prime \prime R}$ Eq.(\ref{ImSigmaReel}) the Fermi factor can now
be neglected compared with the classical limit of the Bose factor, $T/\omega
^{\prime }$. Then the dominant contribution to $\Sigma ^{\prime \prime
R}\left( {\bf k}_F,\omega \right) $ is from classical spin fluctuations $%
T\int \frac{d\omega ^{\prime }}\pi \frac 1{\omega ^{\prime }}\chi
_{sp}^{\prime \prime }=T\chi _{sp}^{\prime }\simeq S_{sp}$ as we see below.
More specifically, we take into account that the integral is peaked near $%
{\bf Q=}\left( \pi ,\pi \right) $ and measure wave vector with respect to
the zone center. For simplicity we consider below the half-filled case $\mu
_0=0$. Then, with the help of $\varepsilon _{{\bf k}+{\bf q+Q}}=-\varepsilon
_{{\bf k}+{\bf q}}$ we approximate the equation for $q_{\Vert }$ Eq.(\ref
{Condition_q_par}) by $v_Fq_{\Vert }=-\left( \omega +\omega ^{\prime
}\right) .$ This gives us for Eq.(\ref{ImSigmaReel}) the approximation 
\begin{equation}
\Sigma ^{\prime \prime R}\left( {\bf k}_F,\omega \right) \approx -\frac{%
UU_{sp}}4\frac 1{2v_F}\int \frac{d^{d-1}q_{\perp }}{\left( 2\pi \right)
^{d-1}}\int \frac{d\omega ^{\prime }}\pi \frac T{\omega ^{\prime }}\chi
_{sp}^{\prime \prime }\left( q_{\perp },q_{\Vert }=-\frac{\omega +\omega
^{\prime }}{v_F};\omega ^{\prime }\right)  \label{SigmaReelRC}
\end{equation}

The dependence of $\chi _{sp}^{\prime \prime }$ on $\omega ^{\prime }$
through $q_{\Vert }=-\left( \omega +\omega ^{\prime }\right) /v_F$ may be
neglected because $q_{\Vert }$ appears only in the combination $\left( \xi
^{-2}+q_{\perp }^2+q_{\Vert }^2\right) $ and in the regime $\omega _{SF}\ll
T $ we have $\omega ^{\prime }/v_F<\omega _{SF}/v_F\sim D\xi ^{-2}/v_F\ll
\xi ^{-1}.$ The latter inequality is generically satisfied when $\xi
^{-1}\ll 1$. Using 
\begin{eqnarray}
T\int \frac{d\omega ^{\prime }}\pi \frac 1{\omega ^{\prime }}\chi
_{sp}^{\prime \prime }\left( q_{\perp },q_{\Vert }=-\frac \omega {v_F}%
;\omega ^{\prime }\right) &=&T\chi _{sp}^{\prime }\left( q_{\perp },q_{\Vert
}=-\frac \omega {v_F};iq_n=0\right) \\
&=&\frac 2{U_{sp}\xi _0^2}\frac T{\xi ^{-2}+q_{\perp }^2+\left( \frac \omega
{v_F}\right) ^2}
\end{eqnarray}
the above equation Eq.(\ref{SigmaReelRC}) for $\Sigma ^{\prime \prime
R}\left( {\bf k}_F,\omega \right) $ reduces precisely to the classical
contribution found using imaginary-time formalism Eq.(\ref{Sigm}). As we saw
in Sec.(\ref{SecRC}), when the condition $\xi >\xi _{th}$ is satisfied, then
this contribution is dominant and leads to $\lim_{T\rightarrow 0}\Sigma
^{\prime \prime R}\left( {\bf k}_F,0\right) \rightarrow \infty .$

\section{Expanded discussion of other approaches}

\label{DiscussionOtherApproaches}

This appendix expands on Sec.(\ref{SecComparisons}) to discuss in detail
various theories, explaining the advantages and disadvantages of each in the
context of the sets of constraints described in Appendices~(\ref
{SubSecConstSingle}) and (\ref{SecConstTwoPart}).

\subsection{Paramagnon theories}

\label{Paramagnon theories}

In standard paramagnon theories,\cite{Stamp}\cite{Enz} the spin and charge
fluctuations are computed by RPA, using either bare or dressed Green's
functions. Then the fluctuations are fed back in the self-energy. In fact
there is a whole variety of paramagnon theories. They are largely
phenomenological. The reader is referred to Ref.\cite{Stamp} for a review.
We concentrate our discussion on recent versions\cite{BulutParamagnon} of
the so-called Berk-Schrieffer formula\cite{Berk}. In this approach, infinite
subsets of diagrams are summed and bare propagators are used in the
calculation of both the susceptibilities and the self-energy, the latter
being given by 
\begin{equation}
\Sigma _\sigma ^{BS}\left( k\right) =Un_{-\sigma }+\frac U4\frac TN%
\sum_q\left[ \left( 3U\chi _{sp}^{RPA}(q)-2U\chi _0(q)\right) +U\chi
_{ch}^{RPA}(q)\right] G_\sigma ^0(k+q).  \label{BerkS}
\end{equation}
The RPA spin and charge susceptibilities have been defined in Eqs.(\ref
{RPASp})(\ref{RPACh}). Comparing with our self-energy formula Eq.(\ref{param}%
), it is clear that here there is no vertex correction. In addition, the
factor of three in front of the spin susceptibility in Eq.(\ref{BerkS}) is
supposed to take into account the presence of both longitudinal and
transverse spin waves and the subtracted term is to avoid double-counting
the term of order $U^2$.

We can now see the advantages and disadvantages of this approach. First,
note that the susceptibilities entering the Berk-Schrieffer formula are the
RPA ones. As we saw in Appendix(\ref{SecSumRules}), these fail to satisfy
both the local Pauli principle and the Mermin-Wagner theorem. Hence,
spurious phase transitions will influence the self-energy in uncontrollable
ways. The collective modes do however satisfy conservation laws since they
are obtained with bare vertices and Green's functions containing a constant
self-energy. The $f$-sum rule Eqs.(\ref{f2}) then is satisfied without
renormalization of the distribution function $n_{{\bf k}}$ because the
zeroth order self-energy is constant. This is all in agreement with the
definition of a conserving approximation for the collective modes.

The high-energy asymptotics of the self-energy sets in at the correct energy
scale $k_n>W$ in this approach, but the second term of the large-frequency
asymptotics is incorrect. Indeed, at large values of $ik_n$, 
\begin{equation}
\lim_{ik_n\rightarrow \infty }\Sigma _\sigma ^{BS}\left( k\right)
=Un_{-\sigma }+\frac U{4ik_n}\frac TN\sum_q\left[ 3U\chi
_{sp}^{RPA}(q)+U\chi _{ch}^{RPA}(q)-2U\chi _0(q)\right] +\ldots
\end{equation}
and the sums can be evaluated as follows using the fluctuation-dissipation
theorem 
\begin{equation}
\frac TN\sum_q\chi _{sp}^{RPA}(q)=2\left\langle n_{\uparrow }n_{\uparrow
}\right\rangle -2\left\langle n_{\uparrow }n_{\downarrow }\right\rangle
\end{equation}
\begin{equation}
\frac TN\sum_q\chi _{ch}^{RPA}(q)=2\left\langle n_{\uparrow }n_{\uparrow
}\right\rangle +2\left\langle n_{\uparrow }n_{\downarrow }\right\rangle -n^2
\end{equation}
\begin{equation}
\frac TN\sum_q\chi _0(q)=n-\frac{n^2}2
\end{equation}
The correlators on the right-hand side take their RPA value so they do not
satisfy the Pauli principle, {\it i.e.} $\left\langle n_{\uparrow
}n_{\uparrow }\right\rangle \neq \left\langle n_{\uparrow }\right\rangle .$
Taking these results together we have 
\begin{equation}
\lim_{ik_n\rightarrow \infty }\Sigma _\sigma ^{BS}\left( k\right)
=Un_{-\sigma }+\frac{U^2}{ik_n}\left[ 2\left\langle n_{\uparrow }n_{\uparrow
}\right\rangle -\left\langle n_{\uparrow }n_{\downarrow }\right\rangle -%
\frac n2\right] +\ldots
\end{equation}
This does not gives the correct asymptotic behavior Eq.(\ref{SelfHaut}) even
if the Pauli principle $\left\langle n_{\uparrow }n_{\uparrow }\right\rangle
=\left\langle n_{\uparrow }\right\rangle $ were satisfied, because $%
\left\langle n_{\uparrow }n_{\downarrow }\right\rangle $ depends on the
interaction $U$.

The paramagnon self-energy Eq.(\ref{BerkS}) also does not satisfy the
consistency requirement Eq.(\ref{sumSigma1}) between self-energy and
collective modes imposed by the equations of motion. To see this we first
note that 
\begin{equation}
\lim_{\tau \rightarrow 0^{-}}\frac TN\sum_k\Sigma _\sigma ^{BS}\left(
k\right) G_\sigma ^{\left( 0\right) }\left( k\right) e^{-ik_n\tau
}=Un_{-\sigma }^2-\frac{U^2}8\frac TN\sum_{{\bf q}}\left[ 3\chi
_{sp}^{RPA}(q)+\chi _{ch}^{RPA}(q)-2\chi _0(q)\right] \chi _0\left( q\right)
\label{sumSBerk}
\end{equation}
Using this expression in the sum-rule Eq.(\ref{sumSigma1}) which relates one
and two-particle correlators and expanding both sides of this sum-rule in
powers of $U$, one finds that it is satisfied only up to order $U^2$. On the
other hand, if one replaces $3\chi _{sp}-2\chi _0$ in Eq.(\ref{BerkS}) by $%
\chi _{sp}$, the sum-rule Eq.(\ref{sumSigma1}) is satisfied to all orders in 
$U$. In our opinion, the problem of enforcing rotational invariance in
approximate theories is highly non-trivial and cannot be solved simply by
adding factor of $3$ in front of $\chi _{sp}$ and then subtracting $2\chi _0$
to avoid double counting. For more detailed discussions see Ref.\cite{Bresil}
and the comments at the end of Sec.(\ref{Single-particle}).

Luttinger's theorem is trivially satisfied if the occupation number is
calculated with the initial constant self-energy since it gets absorbed in
the chemical potential. If the occupation number is calculated with the
Green's function that contains the Berk-Schrieffer self-energy then
Luttinger's theorem is in general violated. It is advisable to use a new
chemical potential.

\subsection{Conserving approximations (FLEX)}

\label{SubSecFLEX}

In the conserving approximation schemes\cite{Baym}, one takes any physically
motivated subset of skeleton diagrams to define a Luttinger-Ward functional $%
\Phi $. Skeleton diagrams contain fully dressed Green's functions and no
self-energy insertions. This functional is functionally differentiated to
generate a self-energy that is then calculated self-consistently since it
appears implicitly in the Green's functions used in the original set of
diagrams. A further functional differentiation allows one to calculate the
irreducible vertices necessary to obtain the collective modes in a way that
preserves Ward identities. If one uses for the free energy the formula 
\begin{equation}
\ln Z=\text{Tr}\left[ \ln \left( -G\right) \right] +\text{Tr}\left( \Sigma
G\right) -\Phi  \label{ZBaym}
\end{equation}
then one obtains thermodynamic consistency in the sense that thermodynamic
quantities obtained by derivatives of the free energy are identical to
quantities computed directly from the single-particle Green's function. For
example, particle number can be obtained either from a trace of the Green's
function or from a chemical potential derivative of the free energy. In this
scheme, Luttinger's theorem is satisfied as long as perturbation theory
converges since then any initial guess for the Luttinger-Ward functional
will satisfy Luttinger's theorem.

FLEX refers to a particular choice of diagrams for $\Phi .$ This choice
leads to the following self-consistent expression for the self-energy 
\begin{equation}
\Sigma _\sigma ^{BS}\left( k\right) =Un_{-\sigma }+\frac U4\frac TN%
\sum_q\left[ \left( 3U\tilde{\chi}_{sp}^{RPA}(q)-2U\tilde{\chi}_0(q)\right)
+U\tilde{\chi}_{ch}^{RPA}(q)\right] G_\sigma (k+q).  \label{BerkSFL}
\end{equation}
This expression for the self-energy does not contain vertex corrections,
despite the fact that, contrary to the electron-phonon case, Migdal's
theorem does not apply here. We have explained in detail in Sec.(\ref
{SubSecWhyFLEX}) why this may lead to qualitatively wrong results, such as
the absence of precursors of antiferromagnetic bands and of the pseudogap in 
$A(\vec{k}_F,\omega )$ in two dimensions.

Another drawback of this approach is that it does not satisfy the Pauli
principle in any form, either local or through crossing symmetry\cite
{StampThese}. Indeed, one would need to include all exchange diagrams to
satisfy it. In practice this is never done. In the same way that there is
nothing to constrain the value of $\left\langle n_{\uparrow }n_{\uparrow
}\right\rangle $ obtained by the fluctuation-dissipation theorem to be equal
to $\left\langle n_{\uparrow }\right\rangle $, there is nothing to
explicitly constrain the value of $\left\langle n_{\uparrow }n_{\downarrow
}\right\rangle $. Nevertheless, the Mermin-Wagner theorem is believed to be
satisfied in FLEX because the feedback through the self-energy tends to
prevent the divergence of fluctuations in low dimension.\cite{BickersPrive},%
\cite{Serene}. Physically however, this seems to be an artificial way of
satisfying the Mermin-Wagner theorem since this theorem should be valid even
in localized spin systems where single-particle properties are negligibly
influenced by thermal fluctuations. We also point out that the proof of the
Mermin-Wagner theorem in $n\rightarrow \infty $ models implies that the
finite temperature phase transition in two dimensions is not simply removed
by thermal fluctuations, but that it is replaced by a crossover to the
renormalized classical regime with exponentially growing susceptibility. The
fact that the conserving susceptibility in FLEX does not show such behavior%
\cite{Serene} means that FLEX is actually inconsistent with the generic
phase space arguments responsible for the absence of finite-temperature
phase transition in two dimensions. The case of one dimension also suggests
that collective modes by themselves should suffice to guarantee the
Mermin-Wagner theorem without feedback on single-particle properties.
Indeed, in one dimension one shows by diagrammatic methods (parquet
summation or renormalization) that the zero-temperature phase transition is
prohibited at the two-particle level even {\it without} self-energy effects%
\cite{geo}.

Although, the second-order diagram is included correctly in FLEX, it does
not have the correct coefficient in the $1/ik_n$ expansion of the
self-energy. More importantly, the high-frequency behavior sets-in too late
to give the Hubbard bands, as we have explained in Sec.(\ref{SubSecWhyFLEX}%
). We have also seen a case where FLEX, as judged from comparisons with
Monte Carlo simulations (Fig.(1a) of Ref.\cite{Vilk2}), does not reproduce
the results of second-order perturbation theory even when it is a good
low-energy approximation.

One of the inconsistencies of conserving approximations that is seldom
realized, is that the self-energy is inconsistent with the collective modes.
In other words, the consistency formula Eq.(\ref{Consistency}) is not
satisfied in the following sense. The explicit calculation of $\Sigma G$
leads to an estimate of{\it \ }$U\left\langle n_{\uparrow }n_{\downarrow
}\right\rangle $ that differs from the one obtained by applying the
fluctuation-dissipation theorem to the {\it conserving }spin and charge
susceptibilities.

\subsection{Pseudo-potential parquet approach}

In the parquet approach, one enforces complete antisymmetry of the four
point function by writing down fully crossing-symmetric equations for these.
There are three irreducible vertices, namely one for the particle-particle
channel, and one for each of the two particle-hole channels. They obey the
so-called parquet equations.\cite{MartinDominicis} The Green's functions are
dressed by a self-energy which itself contains the four point function. In
this way, self-consistency between one-particle and two-particle quantities
is built-in. Solutions are possible for the one-impurity problem\cite
{Gavoret} and in one-dimension\cite{geo}. However, to solve the parquet
equations in higher dimension with presently available computing power is
impossible. Bickers {\it et al.}\cite{parquet}\cite{FLEX-parquet} have
formulated the parquet equations as a systematic improvement over FLEX and
have devised a way to do practical calculations by introducing so-called
pseudo-potentials. Since the main computational difficulty is in keeping the
full momentum and frequency dependence of the four point functions entering
the calculation of the self-energy, this is where the various fluctuations
channels are approximated by RPA-like forms Eq.(\ref{RPApseudo}) but with
fully dressed propagators and an effective interaction (pseudo-potential)
instead of $U$. A different strategy is under development.\cite{BickersPrive}
The criticism of the present section applies only to the current
pseudo-potential parquet approach.\cite{parquet}\cite{FLEX-parquet}

It can be seen that one drawback of this approach at the physical level is
that the use of constant effective interactions with dressed single-particle
propagators means that the fluctuations used in the calculation of the
self-energy do not satisfy conservation laws, as we just demonstrated in Sec(%
\ref{SecConstTwoPart}). Furthermore, the pseudopotentials are determined by
asking that the susceptibilities extracted from the four-point functions in
the parquet equations match the corresponding RPA-pseudo-potential
susceptibility at only one wave vector and frequency. The choice of this
matching point is arbitrary: Should the match be done for the typical, the
average, or the maximal value of the susceptibility in the Brillouin zone?

As we have seen in Sec.(\ref{SecFLEX}), even if the expression for the
self-energy in this approach explicitly has the second-order perturbation
theory diagram in it, this is not sufficient to ensure that the correct high
frequency asymptotic behavior starts at the appropriate frequency scale $%
ik_n\sim W$. Nevertheless, in many cases the results of the calculations
performed with this approach are not so different from second-order
perturbation theory, as can be seen from Fig.(1) of Ref.(\cite{Vilk2}).

Going rapidly through the rest of our list of properties, we see that the
consistency requirement $\Sigma _\sigma \left( 1,\overline{1}\right)
G_\sigma \left( \overline{1},1^{+}\right) =U\left\langle n_{\uparrow
}n_{\downarrow }\right\rangle $ is at least approximately built-in by
construction. Concerning the local-moment sum-rule and the Mermin-Wagner
theorem, it has been shown that the so-called ``basic'' parquet equations
should have the same critical behavior as the leading term in the $1/N$
expansion,\cite{Bickers1/N} and hence should satisfy the Mermin-Wagner
theorem.\cite{BickersPrive} The pseudo-potentials should not affect the
self-consistency necessary to satisfy the Mermin-Wagner theorem but the fact
that they are matched at a single point might introduce difficulties,
especially if the wave-vector at which $\chi _{sp}$ becomes unstable is
unknown from the start. As far as the Pauli principle is concerned, it
should be at least approximately satisfied both locally and in momentum
space. Nothing however in the approach enforces conservation laws.

\subsection{Present approach}

The role of the above sum-rules in our approach has been discussed in detail
in the main text. Here we will discuss only a few additional points.

If we concentrate on the ${\bf q}=0$ properties, our spin and charge
correlations behave as a special case of the ``local Fermi liquid'' defined
in Ref.\cite{Bedell}. A ``local Fermi liquid'' is a description of ${\bf q}%
=0 $ properties that applies when the self-energy, and consequently
irreducible vertices, depend only on frequency, not on momentum. In a local
Fermi liquid there are only two Landau parameters, which in our case are $%
F_0^a=-U_{sp}\chi _0\left( 0^{+},0\right) /2$ and $F_0^s=U_{ch}\chi _0\left(
0^{+},0\right) /2.$ Unitarity and the forward scattering sum rule, if valid,
imply that there is no ferromagnetism in the repulsive case\cite{Bedell}, as
we have found. One can check explicitly that the forward scattering sum rule
is satisfied to within about $15\%$ in our usual Monte Carlo parameter
range. However, as discussed in Appendix (\ref{Lat}), the forward scattering
sum-rule refers only to wave vectors on the Fermi surface, not to the local
version of the Pauli principle. Furthermore, the validity of this sum rule
has been questioned.\cite{Chitov} The effective mass at this level of
approximation is the bare one, as in a transitionally invariant {\it local}
Fermi liquid.\cite{Bedell} Recall however that our microscopic calculations
are not phenomenological: they explicitly give a value for the Landau
parameters. Also, our results extend well beyond the ${\bf q}=0$ quantities
usually considered in Fermi liquid theory.

The quasi-particle weight $Z$ calculated with $\Sigma _\sigma ^{\left(
1\right) }$ can differ substantially from the initial one. This means that
if we were to calculate the susceptibility with the corresponding frequency
and momentum dependent irreducible vertices $\Gamma ^{\left( 1\right) }$
there would be sizeable compensation between vertices and self-energy
because our calculations with $\Sigma _\sigma ^{\left( 0\right) }$ ($Z=1$)
and constant renormalized vertices already gave excellent agreement with
Monte Carlo simulations.

Finally, consider the high-frequency asymptotics. Since we use bare
propagators, the high-frequency asymptotics comes in at the appropriate
frequency scale, namely $ik_n\sim W$ and the Hubbard bands do exist in our
theory. However, the coefficient of proportionality in front of the
asymptotic form $1/ik_n$ is incorrect. Using Eqs.(\ref{param}),(\ref
{SusSpinSum}),(\ref{SusChargeSum}) we can write the high-frequency
asymptotics in the following form 
\begin{equation}
\lim_{ik_n\rightarrow \infty }\Sigma _\sigma \left( {\bf k,}ik_n\right)
=Un_{-\sigma }+\frac U{ik_n}\left[ \left( \frac{U_{sp}+U_{ch}}2\right)
\left\langle n_{-\sigma }^2\right\rangle -U_{ch}n_{-\sigma }^2+\left( \frac{%
U_{sp}-U_{ch}}2\right) \left\langle n_{\uparrow }n_{\downarrow
}\right\rangle \right] +\ldots  \label{OG}
\end{equation}
This form is useful to understand what is necessary to obtain the
quantitatively correct high-frequency behavior. Indeed, one would recover
the exact result Eq.(\ref{SelfHaut}), if one were to take into account that:
i) the irreducible vertices become equal to the bare one $U$ at
high-frequencies; ii) the local Pauli principle $\left\langle \hat{n}%
_{-\sigma }^2\right\rangle =n_{-\sigma }$ is satisfied. Contrary to most
other approaches, our theory does satisfy the local Pauli principle Eq.(\ref
{Pauli}) exactly. However, since our irreducible vertices are constant and
tuned to describe the low energy physics, we violate the first of the above
requirements. It is thus clear that for a correct quantitative description
of both the low energy physics and the Hubbard bands one needs to work with
frequency-dependent irreducible vertices.


\begin{references}
\bibitem{NoteSelf0}  $\Sigma _\sigma ^{\left( 0\right) }=\Gamma _{\sigma
-\sigma }G_{-\sigma }\left( \overline{1},\overline{1}^{+}\right) +\Gamma
_{\sigma \sigma }G_\sigma \left( \overline{1},\overline{1}^{+}\right) .$
Note that at equilibrium, this initial guess for the self-energy has a
Hartree-Fock from but with $U$ replaced by $U_{ch}$, clearly only a
low-frequency approximation to the self-energy.
\end{references}
\end{document}